\documentclass[12pt]{article}
\usepackage{lineno}
\usepackage{amsbsy,amsmath,amsthm,latexsym,amssymb}
\usepackage{graphicx,stmaryrd,sectsty}
\usepackage{setspace}
\usepackage[font=small,labelfont=bf]{caption}
\usepackage{verbatim}
\usepackage{graphicx}
\usepackage{caption}
\usepackage{subcaption}
\usepackage{float,color}
\newcommand{\B}{{Ba\v zant}}
\newcommand{\bc}{\begin{center}}
\newcommand{\ec}{\end{center}}
\newcommand{\bfr}{\begin{flushright}}
\newcommand{\efr}{\end{flushright}}

\newcommand{\no}{\noindent}
\newcommand{\be}{\begin{enumerate}}
\newcommand{\ee}{\end{enumerate}}
\newcommand{\bi}{\begin{itemize}}
\newcommand{\ei}{\end{itemize}}
\newcommand{\bd}{\begin{description}}
\newcommand{\ed}{\end{description}}
\newcommand{\beq}{\begin{equation}}
\newcommand{\eeq}{\end{equation}}
\newcommand{\bea}{\begin{eqnarray}}
\newcommand{\eea}{\end{eqnarray}}

\newcommand{\bfi}{\begin{figure}}
\newcommand{\efi}{\end{figure}}
\newcommand{\bay}{\begin{array}{l}}
\newcommand{\eay}{\end{array}}

\def\mb#1{\mbox {\boldmath {$#1$}}} 

\newcommand{\mbf}{\mathbf}
\newcommand{\cref}[1]{(\ref{#1})}   

\sectionfont{\normalsize}
\subsectionfont{\normalsize}
\subsubsectionfont{\normalsize}

\begin{document}

\begin{titlepage}
\clearpage\thispagestyle{empty}
\noindent
\hrulefill
\begin{figure}[h!]
\centering
\includegraphics[width=2 in]{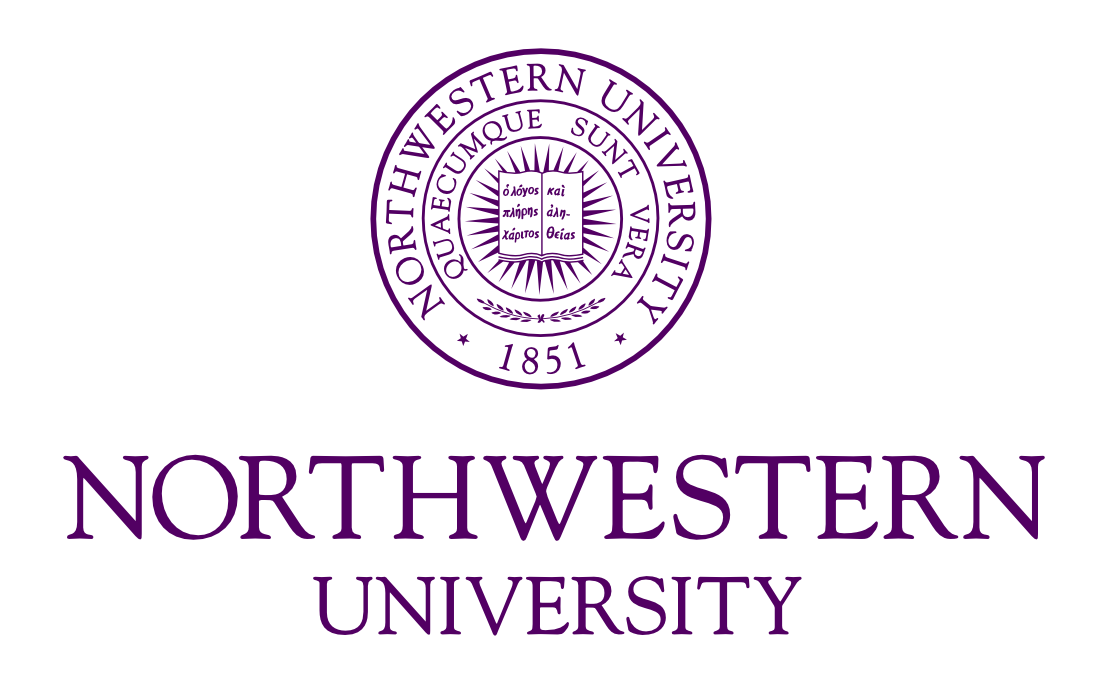}
\end{figure}
\begin{center}
{
{\bf Center for Sustainable Engineering of Geological and
Infrastructure Materials (SEGIM)} \\ [0.1in]
Department of Civil and Environmental Engineering \\ [0.1in]
McCormick School of Engineering and Applied Science \\ [0.1in]
Evanston, Illinois 60208, USA
}
\end{center} 
\hrulefill \\ \vskip 2mm
\vskip 0.5in
\begin{center}
{\large {\bf \uppercase{Asymptotic Expansion Homogenization of Discrete Fine-Scale Models with Rotational Degrees of Freedom for the Simulation of Quasi-Brittle Materials}}}\\[0.5in]
{\large {\sc Roozbeh Rezakhani, Gianluca Cusatis}}\\[0.75in]
{\sf \bf SEGIM INTERNAL REPORT No. 15-09/047A}\\[0.75in]
\end{center}
\noindent {\footnotesize {{\em Submitted to Journal of Mechanics and Physics of Solids \hfill October 2014} }}
\end{titlepage}

\newpage
\clearpage \pagestyle{plain} \setcounter{page}{1}

\begin{center}
{\large {\bf Asymptotic Expansion Homogenization of Discrete Fine-Scale Models with Rotational Degrees of Freedom for the Simulation of Quasi-Brittle Materials}}
\\[7mm]  {\bf     
Roozbeh Rezakhani \footnote{Research Assistant, Northwestern University, CEE Department, 2145 N Sheridan Rd
Evanston, IL 60208, USA.}
and  
Gianluca Cusatis \footnote{Corresponding author. Email: g-cusatis@northwestern.edu. Associate Professor, Northwestern University, CEE Department, 2145 N Sheridan Rd
Evanston, IL 60208, USA.}
}
\end{center} 

{\small \no {\bf   Abstract}: Discrete fine-scale models, in the form of either particle or lattice models, have been formulated successfully to simulate the behavior of quasi-brittle materials whose mechanical behavior is inherently connected to fracture processes occurring in the internal heterogeneous structure. These models tend to be intensive from the computational point of view as they adopt an ``a priori'' discretization anchored to the major material heterogeneities (e.g. grains in particulate materials and aggregate pieces in cementitious composites) and this hampers their use in the numerical simulations of large systems. In this work, this problem is addressed by formulating a general multiple scale computational framework based on classical asymptotic analysis and that (1) is applicable to any discrete model with rotational degrees of freedom; and (2) gives rise to an equivalent Cosserat continuum. The developed theory is applied to the upscaling of the Lattice Discrete Particle Model (LDPM), a recently formulated discrete model for concrete and other quasi-brittle materials, and the properties of the homogenized model are analyzed thoroughly in both the elastic and inelastic regime. The analysis shows that the homogenized micropolar elastic properties are size-dependent, and they are functions of the RVE size and the size of the material heterogeneity. Furthermore, the analysis of the homogenized inelastic behavior highlights issues associated with the homogenization of fine-scale models featuring strain-softening and the related damage localization. Finally, nonlinear simulations of the RVE behavior subject to curvature components causing bending and torsional effects demonstrates, contrarily to typical Cosserat formulations, a significant coupling between the homogenized stress-strain and couple-curvature constitutive equations.}


\section{Introduction}
Discrete fine-scale models, in the form of either particle or lattice models, have been formulated successfully in the literature to simulate the behavior of a variety of different materials. Their use has become more and more popular in the last few decades due to a number of appealing properties that make them advantageous compared to continuum based formulations. 

The geometry of discrete models is built with reference to the actual internal structure of the material of interest and it consists of ``particles'' connected through either ``contact points'' or ``connecting struts'' (also called ``lattice elements''). This ``a priori'' discretization allows simulating material heterogeneity efficiently in the case of materials - such as  concrete, rock, sea-ice, and toughened ceramics - characterized by hard and stiff inclusions embedded in a more compliant, weak, and brittle, matrix. In addition, the intrinsic particle/lattice spacing automatically provides the formulation with an internal characteristic length which can be made randomly variable if the discrete model is constructed according to the actual random distribution of material heterogeneity.

The degrees of freedom (displacements and rotations) are defined only at a finite number of points -- referred also as ``nodes'' thereinafter --  which, depending on the formulation, may or may not correspond to the partice center of mass or particle centroid. Strain and stress measures are defined at a finite number of points coinciding with the contact points or with some specified points along the connecting struts. The constitutive behavior is formulated through vectorial, as opposed
to tensorial, stress versus strain relationships and stress tractions are supposed to be distributed over either a ``contact area'' or the cross sectional area of the connecting struts (in this paper, this area will be generically referred to as ``facet''). Finally, the classical concepts of equilibrium and compatibility are formulated through algebraic equations, instead of partial differential equations typical of continuum mechanics. One of the main advantages of discrete models is that the discreteness of the formulation permits handling naturally displacement discontinuities arising during damage localization and fracture processes.

Rigid particle models, under the name  of Discrete Element Method (DEM), were first formulated to simulate both natural materials, such as geomaterials \cite{Cundall-1,Serrano-1,Cundall-3,Kawai-1}, as well as man-made materials like concrete \cite{Zubelewicz-3,Plesha-1,Zubelewicz-4}. A somewhat similar model is the rigid-body-spring model (RBSM), which subdivides the material domain into rigid polyhedral elements interconnected by zero-size springs \cite{Kawai-1,Bolander-1,Bolander-2,Bolander-3}.

Lattice models, pioneered by Hrennikoff \cite{Hrennikoff-1} to solve elastic problems in the pre-computers era, were later developed  by many authors to model fracture in quasi-brittle materials in both 2D  \cite{Schlangen-1},  and 3D \cite{Cusatis-1,Cusatis-2,Lilliu-1,Berton-1}.

More recently, various discrete models, in the form of either lattice or particle models, have been quite successful in simulating concrete materials \cite{Lilliu-1, cusatis-ldpm-1, cusatis-ldpm-2, Leite-1, Kim-1}. For an extensive review of the currently available models for concrete the reader is directed to a recent special issue \cite{CementComposite} collecting several papers covering a wide variety of concrete mechanics phenomena spanning several length scales, from the scale of cement particles to that of reinforced concrete structural members. 

In most applications of interest in practice, fine-scale models lead to fairly large computational systems characterized by a huge computational cost making their practical use rather limited. For example, the full-scale computational analysis of an average concrete bridge would require millions of degrees of freedom or the simulation of a rock formation would require billions of degrees of freedom. The solution of such large problems, although possible in principle with large super computer clusters, is unimaginable in everyday engineering practice. For this reason, many studies have been devoted to finding optimal and rigorous approaches for multiscale  computation. 

Among different multiscale techniques available in the literature \cite{Galvanetto-1}, the ones based on homogenization theory have been widely used over the past decades. The homogenization theory relies on two main assumptions. The first is the existence of a certain volume of material, the so called Representative Volume Element (RVE) or Unit Cell (UC), carrying a complete description of the internal material structure \cite{Gitman-1,Kouznetsova-1}. The second is that the size of such a volume is much smaller than the size of the overall solid volume under consideration. The latter is also known as the ``scale separation'' assumption. 


Hill \cite{Hill-1}, Eshelby \cite{Eshelby-1}, Hashin and Strikman \cite{Hashin-1} pioneered analytical homogenization techniques which were developed later by other authors \cite{Christensen-1,Nemat-Nasser-1}. Analytical homogenization is able to reasonably approximate material properties when the exact solution of the boundary value problem associated with the RVE problem can be obtained. However, in this approach, elastic behavior, small strains, and relatively simple internal structure are the limiting assumptions typically adopted. When complicated heterogeneous structures are considered, or constitutive behavior of constituents are nonlinear, other homogenization techniques \cite{Lopez-Pamies-1, Lopez-Pamies-2} needs to be considered.                  

To overcome these difficulties, computational homogenization is often used in the literature  \cite{Smith-1,Feye-1,Kouznetsova-1,Miehe-1}. In this approach, a single RVE is assigned to each calculation point (e.g. gauss point in a Finite Element mesh) in the macro domain and at each step of the nonlinear analysis, macro-strain increments are imposed as essential boundary conditions to the RVE. The solution of the RVE boundary value problem is then averaged for the calculation of the associated macroscopic stress tensor. Since no assumption is made for the macroscopic constitutive law, this method can be used for materials featuring extremely nonlinear behavior.

A somewhat similar but more mathematically rigorous homogenization technique is the so-called Asymptotic Expansion Homogenization (AEH)
that uses the asymptotic expansion of the displacement field based on a length parameter representing the ratio between the length scale of material heterogeneity and the macroscopic length scale. Starting from this expansion hierarchical boundary value problems are obtained at different scales. This approach can easily handle problems with multiple (more than 2) scales in both space and time \cite{Fish-3}; it does not make assumptions on the character of the macroscopic constitutive equations; and its implementation in computer codes is relatively simple. 

Within the extensive literature on AEH, remarkable is the work of the following authors. Hassani \cite{Hassani-1, Hassani-2} investigated formulation of homogenization theory and topology optimization and its numerical application to materials with periodic microstructure. Chung \cite{Chung-1} presented detailed derivation of multiple scale formulation for elastic solids.  Fish employed this approach to study elastic as well as elasto-plastic composites \cite{Fish-2}. Ghosh \cite{Ghosh-1} adopted MH along with Voronoi Cell Finite Element Method (VCFEM) to study the behavior of composites with random meso-structure \cite{Ghosh-2}. More recently, Fish \cite{Fish-3} introduced the Generalized Mathematical Homogenization (GMH) to derive continuum constitutive equations starting from Molecular Dynamics (MD). 

All the aforementioned work is relevant to Cauchy continuum formulations. However, homogenization schemes were also used for the multiscale analysis of Cosserat continuum models, in which an independent rotation field appears in addition to the displacement field. Feyel \cite{Feye-1} built a homogenization scheme to couple a Cauchy continuum formulation at the micro-scale giving rise to a Cosserat continuum formulation at the macro-scale. Asymptotic homogenization technique was employed by Forest \cite{Forest-1} for upscaling elastic Cosserat solids. In this work, the author studied various types of asymptotic expansions for the displacement and rotation fields and investigated their effect on the resulting macroscopic continuum behavior. Results of this investigation, showed that the nature of the homogenized continuum depends on the ratio of the Cosserat characteristic length of constituents, size of heterogeneity and typical size of the structure. 


Chan et al. \cite{Chan-1} derived the governing constitutive equations for strain gradient elasticity for both homogeneous and functionally graded materials using the strain energy density function and the related definitions of the stress fields. They showed that additional terms appear in the equations that are related to the strain gradient nonlocality and the interaction between material nonhomogeneity. Bardenhagen et al. \cite{Bardenhagen-1} obtained a nonlinear higher order gradient continuum representation of discrete periodic micro-structures by means of an energy approach. The developed model was then employed to investigate the existence and stability of localization bands and their relationship to the model loss of ellipticity. Finally, homogenization of discrete atomic models into equivalent continuum can be found in publications where the authors exploited asymptotic analysis techniques \cite{Caillerie-1} and the mathematical $\Gamma$-convergence method \cite{Braides-1}.

The present study derives a general multiscale homogenization scheme suitable for upscaling materials whose fine-scale behavior can be successfully approximated through the use of discrete models featuring both translational and rotational degrees of freedom.

\section{The Fine-Scale Problem}
\begin{figure}[t]
\centering 
{\includegraphics[width=\textwidth]{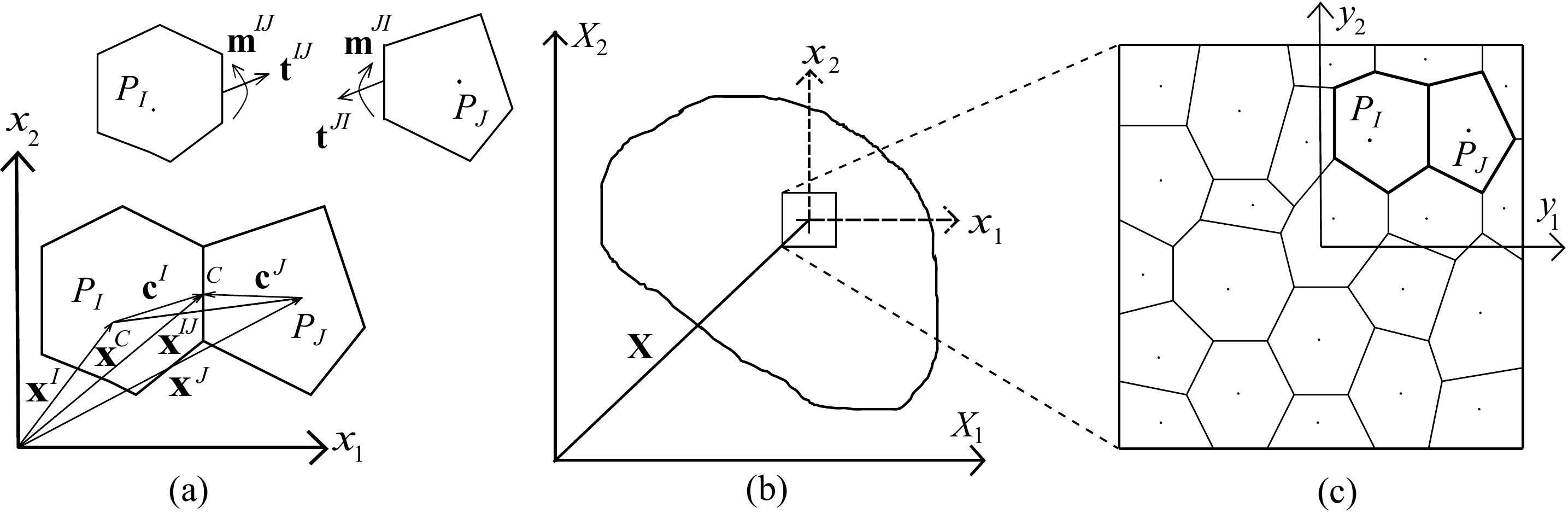}}
\caption{Geometrical explanation of the two-scale problem: (a) Geometry of two neighboring particles. (b) Macro material domain. (c) Meso-scale domain with material heterogeneity.}
\label{TwoScaleAnalysis}
\end{figure}

With reference to Figure \ref{TwoScaleAnalysis}a, let us consider the interaction of two adjacent particles, $I$ and $J$, sharing a generic facet. If one limits the analysis to the case of small strains and displacements -- which is a reasonable assumption in absence of large plastic deformation prior to fracture as observed in brittle and quasi-brittle materials -- meaningful measures of deformation \cite{cusatis-ldpm-1} can be defined as

\begin{equation} \label{eps} 
\epsilon^{IJ}_{\alpha}=\frac{1}{r} \left(\mathbf{U}^J + \mb \Theta^{J} \times \mathbf{c}^{J} - \mathbf{U}^I - \mb{\Theta}^I \times \mathbf{c}^I \right) \cdot \mathbf{e}^{IJ}_{\alpha} 
\end{equation}

and

\begin{equation} \label{curvature} 
\chi^{IJ}_{\alpha}=\frac{1}{r} \left( \mb{\Theta}^{J} - \mb{\Theta}^I  \right) \cdot \mathbf{e}^{IJ}_{\alpha}
\end{equation}

where $\epsilon^{IJ}_{\alpha}=$ facet strains; $\chi^{IJ}_{\alpha}=$ facet curvatures; $r=|\mathbf{x}^{IJ}|$; $\mathbf{x}^{IJ}=\mathbf{x}^J-\mathbf{x}^I $ is the vector connecting the particle nodes $P_I$ and $P_J$;  $\mathbf{e}^{IJ}_{\alpha}$ ($\alpha=N,M,L$) are unit vectors defining a facet Cartesian system of reference such that $\mathbf{e}^{IJ}_{N}=$ is orthogonal to the facet and $\mathbf{e}^{IJ}_{N} \cdot \mathbf{x}^{IJ} >0$; $\mathbf{U}^I$, $\mathbf{U}^J$ = displacement vectors of node $P_I$ and $P_J$; $\mb{\Theta}^{I}$, $\mb{\Theta}^{J}$ = rotation vectors of node $P_I$ and $P_J$; and $\mathbf{c}^{I}$, $\mathbf{c}^{J}$ = vectors connecting nodes $P_I$ and $P_J$ to the facet centroid, see Fig. \ref{TwoScaleAnalysis}a. It must be observed here that displacements and rotations are assumed to be independent variables.

For given strain and curvature vectors, a vectorial constitutive equation provide stress, $\mathbf{t}^{IJ}$, and couple, $\mathbf{m}^{IJ}$, tractions on each facet. Formally one can write $\mathbf{t}^{IJ} = t_{{\alpha}}(\epsilon_{N}, ...) \mathbf{e}^{IJ}_{\alpha}$ and $\mathbf{m}^{IJ} = m_{{\alpha}}(\chi_{N}, ...) \mathbf{e}^{IJ}_{\alpha}$  where, in general, summation rule applies over $\alpha$. As an example, the elastic behavior can be formulated through the following equations

\begin{equation} \label{elastic}
t_{\alpha} = E_\alpha \epsilon_{\alpha};  \hspace{0.25 in}  m_{\alpha} = W_{\alpha} \chi_{\alpha}= E_\alpha \ell_\alpha^2 \chi_{\alpha}; \hspace{0.25 in} (\alpha=N,M,L)
\end{equation}

in which each traction component is proportional to the associated strain or curvature (summation rule does not apply); and $E_\alpha$, $W_\alpha$ are fine-scale elastic constants which are related by a characteristic length $\ell_\alpha$. An example of nonlinear facet constitutive equations is reported in Appendix \ref{LDPM}, Section \ref{LDPM-Constitutive}, with reference to the so-called Lattice Discrete Particle Model (LDPM)  that will be considered in the numerical examples.

Finally, the computational discrete fine-scale framework is completed by imposing the equilibrium of each single particle subject to the effect of all surrounding particles. Translational and rotational dynamic equilibrium equations read

\begin{equation} \label{motion-1}
 M_u^I\ddot{\mathbf U}^I +\mathbf{M}_{u \theta}^I\ddot{\mb \Theta}^I - V^I \mathbf{b}^0=\sum_{\mathcal{F}_I} A \mathbf{t}^{IJ}
\end{equation}

and 

\begin{equation}\label{motion-2}
\mathbf{M}_{\theta}^I\ddot{\mb \Theta}^I = \sum_{\mathcal{F}_I}  A ( \mathbf{w}^{IJ} +\mathbf{m}^{IJ})
\end{equation}

where $\mathbf{w}^{IJ} = \mathbf c^I \times \mathbf {t}^{IJ}$ is the moment of the traction $\mathbf {t}^{IJ}$ with respect to the particle node $P_I$; $\mathcal{F}_I$ is the set of facets surrounding node P$_I$ and obtained by collecting all the facets associated with each node pair $(I,J)$; $A$ = facet area; superimposed dots represent time derivatives; $V^I$ is the particle volume;  $\mathbf{b}^0$ is the body force vector; $M_u^I$ = mass of node $P_I$; and $\mathbf{M}_{u \theta}^I$, $\mathbf{M}_{\theta}^I$ = moment of inertia tensors. It is worth observing that $\mathbf{M}_{u \theta}^I =  \mathbf{0}$ and $\mathbf{M}_{\theta}^I = M_{\theta}^I \mathbf{I}$ if the particle node is the particle center of mass; the axes of the system of reference are parallel to the particle principal axes of inertia; and the principal moments of inertia are the same in all directions. These conditions, although applicable only to a limited number of cases (e.g. spherical particles), do not reduce the conceptual generality of the derivation that will be presented in this paper and will be assumed thereinafter for simplicity.

\section{ Asymptotic Expansion Homogenization } \label{HomogTheory}
In this section, the two-scale homogenization of the general fine-scale problem introduced in the previous section is pursued by means of the approach proposed in Ref.  \cite{Fish-3}. In the original formulation only central forces were assumed to act on the particles and, consequently, the rotational equilibrium equation was not considered. 

\subsection{Two Scale Approximation and Asymptotic Expansions}
In order to perform a two-scale asymptotic expansion homogenization, a periodic discrete system, composed by a number of adjacent RVEs, is considered in this section. In Figure \ref{TwoScaleAnalysis}b, the generic macroscopic material domain and the corresponding global coordinate system $\bf X$ are shown. At any point in the macroscopic domain, two separate length scales and the corresponding local coordinate systems, $\bf x$ and  $\bf y$, are introduced to represent (1) the macroscopic domain, in which the problem is defined as homogeneous continuum with no detail of material heterogeneity, and (2) the meso-scale domain, in which heterogeneity is modeled by the discrete meso-scale model. Vector $\mbf X$, as shown in the figure, is the vector connecting the origin of the global macroscopic coordinate system  to the mass center of a generic RVE. In Figure \ref{TwoScaleAnalysis}c, a zoomed view of the macroscopic material point is shown in the local meso-scale coordinate system $\bf{y}$, in which a representative volume of heterogeneous material is depicted. One should consider that in Figure \ref{TwoScaleAnalysis}a, particles $I$ and $J$ are shown in the local macroscopic coordinate system $\bf x$. Therefore, they should be plotted in smaller size compared to Figure \ref{TwoScaleAnalysis}c, but this was not done for the sake of clarity. If the separation of scales exists, one can write the following relationship linking macro and meso local coordinate systems    
\begin{eqnarray}\label{scale-link-1}
\mathbf{x}=\eta \mathbf{y};  \hspace{0.25 in}   0< \eta <<1
\end{eqnarray}

where $\eta$ is a very small positive scalar. In addition, the displacement of a generic node P$_I$,  $\mathbf{U}^I = \mathbf{u}(\mathbf{x}^I, \mathbf{y}^I)$, can be approximated by means of the following asymptotic expansion

\begin{equation}
\mathbf{u}(\mathbf x, \mathbf y) \approx \mathbf u^0(\mathbf x, \mathbf y)+\eta \mathbf u^1(\mathbf x, \mathbf y) 
\label{disp-expansion}
\end{equation}

where only terms up to order $\mathcal{O}(\eta)$ are considered. Functions $\mathbf  u^0(\mathbf x, \mathbf y)$, and $\mathbf  u^1(\mathbf x, \mathbf y)$ are continuous with respect to $\mathbf{x}$ and discrete (i.e. defined only at finite number of points) with respect to $\mathbf{y}$.

In order to define the asymptotic expansion for rotations, it is convenient first to postulate the existence of a continuous displacement-like field $\mbf{d}^\eta (\mathbf x)$ such that $2 \mbf \Theta^I= \mb  \nabla \times \mbf{d}^\eta|_{\mathbf x =\mathbf x^I}$. If $\mbf{d}^\eta (\mathbf x)$ is replaced by a two-scale approximation similar to the one in Equation \ref{disp-expansion}, one can write, ${\boldsymbol \Theta}^I = \mb{\theta}(\mathbf{x}^I, \mathbf{y}^I)$, and 

\begin{equation}
\mb \theta (\mathbf x, \mathbf y)  \approx \eta^{-1} \mb \omega^{0}(\mathbf x, \mathbf y) + \mb \varphi^0(\mathbf x, \mathbf y)+ \mb \omega^{1}(\mathbf x,\mathbf y)+\eta \mb\varphi^{1}(\mathbf x, \mathbf y) 
\label{rot-expansion}
\end{equation}

where $2 \boldsymbol\omega^{0} = \boldsymbol \nabla_y \times \mbf{d}^0$;  $2 \boldsymbol\varphi^{0} = \boldsymbol  \nabla_x \times \mbf{d}^0$;  $2 \boldsymbol \omega^{1} = \boldsymbol  \nabla_y \times \mbf{d}^1$;  $2 \boldsymbol \varphi^{1} = \boldsymbol  \nabla_x \times \mbf{d}^1$; and subscripts $x$ and $y$ identify the nabla operator in the coarse- and fine-scale, respectively. Thus, $\mb\omega^{0}$, $\mb\omega^{1}$ should be interpreted as rotations in the fine-scale whereas $\mb\varphi^{0}$, $\mb\varphi^{1}$ as the corresponding coarse-scale rotations. It is worth observing here that, contrarily to the expansion of displacements, the asymptotic expansion for rotations features a term of order $\mathcal{O}(\eta^{-1})$  and two distinct terms of order $\mathcal{O}(1)$. 

In the macroscopic coordinate $\mathbf{x}$, the difference in position between nodes $P_I$ and $P_J$ can be considered as infinitesimal. Hence, in order to obtain the asymptotic expansion of strains and curvatures, it is convenient first to obtain the Taylor series expansion of displacement and rotation at nodes $P_J$ around point $P_I$ of coordinate $\mathbf{x}^I$ in the local coordinate system $\bf x$. By assuming that the displacement and rotation fields in Equations \ref{disp-expansion} and \ref{rot-expansion}, are continuous and differentiable with respect to $\mathbf{x}$, one can write 

\begin{eqnarray}\label{taylor-1-J}
\begin{aligned}
U_i^J=u_i(\mathbf{x}^J,\mathbf{y}^J) = u_i^J+u^J_{i,j} \,x^{IJ}_j + \frac{1}{2}u^J_{i,jk}\, x^{IJ}_j x^{IJ}_k +\cdots
\end{aligned}
\end{eqnarray}
\begin{eqnarray}\label{taylor-2-J}
\begin{aligned}
\Theta_i^J=\theta_i(\mathbf{x}^J,\mathbf{y}^J)= \theta_i^J+\theta_{i,j}^J x^{IJ}_j +\frac{1}{2} \theta^J_{i,jk}\, x^{IJ}_j x^{IJ}_k +\cdots
\end{aligned}
\end{eqnarray}

where $u_i^J=u_i(\mathbf{x}^I,\mathbf{y}^J)$; $ u^J_{i,j} = \partial u_i/\partial x_j (\mathbf{x}^I,\mathbf{y}^J)$; $u^J_{i,jk} =\partial^2 u_i/\partial x_j \partial x_k (\mathbf{x}^I,\mathbf{y}^J)$; $\theta_i^J= \theta_i(\mathbf{x}^I,\mathbf{y}^J)$; $ \theta^J_{i,j} = \partial \theta_i/\partial x_j (\mathbf{x}^I,\mathbf{y}^J)$; $\theta^J_{i,jk} =\partial^2 \theta_i/\partial x_j \partial x_k (\mathbf{x}^I,\mathbf{y}^J)$; $x^{IJ}_j$ is a vector connecting node $P_I$ to node $P_J$ in the $\mathbf{x}$ space. By substituting Equations \ref{disp-expansion} and \ref{rot-expansion}  into Equation \ref{eps}, and using the Taylor expansion of displacement and rotation of node $P_J$ around node $P_I$ (Equations \ref{taylor-1-J} and \ref{taylor-2-J}) one obtains the multiple scale definition of facet strains (see Appendix \ref{exp-strains-details} for details)

\begin{equation}\label{eps-expansion}
\epsilon_{\alpha}=\eta^{-1} \epsilon_{\alpha}^{-1} + \epsilon_{\alpha}^0 + \eta \epsilon_{\alpha}^1
\end{equation}

where

\begin{equation}\label{eps-expansion-minus}
\epsilon_{\alpha}^{-1} = \bar{r}^{-1} \bigg[  u_i^{0J} - u^{0I}_i + \varepsilon_{ijk} \omega_j^{0J} \bar c_{k}^{J} - \varepsilon_{ijk} \omega_j^{0I} \bar c_{k}^{I} \bigg]e^{IJ}_{\alpha i} 
\end{equation}
\begin{equation}\label{eps-expansion-zero}
\epsilon_{\alpha}^0 =\bar{r}^{-1} \bigg[  u_i^{1J} + u^{0J}_{i,j} y^{IJ}_j - u^{1I}_i + \varepsilon_{ijk} \bigg( \varphi_j^{0J} + \omega_j^{1J} + \omega_{j,m}^{0J} y^{IJ}_m  \bigg) \bar c_{k}^{J} -  \varepsilon_{ijk} \bigg( \varphi_j^{0I} + \omega_j^{1I} \bigg) \bar c_{k}^{I} \bigg] e^{IJ}_{\alpha i} 
\end{equation}
\begin{equation}\label{eps-expansion-plus}
\begin{split}
\epsilon_{\alpha}^1 = \bar{r}^{-1} \bigg[  u^{1J}_{i,j} y^{IJ}_j + \frac{1}{2}u^{0J}_{i,jk} y^{IJ}_j y^{IJ}_k + \varepsilon_{ijk} \bigg( \varphi_j^{1J} + \varphi_{j,m}^{0J} y^{IJ}_m + \omega_{j,m}^{1J} y^{IJ}_m + \frac{1}{2} \omega^{0J}_{j,mn} y^{IJ}_m y^{IJ}_n \bigg) \bar c_{k}^{J} - \varepsilon_{ijk} \varphi_j^{1I} \bar c_{k}^{I} \bigg] e^{IJ}_{\alpha i} 
\end{split}
\end{equation}

In the previous equations, $\varepsilon_{ijk}$ is the Levi-Civita permutation symbol and length type variables have been changed into their $\mathcal{O}(1)$ counterparts by using Equation \ref{scale-link-1}: $r=\eta \bar{r}$, $c_{k}^I=\eta \bar{c}_{k}^I$, $c_{k}^J=\eta \bar{c}_{k}^J$. 

Similarly, multiple scale definition of facet curvature can be calculated as (see Appendix \ref{exp-strains-details} for details)

\begin{equation}\label{curv-expansion}
\eta \chi_{\alpha}=\eta^{-1} \psi_{\alpha}^{-1} + \psi_{\alpha}^{0} + \eta \psi_{\alpha}^1
\end{equation}

where

\begin{equation}\label{curv-expansion-minus2}
\psi_{\alpha}^{-1} =\bar{r}^{-1} \bigg[ \omega_i^{0J} - \omega_i^{0I} \bigg] {e}^{IJ}_{\alpha i}
\end{equation}
\begin{equation}\label{curv-expansion-minus1}
\psi_{\alpha}^{0} = \bar{r}^{-1} \bigg[ \varphi_i^{0J}+ \omega_i^{1J} + \omega_{i,j}^{0J} y^{IJ}_j - \varphi_i^{0I} - \omega_i^{1I} \bigg] {e}^{IJ}_{\alpha i}
\end{equation}
\begin{equation}\label{curv-expansion-zero}
\psi_{\alpha}^1 = \bar{r}^{-1} \bigg[ \varphi_i^{1J} + \omega_{i,j}^{1J} y^{IJ}_j + \varphi_{i,j}^{0J} y^{IJ}_j + \frac{1}{2} \omega_{i,jk}^{0J} y^{IJ}_j y^{IJ}_k - \varphi_i^{1I} \bigg] {e}^{IJ}_{\alpha i}
\end{equation}

It is worth noting that in this section as well as in the rest of the paper superscript $IJ$ has been dropped when the permutation of $I$ and $J$ is not associated with a sign change. 

\subsection{Multiple-Scale Equilibrium Equations}

In order to obtain the correct scale separation of the governing equations, a rescaling of the discrete equilibrium equations needs to be performed. For the sake of simplicity, and since only quasi-static problems are concerned in the current research, it is assumed $\mathbf M^I_{u \theta} = 0$ and $\mathbf{M}_{\theta}^I = M_{\theta}^I \mathbf{I}$ on the left hand side of Equation \ref{motion-1}. Rescaling is pursued by assuming that the material density, mass per unit volume, is of order zero: $\rho \sim \mathcal{O}(1)$, which along with the displacement asymptotic expansion implies that the left-hand-side of Equation \ref{motion-1} is $\sim \mathcal{O}(\eta^3)$. By dividing both sides of Equation \ref{motion-1} by $\eta^3$, and considering that all length variables should be considered $\sim \mathcal{O}(\eta^1)$, one obtains
 
\begin{equation} \label{motion-1-rescaled}
\bar{M}_u^I\ddot { {\mathbf u}}^I - \bar{V}^I \mathbf{b}^0=\eta^{-1}\sum_{\mathcal{F}_I}{\bar{A}\, t_{\alpha} {\mathbf e}_\alpha^{IJ}}
\end{equation}

where $\bar{M}_u^I = M_u^I/\eta^3$, $\bar{V}^I=V^I/\eta^3$  , $\bar{A} = A/\eta^2$ are all quantities $\sim \mathcal{O}(1)$. For reason of dimensionality, body forces $b^{0}_i$ can be always assumed to be proportional to gravity $\rho g$ and, consequently, they can be considered $\mathcal{O}(1)$ quantities as well.

One can rescale the rotational equation in a similar fashion by recognizing that, according to the previous discussion, the rotational moment of inertia is $\sim \mathcal{O}(\eta^5)$. Dividing both sides of Equation \ref{motion-2} by $\eta^4$ one obtains

\begin{equation}\label{motion-2-rescaled} 
\eta \bar {M_\theta^I}\ddot{\mb \theta}^I=\eta^{-1} \sum_{\mathcal{F}_I} \bar{A}\, (\eta^{-1} w_{\alpha}{\mathbf e}_\alpha^{IJ}  + \eta^{-1} m_{\alpha}{\mathbf e}_\alpha^{IJ})
\end{equation}

where $\bar{M}_\theta^I = M_\theta^I/\eta^5$ is $\sim \mathcal{O}(1)$.


In the elastic regime one can write: $t_{\alpha} = \eta^{-1} t^{-1}_{\alpha} +  t^{0}_{\alpha} + \eta t^{1}_{\alpha}$; where $t^{(\cdot)}_{\alpha} = E_\alpha \epsilon^{(\cdot)}_\alpha$, and $E_\alpha$ is assumed to be $\sim \mathcal{O}(1)$. In addition, $q_\alpha = \eta^{-1} m_{\alpha} = \eta^{-1} q^{-1}_{\alpha} + q^{0}_{\alpha} + \eta q^{1}_{\alpha}$ in which $q^{(\cdot)}_{\alpha} =\bar{W}_\alpha \psi^{(\cdot)}_\alpha$; $\bar{W}_\alpha=E_\alpha {\bar{\ell}}^{2}_\alpha$; and $\bar{\ell} = \ell / \eta$. Finally, $p_{\alpha} = \eta^{-1}w_{\alpha} = \eta^{-1} p^{-1}_{\alpha} +  p^{0}_{\alpha} + \eta p^{1}_{\alpha}$ where $p^{(\cdot)}_{\alpha} {\mathbf e}_\alpha^{IJ} = \bar{\mathbf{c}}^I \times t^{(\cdot)}_{\alpha} {\mathbf e}_\alpha^{IJ}$. Since $w_\alpha$ and $m_\alpha$ are moments, it is reasonable that the asymptotic expansion of those variables divided by $\eta$ is similar to the one for tractions $t_\alpha$, considering that length type variables are considered to be $\sim \mathcal{O}(\eta)$.

Introducing these traction expressions along with the asymptotic definition of displacement and rotation fields Equations \ref{disp-expansion} and \ref{rot-expansion} (which also imply $\eta \ddot{\mb \theta}^{0I} =\ddot{\mb \omega}^{0I}+\mathcal{O}(\eta)$), into the rescaled equilibrium equations leads to 

\begin{equation} \label{motion-1-sep}
\eta^{-2} \sum_{\mathcal{F}_I}{\bar{A}\, t^{-1}_{\alpha} {\mathbf e}_\alpha^{IJ}}+ \eta^{-1} \sum_{\mathcal{F}_I}{\bar{A}\, t^{0}_{\alpha} {\mathbf e}_\alpha^{IJ}} + \sum_{\mathcal{F}_I}{\bar{A}\, t^1_{\alpha} {\mathbf e}_\alpha^{IJ}} - \bar{M}_u^I\ddot { {\mathbf u}}^{0I} + \bar{V}^I \mathbf{b}^0+\mathcal{O}(\eta) = \mathbf{0}
\end{equation}

and

\begin{eqnarray}\label{motion-2-sep}
\begin{aligned}
\eta^{-2}\sum_{\mathcal{F}_I} \bar{A}\, (p^{-1}_{\alpha}{\mathbf e}_\alpha^{IJ} + q^{-1}_{\alpha}{\mathbf e}_\alpha^{IJ}) + \eta^{-1}\sum_{\mathcal{F}_I} \bar{A}\, (p^{0}_{\alpha}{\mathbf e}_\alpha^{IJ} + q^{0}_{\alpha}{\mathbf e}_\alpha^{IJ})  \\ - \bar {M_\theta^I}\ddot{\mb \omega}^{0I}  + \sum_{\mathcal{F}_I} \bar{A}\, (p^{1}_{\alpha}{\mathbf e}_\alpha^{IJ} + q^{1}_{\alpha}{\mathbf e}_\alpha^{IJ}) + \mathcal{O}(\eta) = \mathbf{0}
\end{aligned}
\end{eqnarray}

in which terms of different orders are gathered together. The multiple scale equations reported above can also be used for nonlinear constitutive equations provided that facet tractions and facet moments can be expressed through the multiple scale decomposition exploited above. It will be shown later in the paper that this can be indeed achieved under some reasonable assumptions.

\subsection{The RVE Problem}\label{RVEproblem}
Let's first consider the equilibrium equations at the $\mathcal{O}(\eta^{-2})$ scale. From Equations \ref{motion-1-sep} and \ref{motion-2-sep}, it is evident that the $\mathcal{O}(\eta^{-2})$ equilibrium equations represent the equilibrium of all particles in the RVE subjected to the stress tractions $t_\alpha^{-1}$ and the moment tractions $q_\alpha^{-1}$ and without any applied external load. Consequently, solution of the $\mathcal{O}(\eta^{-2})$  problem implies $t_\alpha^{-1}=0$ and $q_\alpha^{-1}=0$, which in turn, leads to $\epsilon_{\alpha}^{-1}=0$  and $\psi_{\alpha}^{-1}=0$. By taking into consideration the definitions of $\epsilon^{-1}_{\alpha}$ and $\psi^{-1}_{\alpha}$ (Equations \ref{eps-expansion-minus} and \ref{curv-expansion-minus2}) such result indicates that the $\mathcal{O}(\eta^{-2})$  problem represents a rigid body rototranslation of the RVE. This can be expressed as 

\begin{equation}\label{U0}
u_i^0(\mathbf{X}, \mathbf{y}) = v_i^0(\mathbf{X}) + \varepsilon_{ijk} y_k \omega_j^{0}(\mathbf{X}) 
\end{equation}

in which the fields $\mathbf{v}^0$ and $\mb \omega^{0}$ are only dependent on macroscopic coordinate system $\bf X$, i.e. these quantities varies smoothly in the macro-scale material domain; they do not change within the RVE domain; and they can be calculated when kinematic boundary conditions are specified for the $\mathcal{O}(\eta^{-2})$  problem. These boundary conditions must describe the physical fact that the RVE is attached to a point in the macroscopic continuum.  Hence, $\mathbf{v}^0$ must correspond to the macroscopic displacement field, and $\mb \omega^{0}$ must be equal to the macroscopic rotation field: $\mb \varphi^{0} = \mb \omega^{0}$. Since $\mb \omega^{0}$ is constant over the RVE, then $\mb \varphi^{0}$ is also constant in the RVE. 



On the basis of Equation \ref{U0} and the discussion above, one can rewrite the $\mathcal{O}(1)$ strains and curvatures as (See Appendix \ref{Revised-strain-curvature} for details) 

\begin{equation}\label{eps-expansion-zero'}
\epsilon_{\alpha}^0 =\bar{r}^{-1} \left( u_i^{1J} - u^{1I}_i + \varepsilon_{ijk}  \omega_j^{1J} \bar c_{k}^{J} - \varepsilon_{ijk} \omega_j^{1I} \bar c_{k}^{I}\right) {e}^{IJ}_{\alpha i} + P^\alpha_{ij} \left(\gamma_{ij}  
+ \varepsilon_{jmn}  \kappa_{im} y_{n}^{c} \right) 
\end{equation}
\begin{equation}\label{curv-expansion-minus1'}
\psi_{\alpha}^{0} = \bar{r}^{-1} \left( \omega_i^{1J}- \omega_i^{1I} \right) {e}^{IJ}_{\alpha i}+ P^\alpha_{ij} \kappa_{ij} 
\end{equation}

where $\gamma_{ij} = v^{0}_{j,i} - \varepsilon_{ijk} \omega_k^{0}$, $\kappa_{ij}=\omega_{j,i}^{0}$ are the macroscopic Cosserat strain and curvature tensors, respectively. The vector $\mbf{y}^c$ is the position vector of the centroid of the common facet between particle $I$ and $J$ and $P^\alpha_{ij} = n^{IJ}_i e^{IJ}_{\alpha j}$ is a projection operator. Comparing the first term of Equation \ref{eps-expansion-zero'} with Equation \ref{eps}, it can be concluded that this term is the lower scale definition of the three components of the facet strains (one normal and two tangential) written in terms of fine-scale displacements and rotations $u^1$ and $\omega^1$. The second term of Equation \ref{eps-expansion-zero'}, $P^\alpha_{ij} \left(\gamma_{ij} + \varepsilon_{jmn}  \kappa_{im} y_{n}^{c} \right)$, is the projection of macroscopic Cosserat strain and curvature tensors on each facet. Similarly, Equation \ref{curv-expansion-minus1'} shows that the $\mathcal{O}(1)$ curvature includes a fine-scale term (see Equation \ref{curvature}), which depends on fine-scale rotation term $\omega^1$, and a coarse-scale term corresponding to the projection of macroscopic curvature tensor on each facet. Therefore, Equations \ref{eps-expansion-zero'} and \ref{curv-expansion-minus1'} express the $\mathcal{O}(1)$ facet strains and curvatures as the sum of their fine-scale counterparts and the projection of macroscopic strain and curvature tensors onto the facet level. It is worth nothing that the projection operator $P^\alpha_{ij}$ corresponds exactly to the one used in the microplane model \cite{Bazant-2, xinwei-1} if $ e^{IJ}_{N i}\equiv n^{IJ}_i $, i.e. the discrete model is formulated in such a way the facets are orthogonal to the associated lattice struts. In addition, it must be noted that the term $\varepsilon_{jmn}  \kappa_{im} y_{n}^{c}$ transforms the macroscopic curvature tensor, which is constant over the RVE, to different strain values at different positions $y^c_n$ inside the RVE, which is then projected on the facets through the operator $P^\alpha_{ij}$. Expanding this term for different components of curvature tensor, it can be shown that it perfectly corresponds to the strain field generated by curvatures in classical beam theories.
 
Strains and curvatures of order $\mathcal{O}(\eta)$ can also be rewritten by taking into account Equation \ref{U0}. One gets

\begin{equation}\label{eps-expansion-plus'}
\begin{aligned}
\epsilon_{\alpha}^1 = \bar{r}^{-1} \bigg[& u^{1J}_{i,j} y^{IJ}_j + \varepsilon_{ijk} \varphi_j^{1J} \bar c_{k}^{J} + \varepsilon_{ijk} \omega_{j,m}^{1J} y^{IJ}_m \bar c_{k}^{J} - \varepsilon_{ijk} \varphi_j^{1I} \bar c_{k}^{I} \\
&+ \frac{1}{2}v^{0}_{i,jk} y^{IJ}_j y^{IJ}_k + \frac{1}{2} \varepsilon_{ijk} \omega^{0}_{j,mn} y^{IJ}_m y^{IJ}_n  y_{k}^{c} + \varepsilon_{ijk} \omega_{j,m}^{0} y^{IJ}_m \bar c_{k}^{J} \bigg] e^{IJ}_{\alpha i} 
\end{aligned}
\end{equation}
\begin{equation}\label{curv-expansion-zero'}
\psi_{\alpha}^1 = \bar{r}^{-1} \left[ \varphi_i^{1J} + \omega_{i,j}^{1J} y^{IJ}_j - \varphi_i^{1I} + \omega_{i,j}^{0} y^{IJ}_j + \frac{1}{2} \omega_{i,jk}^{0J} y^{IJ}_j y^{IJ}_k \right] {e}^{IJ}_{\alpha i}
\end{equation}

Detailed mathematical derivation of Equations \ref{eps-expansion-zero'} through \ref{curv-expansion-zero'} is provided in Appendix \ref{Revised-strain-curvature}. 

In the previous derivation, where linear elastic behavior was assumed, the equilibrium equations at the $\mathcal{O}(\eta^{-2})$ scale were shown to represent the rigid body motion conditions for the RVE and, consequently, they led to zero strains, $\epsilon_{\alpha}^{-1}$, curvatures, $\psi_{\alpha}^{-1}$, tractions,  $t_\alpha^{-1}$, and moments, $p_\alpha^{-1}$, and $q_\alpha^{-1}$, at the $\mathcal{O}(\eta^{-1})$ scale. These conditions can be reasonably assumed \emph{a priori} in the case of nonlinear material behavior. In this case case one may write  $t_\alpha = t_\alpha(\epsilon_{\beta}^{0}+\eta \epsilon_{\beta}^{1})$; $p_\alpha = p_\alpha(\epsilon_{\beta}^{0}+\eta \epsilon_{\beta}^{1})$, and $q_\alpha = q_\alpha(\eta^{-1}\psi^0_\beta+\psi^1_\beta)$ in which $\alpha,~\beta = N,M,L$. Since $\eta$ is a small quantity, one can also write the Taylor expansion of $t_\alpha$ and $p_\alpha$ around the $\mathcal{O}(1)$ component of strain and the Taylor expansion of $q_\alpha$ around the $\mathcal{O}(\eta^{-1})$ component of curvature:

\begin{equation}\label{TaylorExpansions}
\begin{split}
& t_\alpha = t_\alpha(\epsilon_{\beta}^{0}+\eta \epsilon_{\beta}^{1}) = t_\alpha(\epsilon_{\beta}^{0}) + \eta \frac{\partial t_\alpha(\epsilon_{\beta}^{0})}{ \partial \epsilon^0_\gamma } \epsilon_\gamma^1 \\
& p_\alpha = p_\alpha(\epsilon_{\beta}^{0}+\eta \epsilon_{\beta}^{1}) = p_\alpha(\epsilon_{\beta}^{0}) + \eta \frac{\partial p_\alpha(\epsilon_{\beta}^{0})}{ \partial \epsilon^0_\gamma } \epsilon_\gamma^1 \\
& q_\alpha = q_\alpha(\eta^{-1}\psi^0_\beta+\psi^1_\beta) = q_\alpha(\eta^{-1}\psi^0_\beta) + \eta \frac{\partial q_\alpha(\eta^{-1}\psi^0_\beta)}{ \partial \psi^0_\gamma}\psi_\gamma^1
\end{split}
\end{equation}

which can be rewritten as  $t_{\alpha} = t^{0}_{\alpha} + \eta t^{1}_{\alpha}$; $p_{\alpha} = p^{0}_{\alpha} + \eta p^{1}_{\alpha}$; $q_{\alpha} = q^{0}_{\alpha} + \eta q^{1}_{\alpha}$, with the following conditions

\begin{equation} \label{ZeroOne-Terms-Def}
\begin{gathered}
t_\alpha^0=t_\alpha(\epsilon^0_\beta);  ~~~  p_\alpha^0=p_\alpha(\epsilon^0_\beta);  ~~~ q_\alpha^0=q_\alpha(\eta^{-1} \psi^0_\beta); \\
t_\alpha^1=\frac{\partial t^0_\alpha}{ \partial \epsilon^0_\gamma } \epsilon_\gamma^1; ~~~ p_\alpha^1=\frac{\partial p^0_\alpha}{ \partial \epsilon^0_\gamma } \epsilon_\gamma^1; ~~~ q_\alpha^1=\frac{\partial q^0_\alpha}{ \partial \psi^0_\gamma }\psi_\gamma^1
\end{gathered}
\end{equation}

This demonstrates that Equations \ref{motion-1-sep}, and \ref{motion-2-sep} are valid also in the case of nonlinear material behavior under the assumption that traction and moments at the $\mathcal{O}(\eta^{-1})$ scale are zero as required, in the linear case, by the rigid body motion of the RVE.

The RVE problem is governed by the $\mathcal{O}(\eta^{-1})$ terms in Equations \ref{motion-1-sep} and \ref{motion-2-sep}. Considering those terms and scaling back all the variables, one can write the $\mathcal{O}(\eta^{-1})$ equations as

\begin{equation}\label{RVE-1}
\sum_{\mathcal{F}_I}{{A}\, t^{0}_{\alpha} {\mbf e}_\alpha^{IJ}} = 0;  \hspace{0.25 in} \sum_{\mathcal{F}_I} {A}\, (\mbf {c}^I \times t^{0}_{\alpha} \mbf {e}_\alpha^{IJ} + m^{0}_{\alpha}{\mbf e}_\alpha^{IJ}) = 0
\end{equation}


Equations \ref{RVE-1} are force and moment equilibrium equations of each single particle inside the RVE subjected to $\mathcal{O}(1)$ facet traction $t_\alpha^0$ and moment $m_\alpha^0$ vectors, which, in turn, are functions of $\epsilon^0_\alpha$ and $\psi^0_\alpha$, consisting of a coarse-scale and a fine-scale term (see Equations \ref{eps-expansion-zero'} and \ref{curv-expansion-minus1'}). 
In other words, Equations \ref{RVE-1} state that the macroscopic strain, $\gamma_{ij} = v^{0}_{j,i} - \varepsilon_{ijk} \omega_k^{0}$, and curvature, $\kappa_{ij}=\omega_{j,i}^{0}$, tensors should be applied on all RVE facets as negative eigenstrains, and the fine-scale solution, in terms of displacements $u^1_i$ and rotations $\omega^1_i$ of each particle, must be calculated satisfying its force and moment equilibrium equations, while periodic boundary conditions are enforced on the RVE. The solution of the equilibrium equations also provides facet traction $t_\alpha^0$ and moment $m_\alpha^0$ vectors that are later used to compute the macroscopic stress and couple tensors.

\subsection{The Macroscopic Problem} \label{macro-derivation}
Finally, let us consider the $\mathcal{O}(1)$ equilibrium equations in Equations \ref{motion-1-sep} and \ref{motion-2-sep}. The $\mathcal{O}(1)$ translational equilibrium equation for each particle in the RVE reads

\begin{equation} \label{macro-1}
{M}_u^I \ddot {u}_i^{0I} = \eta \sum_{\mathcal{F}_I}{{A} \frac {\partial {t}^{IJ}_{i}}{\partial \epsilon^0_{\alpha}} \epsilon^1_{\alpha}} + {V}^I {b}^{0}_i
\end{equation}

where all the variables have been scaled back in the original system of reference, and $t_i^{IJ}=t^0_\beta e_{\beta i}^{IJ}$.
By using Equation \ref{U0} and by averaging the contribution of all particles in the RVE, one can write (see Appendix \ref{MacroEquil-Derivation} for details)

\begin{equation} \label{macro-1-1-averaged}
\rho_u \ddot {v}_i^{0} = \frac{1}{V_0}\sum_I \sum_{\mathcal{F}_I}{\eta A \frac {\partial {t}^{IJ}_{i}}{\partial \epsilon^0_{\alpha}} \epsilon^1_{\alpha}} + b_i
\end{equation}

where $V_0$ is the volume of the RVE; $\rho_u=\sum_I {M}^I_u/V_0$ is the mass density of the macroscopic continuum; $b_i=b^0_i (1 - \phi)$;  and $\phi= 1 -\sum_I {V}^I/V_0$ is the porosity of the macroscopic continuum. Equation \ref{macro-1-1-averaged} was derived under the assumption that $\sum_I {M}_u^I y^{I}_i=0$, which corresponds to the assumption that the local system of reference is the mass centroid of the particle system within the RVE.

Before proceeding with the derivations, let's take a closer look at the definition of $\epsilon^{1}_{\alpha}$ and the term $(\partial {t}^{IJ}_{i}/\partial \epsilon^0_{\alpha}) \epsilon^1_{\alpha}$ on the RHS of Equation \ref{macro-1-1-averaged}. Each facet in the material domain is shared between two particles, say $I$ and $J$. Therefore, by summing up the contributions of two adjacent particles, one obtains

\begin{eqnarray} \label{higher-strain-1}
\begin{aligned}
\frac {\partial {t}^{IJ}_{i}}{\partial \epsilon^0_{\alpha}} \epsilon^1_{\alpha} + \frac {\partial {t}^{JI}_{i}}{\partial \epsilon^0_{\alpha}} \epsilon^1_{\alpha}  = & \frac{1}{\bar{r}} \frac {\partial {t}_{i}^{IJ}}{\partial \epsilon^0_{\alpha}} \bigg[ \bigg( u^{1J}_{n,j} y^{IJ}_j + \varepsilon_{njk} \varphi_j^{1J} \bar c_{k}^{J} +  \varepsilon_{njk} \omega_{j,m}^{1J} y^{IJ}_m \bar c_{k}^{J} - \varepsilon_{njk} \varphi_j^{1I} \bar c_{k}^{I} \\
& \hspace{0.5 in} + \frac{1}{2} v^{0}_{n,jk} y^{IJ}_j y^{IJ}_k + \frac{1}{2} \varepsilon_{njk} \omega^{0}_{j,mo} y^{IJ}_m y^{IJ}_o  y_{k}^{c} + \varepsilon_{njk} \omega_{j,m}^{0} y^{IJ}_m \bar c_{k}^{J} \bigg) e^{IJ}_{\alpha n} \bigg] \\
& + \frac{1}{\bar{r}} \frac {\partial {t}_{i}^{JI}}{\partial \epsilon^0_{\alpha}} \bigg[ \bigg( u^{1I}_{n,j} y^{JI}_j + \varepsilon_{njk} \varphi_j^{1I} \bar c_{k}^{I} +  \varepsilon_{njk} \omega_{j,m}^{1I} y^{JI}_m \bar c_{k}^{I} - \varepsilon_{njk} \varphi_j^{1J} \bar c_{k}^{J} \\
& \hspace{0.5 in} + \frac{1}{2} v^{0}_{n,jk} y^{JI}_j y^{JI}_k + \frac{1}{2} \varepsilon_{njk} \omega^{0}_{j,mo} y^{JI}_m y^{JI}_o  y_{k}^{c} + \varepsilon_{njk} \omega_{j,m}^{0} y^{JI}_m \bar c_{k}^{I} \bigg) e^{JI}_{\alpha n} \bigg]
\end{aligned}
\end{eqnarray}

Considering the definition of the vector $\mathbf{y} ^{IJ} = \mathbf{y} ^{J} - \mathbf{y} ^{I}$, one can write $y_m^{IJ} = - y_m^{JI}$ and $\bar c^I_k-\bar c^J_k=y_k^{IJ}$. In addition,  $e^{IJ}_{\alpha i} = -  e^{JI}_{\alpha i}$ and $t_{i}^{IJ} = -t_{i}^{JI}$ hold for each facet. Finally, the sign of $\epsilon^{0}_\alpha$ does not change by interchanging $I$ and $J$ in its definition. This leads to $\partial t_{i}^{IJ}/\partial \epsilon^{0}_\alpha = - \partial t_{i}^{JI}/\partial \epsilon^{0}_\alpha$. Taking all above facts into account, Equation \ref{higher-strain-1} can be written as

\begin{eqnarray} \label{higher-strain-2}
\begin{aligned}
\frac {\partial {t}^{IJ}_{i}}{\partial \epsilon^0_{\alpha}} \epsilon^1_{\alpha} + \frac {\partial {t}^{JI}_{i}}{\partial \epsilon^0_{\alpha}} \epsilon^1_{\alpha}  =  \frac{1}{\bar{r}}\frac {\partial {t}_{i}^{IJ}}{\partial \epsilon^0_{\alpha}} \bigg[ y^{IJ}_m \bigg( & u^{1J}_{n,m} - u^{1I}_{n,m} + \varepsilon_{njk} \omega_{j,m}^{1J} \bar c_{k}^{J} - \varepsilon_{njk} \omega_{j,m}^{1I} \bar c_{k}^{I} \\ 
& + v^{0}_{n,jm} y^{IJ}_j - \varepsilon_{njk} \omega_{j,m}^{0} + \varepsilon_{njk} \omega^{0}_{j,mo} y^{IJ}_o  y_{k}^{c} \bigg) e^{IJ}_{\alpha n} \bigg]  \\
\end{aligned}
\end{eqnarray}

Comparing the expression inside the bracket on the RHS of Equation \ref{higher-strain-2} to the definition of $\epsilon^0_{\alpha}$ in Equation \ref{eps-expansion-zero'}, it can be concluded that

\begin{eqnarray} \label{higher-strain-3}
\begin{aligned}
\frac {\partial {t}^{IJ}_{i}}{\partial \epsilon^0_{\alpha}} \epsilon^1_{\alpha} + \frac {\partial {t}^{JI}_{i}}{\partial \epsilon^0_{\alpha}} \epsilon^1_{\alpha}  =  \frac {\partial {t}_{i}^{IJ}}{\partial \epsilon^0_{\alpha}} \frac {\partial \epsilon^0_{\alpha}}{\partial x_m} y_m^{IJ} = \frac {\partial {t}_{i}^{IJ}}{\partial x_m} y_m^{IJ} 
\end{aligned}
\end{eqnarray}

Therefore, one can average the term $(\partial {t}^{IJ}_{i}/\partial \epsilon^0_{\alpha}) \epsilon^1_{\alpha}$ on each facet and replace it with $1/2(\partial {t}_{i}^{IJ}/\partial x_m) y_m^{IJ}$ in the equilibrium Equations \ref{macro-1-1-averaged}, which can be rewritten as

\begin{equation} \label{macro-1-2-averaged}
\rho_u \ddot {v}_i^{0} = \frac{1}{2V_0} \sum_I \sum_{\mathcal{F}_I}{A} r \frac{\partial {t}^{IJ}_{i}}{\partial x_j} n_j^{IJ} + b_i 
\end{equation}

Finally, by considering that (1) $\partial (t^{IJ}_i n^{IJ}_j )/ \partial x_j = \partial t^{IJ}_i/ \partial x_j n^{IJ}_j + t^{IJ}_i \partial n^{IJ}_j/ \partial x_j$ and (2) $\partial n^{IJ}_j/ \partial x_j=0$ for the periodicity of the problem; and by recalling that $t_i^{IJ}=t^0_\alpha e_{\alpha i}^{IJ}$, one obtains

\begin{equation} \label{macro-eq-cont}
\rho_u \ddot{v}^0_i = \sigma^0_{ji,j} + b_i
\end{equation}

and 

\begin{equation} \label{macro-stress-formula}
\sigma^0_{ij} = \frac{1}{2V_0} \sum_I \sum_{\mathcal{F}_I}{A} r t^0_\alpha P_{i j}^{\alpha}
\end{equation}

Equation \ref{macro-eq-cont} is the classical partial differential equation governing the equilibrium of continua whereas Equation \ref{macro-stress-formula} provides the macroscopic stress tensor by averaging the solution of the RVE problem. It is worth mentioning that Equation \ref{macro-stress-formula} coincides with the virial stress formula for atomistic systems derived in Ref. \cite{Fish-3}, but it is also equivalent to the averaging formula used in the classical microplane model \cite{Bazant-2} formulation and derived through an energetic equivalence. 



The $\mathcal{O}(1)$ moment equilibrium equation is considered next. Since the purpose in this section is to average the equation of motion of all particles inside the RVE and derive the macroscopic equilibrium equation governing the entire RVE, to have a consistent formulation for all particles and RVEs, one must consider the moment of all forces with respect to a fixed point in space.   

For the generic particle $I$, by taking the moment of all forces with respect to the origin of a global macroscopic coordinate system as shown in Figure \ref{TwoScaleAnalysis}b and by considering the results of the $\mathcal{O}(\eta^{-2})$ problem, one can write (see Appendix \ref{MacroEquil-Derivation} for details)

\begin{equation}\label{macro-2} 
{M}_u^I \varepsilon_{ijk} X^{I}_j  \left( \ddot {v}_k^{0} + \varepsilon_{kmn} \eta^{-1} \ddot {\omega}_m^{0} x^{I}_n \right) + \eta^{-1}  {M_\theta^I} \ddot{\omega}_i^{0} = \eta \sum_{\mathcal{F}_I} A \left( {\frac {\partial {w}_{i}^{IJ}}{\partial \epsilon^0_{\alpha}} \epsilon^1_{\alpha}} + \frac {\partial {m}_{i}^{IJ}}{\partial \psi^{0}_{\alpha}} \psi^1_{\alpha} \right) + {V}^I \varepsilon_{ijk} X^{I}_j {b}_k^{0}
\end{equation}

where $X^I_j$ is the position vector of particle $I$ in global coordinate system; $w^{IJ}_i = \varepsilon_{ijk} X^C_j t^{IJ}_k$ is the moment of facet traction with respect to the point $O$; $X^C_j$ is the position vector of the contact point $C$ between the particles $I$ and $J$ in the global coordinate system, and $m_i^{IJ}=m^0_\beta e_{\beta i}^{IJ}$. Also, $x_j^I$  and $x_j^C$ are the position vectors of the particle $I$ and the contact point $C$ with respect to the mass center of the RVE, respectively. 



By summing up the moment equilibrium equations of all particles inside the RVE and dividing by the volume of the RVE, and considering that $X^I_j =  X_j + x^I_j$, one obtains (see Appendix \ref{MacroEquil-Derivation} for details)

\begin{equation}\label{macro-2-averaged-init} 
\frac{1}{V_0} \sum_I {M}_u^I \varepsilon_{ijk} X_j  \ddot {v}_k^{0} + \rho_{im}^{\theta} (\eta^{-1} \ddot{\omega}_m^{0}) = \frac{\eta}{V_0} \sum_I \sum_{\mathcal{F}_I} A \left( {\frac {\partial {w}_{i}^{IJ}}{\partial \epsilon^0_{\alpha}} \epsilon^1_{\alpha}} + \frac {\partial {m}_{i}^{IJ}}{\partial \psi^{0}_{\alpha}} \psi^1_{\alpha} \right)  + \frac{1}{V_0} \sum_I {V}^I \varepsilon_{ijk} X_j {b}_k^{0}
\end{equation}

where $\rho_{im}^{\theta} =\sum_I \left[ M_\theta^I \delta_{im} + M_u^I \varepsilon_{ijk} \varepsilon_{kmn} x_j^I x_n^I \right]/V_0$ is the inertia tensor of the RVE. In deriving Equation \ref{macro-2-averaged-init}, the particle density $M^I/V^I$ was assumed to be constant for all particles; and the local system of reference at the center of the RVE was chosen such that $\sum_I M_u^I x^I_i x^I_j =0$ for any $i \neq j$, i.e. as mentioned earlier in this paper, the axes of the system of reference are principal axes of inertia for the system of particles within the RVE. 



Before moving forward with the derivation, let's first consider the second term on the RHS of Equation \ref{macro-2-averaged-init}. For a facet in the material domain which is shared between particles $I$ and $J$, by summing the contribution of two particles $I$ and $J$ on the term $(\partial {m}_{i}^{IJ}/\partial \psi^{0}_{\alpha}) \psi^1_{\alpha}$ and by considering the definition of $\psi^1_{\alpha}$ (see Equation \ref{curv-expansion-zero'}), one gets

\begin{eqnarray} \label{higher-curv-1}
\begin{aligned}
\frac {\partial {m}_{i}^{IJ}}{\partial \psi^{0}_{\alpha}} \psi^1_{\alpha} + \frac {\partial {m}_{i}^{JI}}{\partial \psi^{0}_{\alpha}} \psi^1_{\alpha} = & \frac{1}{\bar{r}} \frac {\partial {m}_{i}^{IJ}}{\partial \psi^{0}_{\alpha}} \left[ \varphi_i^{1J} + \omega_{i,j}^{1J} y^{IJ}_j - \varphi_i^{1I} + \omega_{i,j}^{0} y^{IJ}_j + \frac{1}{2} \omega_{i,jk}^{0J} y^{IJ}_j y^{IJ}_k \right] {e}^{IJ}_{\alpha i} \\
& + \frac{1}{\bar{r}} \frac {\partial {m}_{i}^{JI}}{\partial \psi^{0}_{\alpha}} \left[ \varphi_i^{1I} + \omega_{i,j}^{1I} y^{JI}_j - \varphi_i^{1J} + \omega_{i,j}^{0} y^{JI}_j + \frac{1}{2} \omega_{i,jk}^{0I} y^{JI}_j y^{JI}_k \right] {e}^{JI}_{\alpha i}
\end{aligned}
\end{eqnarray}

Since the moment stress vector applied on a single facet belonging to two particles $I$ and $J$ are the same in magnitude but opposite in direction, one can write $m_{i}^{IJ} = -m_{i}^{JI}$; and consequently, $\partial m_{i}^{IJ}/\psi^{0}_\alpha = - \partial m_{i}^{JI}/\psi^{0}_\alpha$. In addition, the sign of $\psi^{0}_\alpha$ does not change by interchanging $I$ and $J$ in its definition, and that $y_m^{IJ} = - y_m^{JI}$, $e^{IJ}_{\alpha i} = -  e^{JI}_{\alpha i}$, Equation \ref{higher-curv-1} can be written as

\begin{eqnarray} \label{higher-curv-2}
\begin{aligned}
\frac {\partial {m}_{i}^{IJ}}{\partial \psi^{0}_{\alpha}} \psi^1_{\alpha} + \frac {\partial {m}_{i}^{JI}}{\partial \psi^{0}_{\alpha}} \psi^1_{\alpha} = & \frac{1}{\bar{r}} \frac{\partial {m}_{i}^{IJ}}{\partial \psi^{0}_\alpha} \bigg[ y^{IJ}_j \left(\omega_{n,j}^{1J} + \omega_{n,jk}^{0} y^{IJ}_k - \omega_{n,j}^{1I} \right) \bigg] e^{IJ}_{\alpha n}
\end{aligned}
\end{eqnarray}

If one compares the definition of $\psi^0_{\alpha}$ (see Equation \ref{curv-expansion-minus1'}) to the expression in the bracket on the RHS of Equation \ref{higher-curv-2}, it yields 

\begin{eqnarray} \label{higher-curv-3}
\begin{aligned}
\frac {\partial {m}_{i}^{IJ}}{\partial \psi^{0}_{\alpha}} \psi^1_{\alpha} + \frac {\partial {m}_{i}^{JI}}{\partial \psi^{0}_{\alpha}} \psi^1_{\alpha} = \frac{\partial {m}_{i}^{IJ}}{\partial \psi^{0}_\alpha} \frac{\partial \psi^{0}_\alpha}{\partial x_j} y^{IJ}_j = \frac{\partial {m}_{i}^{IJ}}{\partial x_j} y^{IJ}_j
\end{aligned}
\end{eqnarray}

As a result, one can replace the term $(\partial {m}_{i}^{IJ}/\partial \psi^{0}_{\alpha}) \psi^1_{\alpha}$ in Equation \ref{macro-2-averaged-init}, with the averaged expression derived in the above Equation \ref{higher-curv-3}. Similarly to the derivation relevant to the translational equation of motion, one can replace the term $(\partial {w}_{i}^{IJ}/{\partial \epsilon^0_{\alpha}}) \epsilon^1_{\alpha} $ on the RHS of Equation \ref{macro-2-averaged-init}, by the average value $1/2(\partial {w}_{i}^{IJ}/{\partial x_m}) y_m^{IJ}$ for each facet. Equation \ref{macro-2-averaged-init} can be then rewritten as

\begin{eqnarray}\label{macro-2-3-averaged} 
\begin{aligned}
\rho_{im}^{\theta} (\eta^{-1} \ddot{\omega}_m^{0}) = \frac{\eta}{2V_0}\sum_I \sum_{\mathcal{F}_I} {{A} \bigg( \frac{\partial w^{IJ}_i}{\partial x_j} y^{IJ}_j} + \frac{\partial m^{IJ}_i}{\partial x_j} y^{IJ}_j \bigg) + \frac{1}{V_0} \sum_I \bigg( {V}^I \varepsilon_{ijk} X_j {b}_k^{0} - {M}_u^I \varepsilon_{ijk} X_j  \ddot {v}_k^{0} \bigg)
\end{aligned}
\end{eqnarray} 

Using ${X}^C_j = {X}_j + {x}^C_j$ in the definition of $w^{IJ}_i$ along with the identity equations $\partial (m^{IJ}_i n^{IJ}_j )/ \partial x_j = \partial m^{IJ}_i/ \partial x_j n^{IJ}_j$ and $\partial (w^{IJ}_i n^{IJ}_j )/ \partial x_j = \partial w^{IJ}_i/ \partial x_j n^{IJ}_j$, Equation \ref{macro-2-3-averaged}  can be written as

\begin{eqnarray}\label{macro-2-4-averaged} 
\begin{aligned}
\rho_{im}^{\theta} (\eta^{-1} \ddot{\omega}_m^{0}) = & \frac{\eta}{2V_0}\sum_I \sum_{\mathcal{F}_I} {A} (y^{IJ}_j  \varepsilon_{imk} X_m t^{IJ}_k)_{,j} + \frac{\eta}{2V_0}\sum_I \sum_{\mathcal{F}_I} {A} (y^{IJ}_j \varepsilon_{imk} x^C_m t^{IJ}_k + y^{IJ}_j m^{IJ}_i)_{,j} \\
& + ( \varepsilon_{ijk} X_j {b}_k - \rho_u \varepsilon_{ijk} X_j  \ddot {v}_k^{0} )
\end{aligned}
\end{eqnarray} 

The last term on the RHS of Equation \ref{macro-2-4-averaged} is written considering the fact that $b_k^{0}$ and $\ddot {v}_k^{0}$ are equal for all particles inside the RVE. Furthermore, the first term on the RHS of Equation \ref{macro-2-4-averaged} can be expanded as $(y^{IJ}_j  \varepsilon_{imk} X_m t^{IJ}_k)_{,j} = \varepsilon_{ijk} y^{IJ}_j t^{IJ}_k + \varepsilon_{imk} X_m  (y^{IJ}_j  t^{IJ}_{k})_{,j}$, in which $\partial y^{IJ}_j/ \partial x_j=0$ is used. Therefore, Equation \ref{macro-2-4-averaged} becomes



\begin{eqnarray}\label{macro-2-5-averaged} 
\begin{aligned}
\rho_{im}^{\theta} (\eta^{-1} \ddot{\omega}_m^{0}) = & \frac{1}{2V_0}\sum_I \sum_{\mathcal{F}_I} {A} \varepsilon_{ijk} x^{IJ}_j t^{IJ}_k+ \frac{1}{2V_0}\sum_I \sum_{\mathcal{F}_I} {A} (x^{IJ}_j \varepsilon_{imk} x^C_m t^{IJ}_k + x^{IJ}_j m^{IJ}_i)_{,j} \\
& +  \varepsilon_{ijk} X_j \bigg( \frac{1}{2V_0}\sum_I \sum_{\mathcal{F}_I} A (x^{IJ}_m  t^{IJ}_{k})_{,m} + {b}_k^{0} - \rho_u \ddot {v}_k^{0} \bigg)
\end{aligned}
\end{eqnarray} 

The last term on the RHS of Equation \ref{macro-2-5-averaged} is the moment of the translational equilibrium equation of the RVE (see Equation \ref{macro-1-2-averaged}) around the origin of the macroscopic global coordinate system; therefore, it is equal to zero.  Comparing the first term on RHS of Equation \ref{macro-2-5-averaged} with definition of macroscopic stress tensor of the RVE in Equation \ref{macro-stress-formula}, one can replace it with $\varepsilon_{ijk} \sigma^0_{jk}$. The second term on the RHS of Equation \ref{macro-2-5-averaged} is the divergence of the averaged moment stress tensor of the RVE. The macro-scale rotational equation of motion can be then written as follows


\begin{equation} \label{macro-rotational-final} 
\begin{gathered}
\rho_{\theta ij} (\eta^{-1} \ddot {\omega}_j ^{0}) = {\varepsilon}_{ijk} {\sigma}_{ij}^0 + \frac{\partial \mu^0_{ji}}{\partial x_j}  
\end{gathered}
\end{equation}

where

\begin{equation} \label{macro-momentstress-formula} 
\begin{gathered}
\mu^0_{ij} = \frac{1}{2V_0}\sum_I \sum_{\mathcal{F}_I} {A}r (m_{\alpha}^0 P_{ij}^{\alpha}  + t_{\alpha}^0 Q_{ij}^{\alpha})
\end{gathered}
\end{equation}

and the matrix $Q_{ij}^{\alpha}$ is defined as $Q_{ij}^{\alpha} = n_i^{IJ} \varepsilon_{jkl} x^C_k e_{\alpha l}^{IJ}$. $\mu^0_{ij}$ is the macroscopic moment stress tensor calculated using the results of RVE analysis, and Equation \ref{macro-rotational-final} corresponds to the classical rotational equilibrium equation of Cosserat continua \cite{Chan-1, Cosserat-1}. According to Equation \ref{macro-momentstress-formula} for macroscopic moment stress tensor and considering that ${x}^C_k = {x}^{I}_k + {c}^I_k$, one can conclude that $\mu^0_{ij}$ consists of three terms: (1) the effect of the facet couple traction $\mbf{m}$; (2) the effect of the moment of the facet stress traction $\mbf{t}$ around the particle node which the facet belongs to, and (3) the effect of the moment of the facet stress traction $\mbf{t}$, transferred to the particle node, around the centroid of the RVE. As result, the moment stress tensor is characterized by three length scales: (1) the facet size, associated to $\mbf{m}$; (2) the particle size or facet spacing; and (3) the size of the RVE.


\section{Numerical Results} \label{NumRes}
The homogenization theory formulated and discussed in the previous sections was implemented in the MARS computational software \cite{mars-1} with the objective of upscaling the Lattice Discrete Particle Model (LDPM). LDPM, formulated, calibrated, and validated by Cusatis and coworkers \cite{cusatis-ldpm-1,cusatis-ldpm-2}, is a meso-scale discrete model which simulates the mechanical interaction of concrete coarse aggregate pieces. LDPM has shown superior capabilities in modeling concrete behavior under dynamic loading \cite{cusatis-Jovanca, cusatis-alonerate}, Alkali Silica Reaction (ASR) deterioration \cite{cusatis-mohammed}, as well as failure and fracture of fiber-reinforced concrete \cite{cusatis-Ed1, cusatis-Ed2}. 

\begin{figure}[h]
        \centering
        \begin{subfigure}[b]{0.45\textwidth}
                \centering
                \includegraphics[scale=0.4]{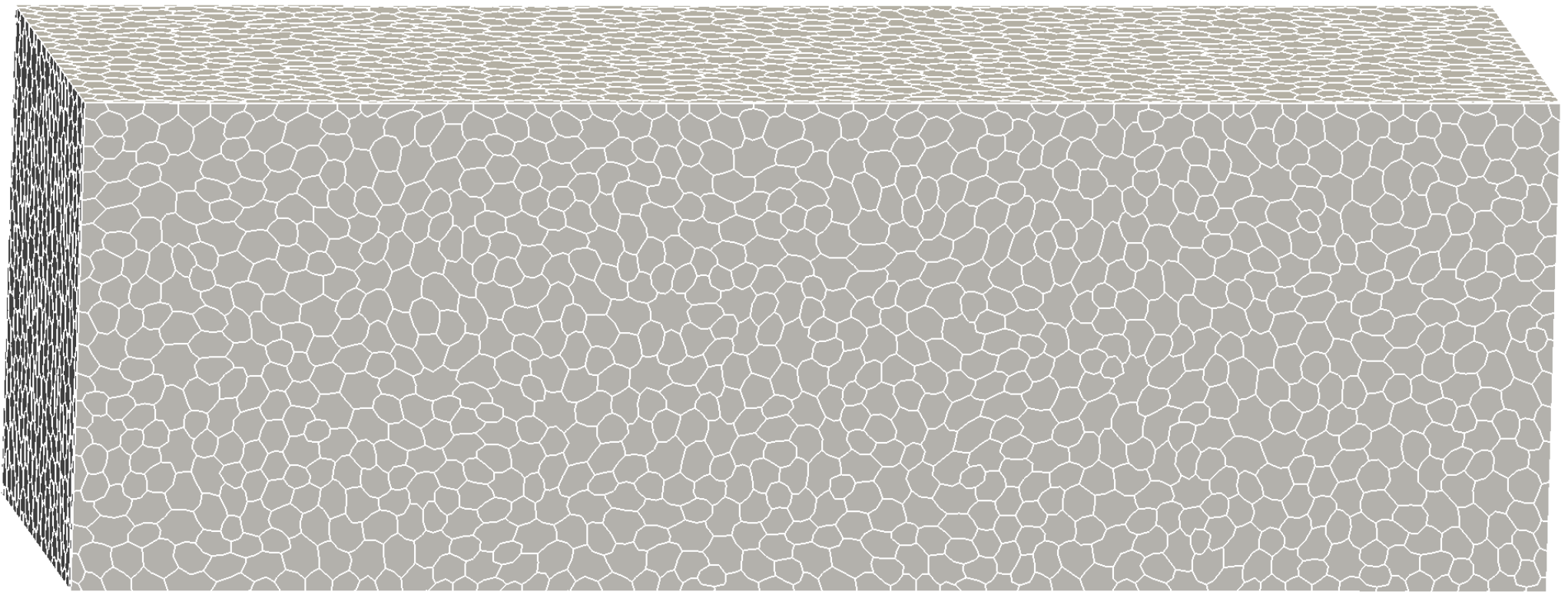}
                \caption{}
        \end{subfigure}
        \begin{subfigure}[b]{0.45\textwidth}
                \centering
                \includegraphics[scale=0.4]{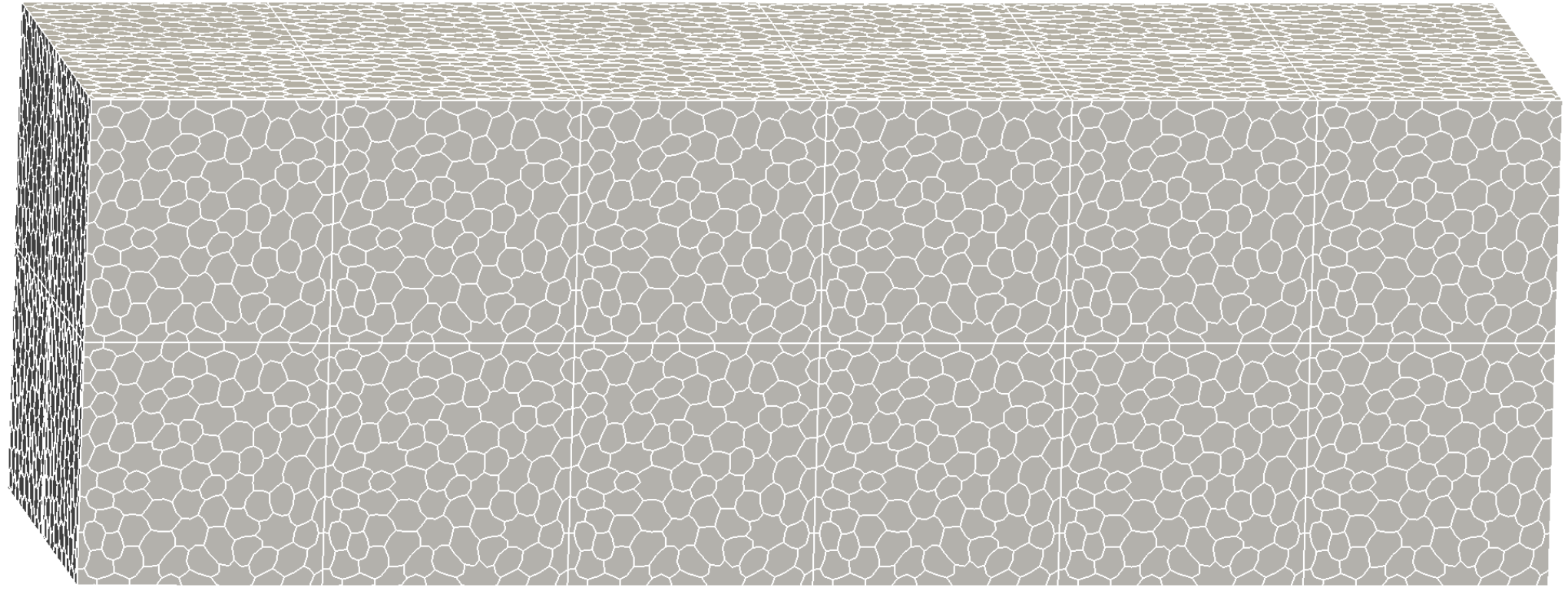}
                \caption{}
        \end{subfigure}
        \caption{Polyhedral particle distribution in a LDPM prism: (a) generic LDPM system, (b) Periodic LDPM system.}
        \label{PeriodicLDPM}
\end{figure}

The complete LDPM  formulation is summarized in Appendix \ref{LDPM}. It is worth mentioning here that the LDPM computational units are polyhedral cells whose construction is anchored to the Delaunay triangulation of the simulated concrete aggregate pieces that are assumed to be spherical and size-graded according to the Fuller size distribution. In the LDPM formulation, each polyhedral cell represents one concrete spherical aggregate piece embedded in the surrounding mortar and the interfaces among the cells represent potential mortar cracks.  Figure \ref{PeriodicLDPM}a shows a typical LDPM system of polyhedral cells and Figure \ref{PeriodicLDPM}b its periodic approximation.


\begin{figure}[t]
\centering 
{\includegraphics[width=0.7\textwidth]{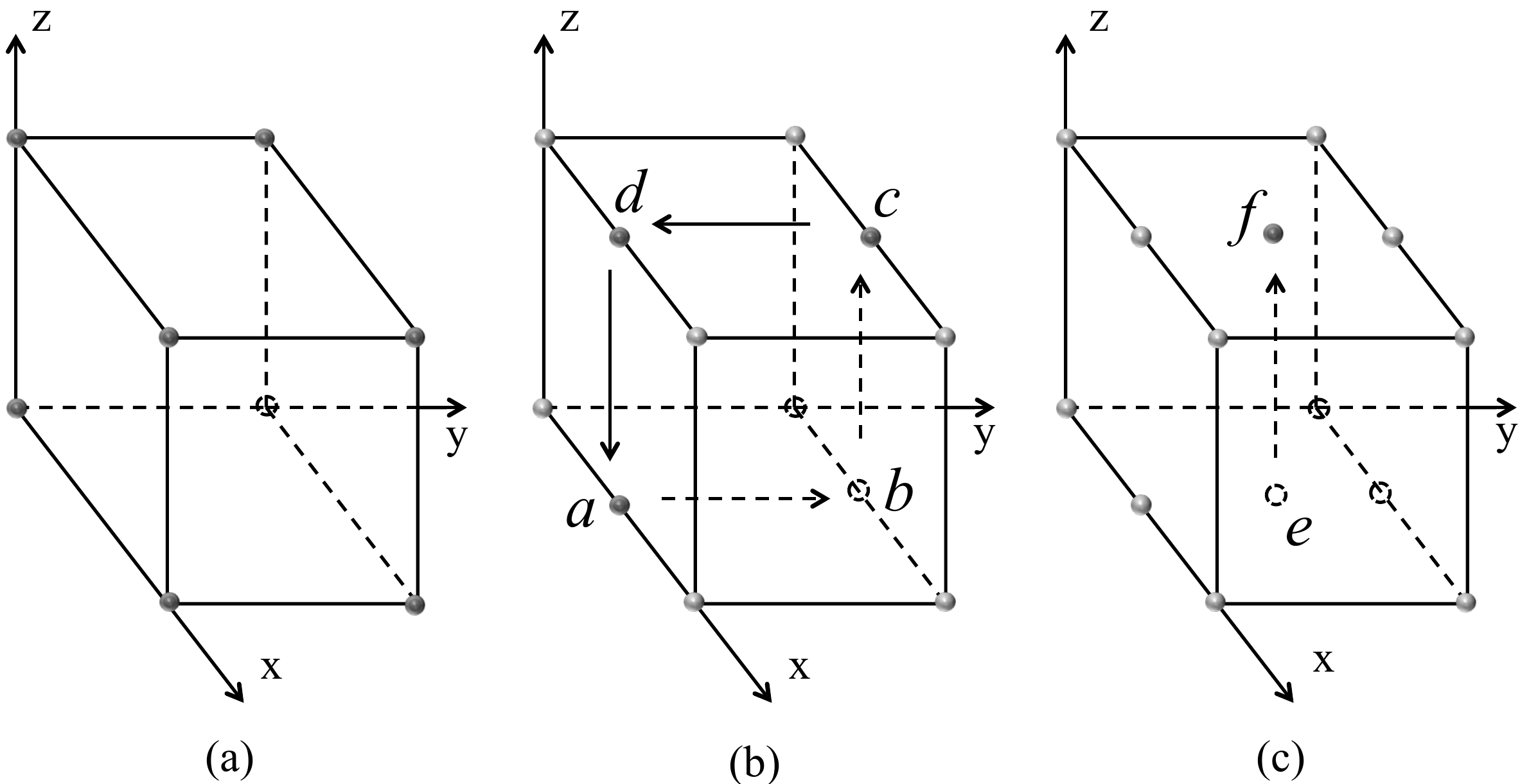}} 
\caption{RVE generation procedure (a) Corner nodes (b) Edge nodes (c) Face nodes}
\label{RVEgen}
\end{figure}

The generic RVE shown in Figure \ref{PeriodicLDPM}b is constructed as follows. 
Eight nodes are created at the vertexes of a cube (Figure \ref{RVEgen}a). Then nodes are randomly placed on a RVE edge parallel to $x$ axis, see node $a$ in Figure \ref{RVEgen}b. Then, these nodes are duplicated on the other three parallel edges along the $x$ axis, see nodes $b, c$, and $d$ in Figure \ref{RVEgen}b. 
Similar procedure is carried out over the edges parallel to $y$ and $z$ axes. Next, the node generation on the RVE surfaces is performed by randomly placing nodes on a cube face with $z$ axis as normal vector, see node $e$ in Figure \ref{RVEgen}c. The same nodes are then duplicated on the opposite RVE faces, see node $f$ in Figure \ref{RVEgen}c. Nodes on parallel cube faces with $x$ and $y$ axes as normal vectors are constructed with the same algorithm. Finally, nodes are placed inside the RVE based on the general LDPM procedure (see Appendix \ref{LDPM} and relevant publications \cite{cusatis-ldpm-1} for details).



As mentioned earlier in this paper the RVE analysis is conducted by imposing periodic boundary conditions. This is obtained by setting the displacements and rotations of the RVE vertexes to be zero and by imposing, through a master-slave constraint, that the periodic edge nodes and face nodes have the same rotations and displacements. 

The overall multiscale numerical procedure adopted in this paper can be summarized as follows.
\begin{itemize}

\item The finite element method is employed to solve the macro-scale homogeneous problem in which external loads and essential BCs are applied incrementally. During each numerical step, strain increments $\Delta{\gamma}_{ij} = \Delta{v}^{0}_{j,i} - \varepsilon_{ijk} \Delta{\varphi}_k^{0}$ and curvature increments $\Delta{\kappa}_{ij}=\Delta{\omega}_{j,i}^{0}$ tensors are calculated at each integration point based on the nodal displacement and rotation increments of the corresponding finite element.

\item The macroscopic strain and curvature increments  are projected into the RVE facets through the proper projection operators: $\Delta{\epsilon}^c_{\alpha} = P^\alpha_{ij} \left(\Delta{\gamma}_{ij}  + \varepsilon_{jmn}  \Delta{\kappa}_{im} y_{n}^{c} \right)$ and $\Delta{\psi}^c_{\alpha} =  P^\alpha_{ij} \Delta{\kappa_{ij}}$. These projected strains and curvatures are imposed, upon sign change, as eigen-strains and eigen-curvatures, $\Delta{\epsilon}^0_{\alpha} =\Delta{\epsilon}^c_{\alpha}+\Delta{\epsilon}^f_{\alpha}=\Delta{\epsilon}^c_{\alpha}- (-\Delta{\epsilon}^f_{\alpha})$  and $\Delta{\psi}^0_{\alpha} =\Delta{\psi}^c_{\alpha}+\Delta{\psi}^f_{\alpha}=\Delta{\psi}^c_{\alpha}- (-\Delta{\psi}^f_{\alpha})$ (See section \ref{RVEproblem}), to the RVE allowing the calculation of the fine-scale solution governed by the fine-scale constitutive equations. 

\item Finally, the fine-scale facet tractions and moments are used to compute, through Equations \ref{macro-stress-formula} and \ref{macro-momentstress-formula}, the macroscopic stresses, $\sigma^0_{ij}$, and couple stresses, 
$\mu_{ij}^0$, for each Gauss point in the FE mesh.

\end{itemize}


\subsection{Elastic RVE Analysis} \label{Elastic Analysis}
This section presents the analysis of the elastic macroscopic behavior of one LDPM RVE. The macroscopic homogenized behavior is analyzed with reference to the classical constitutive equation for Cosserat elasticity, which, in non-dimensional variables, can be written as:

\begin{equation}\label{Cauchy-Constitutive} 
\hat{\sigma}_{ij} =  p_0\hat{\gamma}_{kk} \delta_{ij} + p_1\hat{\gamma}_{(ij)}  +  p_2 \hat{\gamma}_{[ij]}~; ~~~\hat{\mu}_{ij} = q_0 \hat{\kappa}_{kk} \delta_{ij} + q_1 \hat{\kappa}_{(ij)} + q_2 \hat{\kappa}_{[ij]}
\end{equation}  
where $\hat{\sigma}_{ij} = \sigma_{ij} / (2 \mu + \chi)$ and $\hat{\mu}_{ij}= L\mu_{ij} /[(2 \mu+\chi)D^2]$ are the normalized stress and couple tensors, $L$ = characteristic size of the structure of interest, $D$ = size of the RVE;  $\hat{\gamma}_{ij}=\gamma_{ij}$,  $\hat{\kappa}_{ij} =L \kappa_{ij}$ normalized strains and curvatures; $p_0=\lambda/ (2 \mu + \chi) $, $p_1=1$; $p_2 = \chi / (2 \mu+\chi)$;  $q_0=\pi_1/[(2 \mu + \chi)D^2]$, $q_1=(\pi_2+\pi_3)/[(2 \mu + \chi)D^2]$; $q_2=(\pi_2-\pi_3)/[(2 \mu + \chi)D^2]$; $\delta_{ij}$ = kronecker delta; $\mu$, $\lambda$, $\chi$, $\pi_1$, $\pi_2$, and $\pi_3$ are the elastic constants; and the subscript parentheses and brackets represent extraction of the symmetric and antisymmetric, respectively, part of the tensors.

In this section, eight different LDPM RVE sizes $D$= 15, 20, 25, 35, 50, 75, 100, and 150 mm are considered and 5 RVEs, characterized by different placement of the aggregate pieces, is studied for each case. It is worth mentioning that, in LDPM, different spherical aggregate placement inside the RVE yield to different RVE polyhedral particle configurations. The numerical calculations were performed by assuming the concrete mix design and model parameters reported in Appendix \ref{LDPM}. Figure \ref{young-poisson-norm}a shows the homogenized values of $p_0$, $p_2$, and of the normalized Young's modulus defined as $e=E/(2\mu +\chi)=(3 \lambda +2\mu +\chi)/(2 \lambda +2\mu +\chi)$, as function of the RVE size normalized by the maximum spherical aggregate size, $d=D/d_a$. The error bars represent the scatter in the results obtained by simulating 5 different RVEs of the same size but with different realization of spherical aggregate positions inside the RVE. As one can see, the calculated values of the parameters tend to converge to a constant value as the size of the RVE increases and, at the same time, the results become independent of the spherical aggregate distribution inside the RVE. The value of $p_2$ is very close to zero for all RVE sizes and decreases rapidly with respect to the RVE size; this suggests that, for the analyzed fine-scale model, the homogenized stress tensor is symmetric. This result is due to the fact that in the LDPM formulation facet moments are zero, and this leads to facet traction distributions around each particle that have zero moment resultant around the particle node. In Figure \ref{young-poisson-norm}b the homogenized Poisson's ratio is reported based on the equation $\nu = \lambda / (2 \lambda +2\mu +\chi)$ and the calculated asymptotic value, 0.18, corresponds well with the value of 0.175 calculated by exploiting the equivalence between particle models and microplane models \cite{cusatis-ldpm-1}. Finally, Figure \ref{young-poisson-norm}c shows the homogenized parameters, $q_0$, $q_1$, and $q_2$, as a function of the RVE size. These quantities also converge to an asymptotic value and become independent of the RVE spherical aggregate distribution for large enough value of $D/d_a$. By virtue of these results and by recalling the definitions of $q_0$, $q_1$, and $q_2$, it is interesting to note that the macroscopic Cosserat elastic parameters of the homogenized continuum depend quadratically on the RVE size. 


\begin{figure}
        \centering
        \begin{subfigure}[b]{0.32\textwidth}
                \centering
                \includegraphics[width=\textwidth]{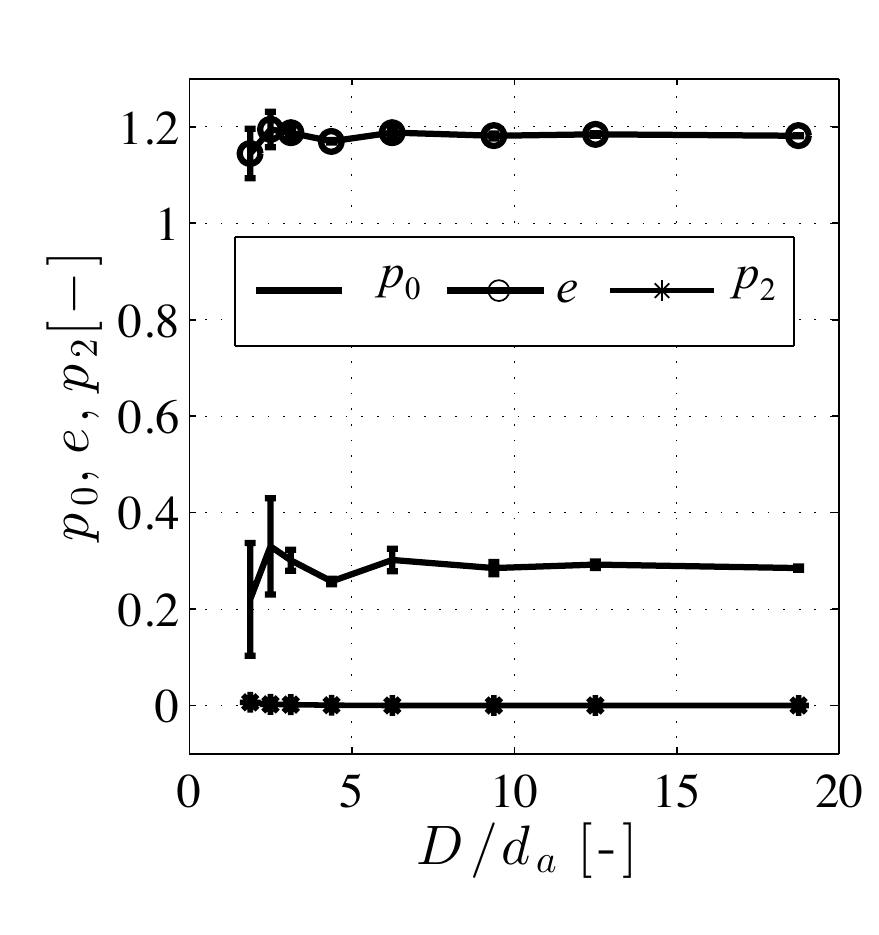}
                \caption{}
                \label{young}
        \end{subfigure}
        \begin{subfigure}[b]{0.32\textwidth}
                \centering
                \includegraphics[width=\textwidth]{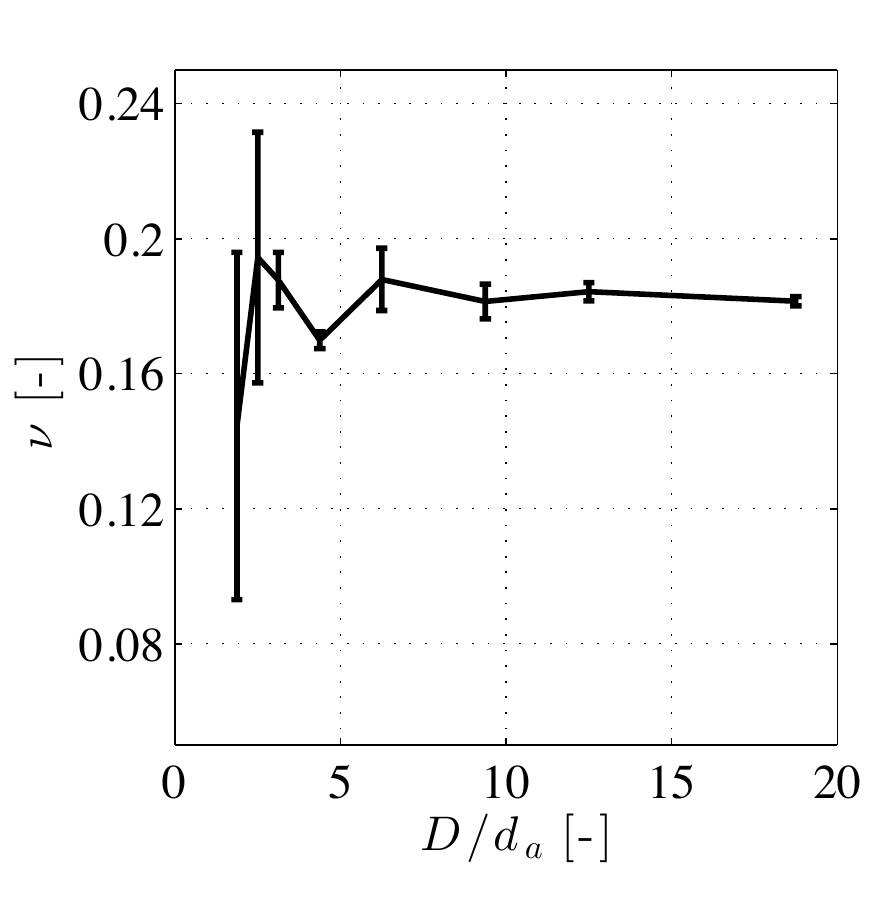}
                \caption{}
                \label{poisson}
        \end{subfigure}
        \begin{subfigure}[b]{0.32\textwidth}
                \centering
                \includegraphics[width=\textwidth]{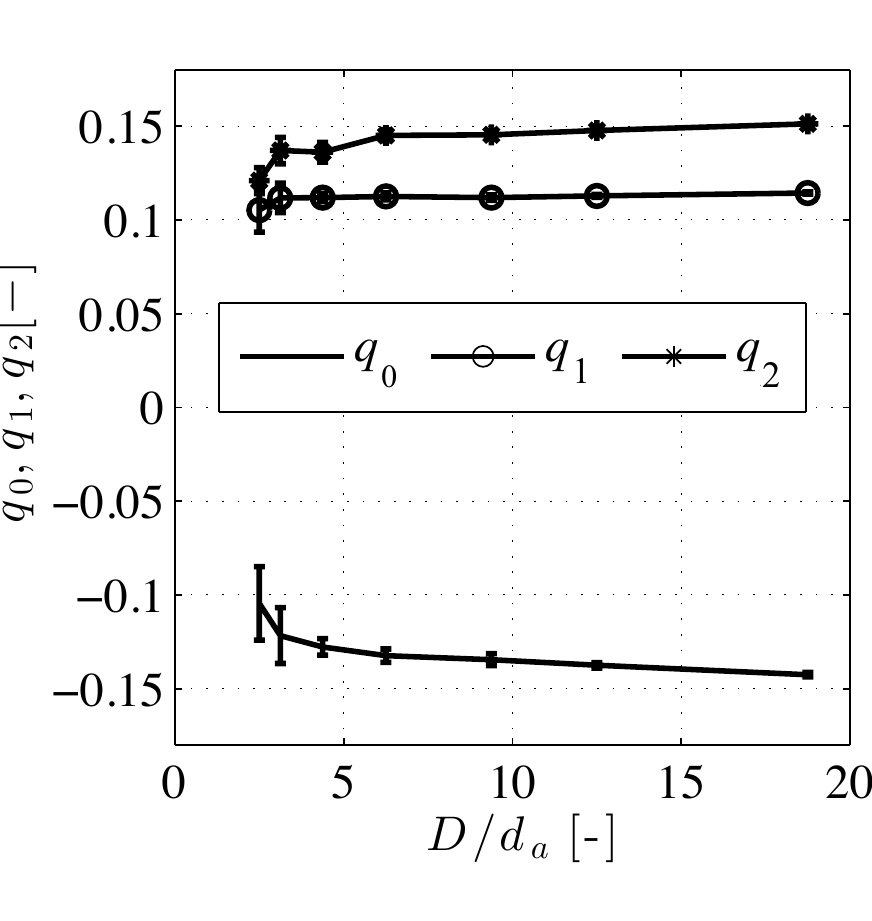}
                \caption{}
                \label{muc}
        \end{subfigure}
        \caption{Variation of elastic normalized effective material properties: (a) $p_0$, $p_2$ and normalized Young modulus $E$. (b) $\nu$ Poisson's ratio. (c) $q_0$, $q_1$ and $q_2$, with respect to the ratio of RVE size to maximum spherical aggregate size.}
        \label{young-poisson-norm}
\end{figure}

\subsection{Nonlinear RVE Analysis}
\begin{figure}[t]
	     \centering 
        \begin{subfigure}[b]{0.1\textwidth}
                \centering
                \includegraphics[width=0.9\textwidth]{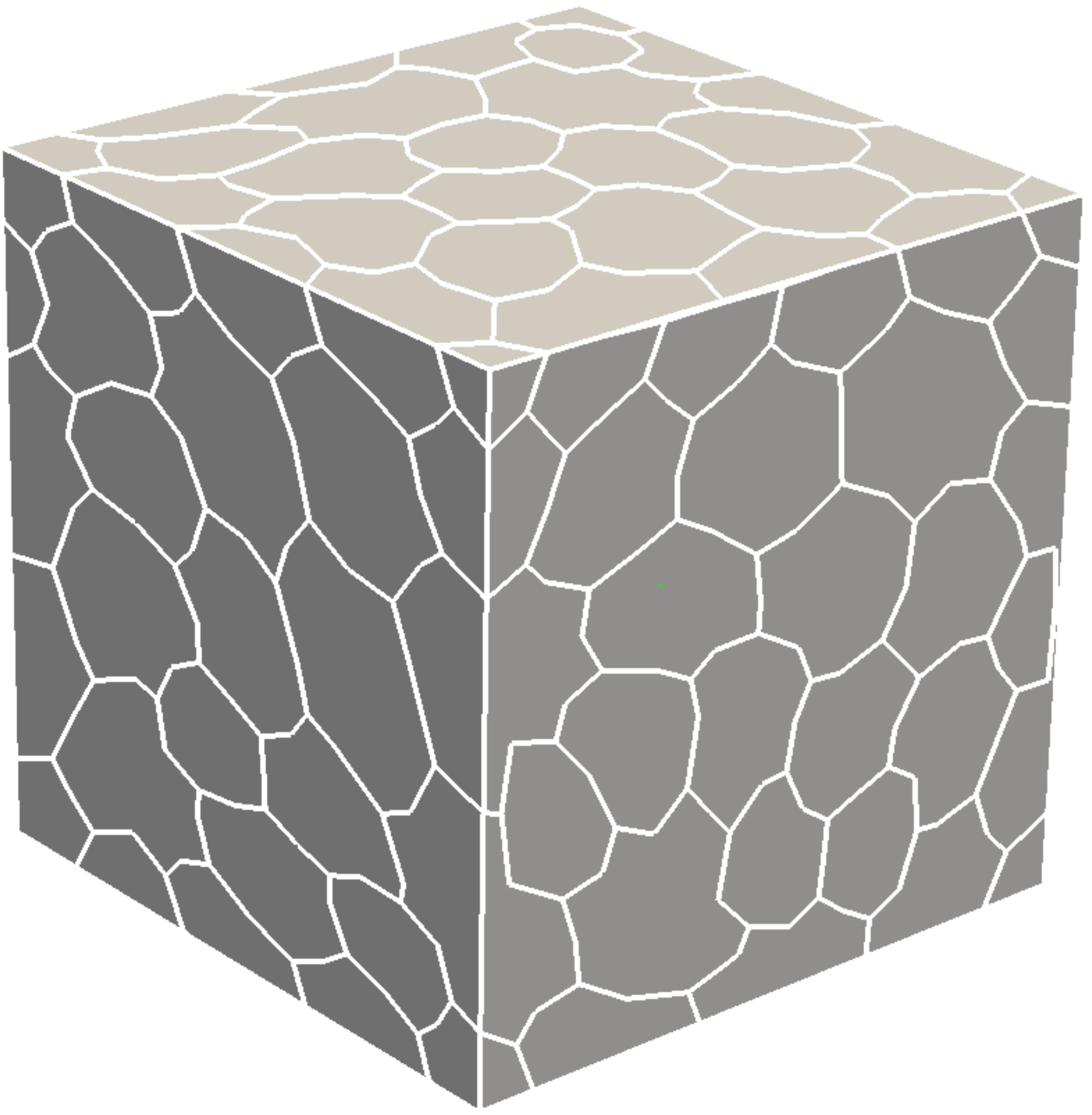}
                \caption{}
        \end{subfigure}
        \begin{subfigure}[b]{0.3\textwidth}
                \centering
                \includegraphics[width=0.55\textwidth]{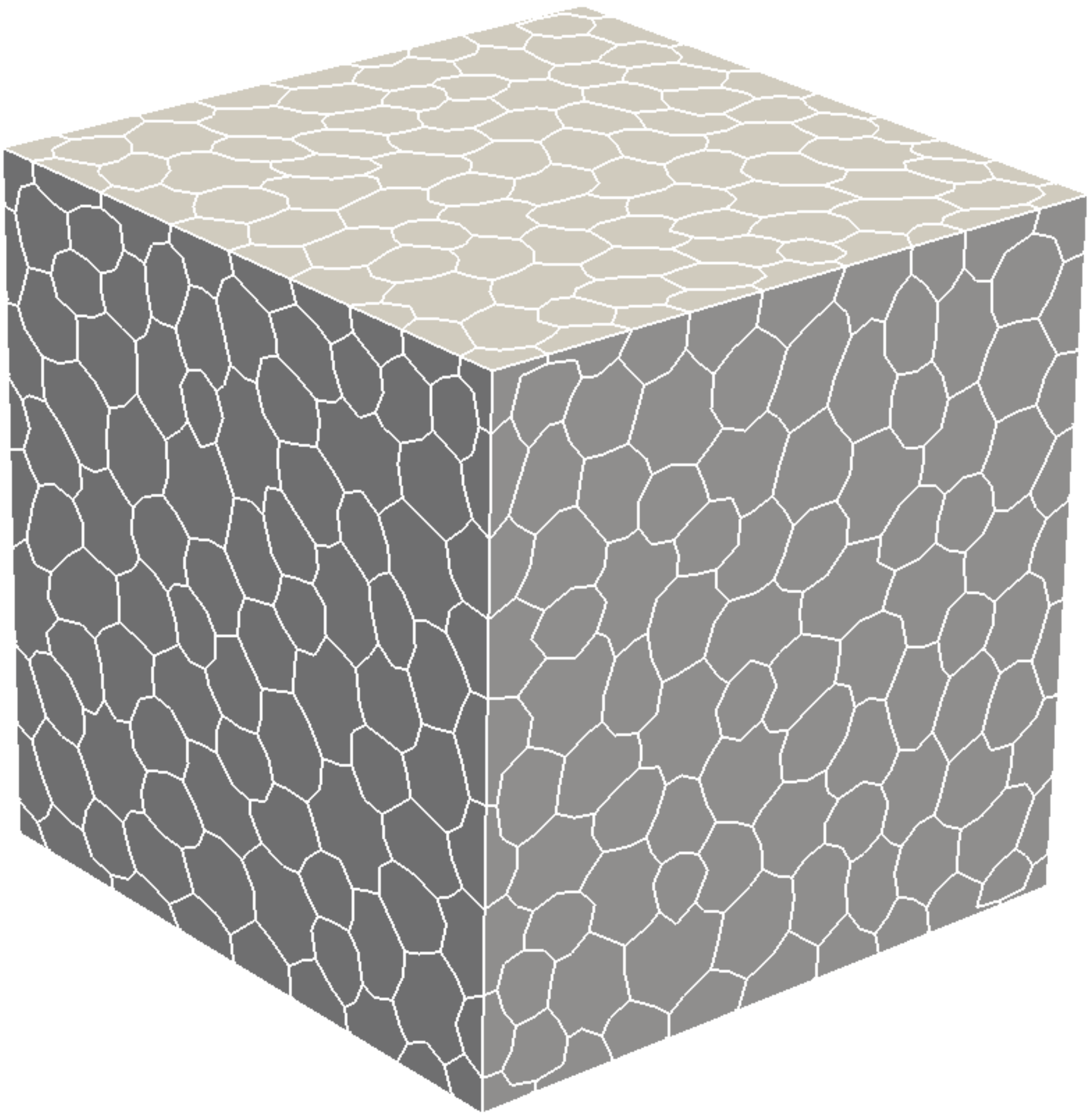}
                \caption{}
        \end{subfigure}
        \begin{subfigure}[b]{0.3\textwidth}
                \centering
                \includegraphics[width=1\textwidth]{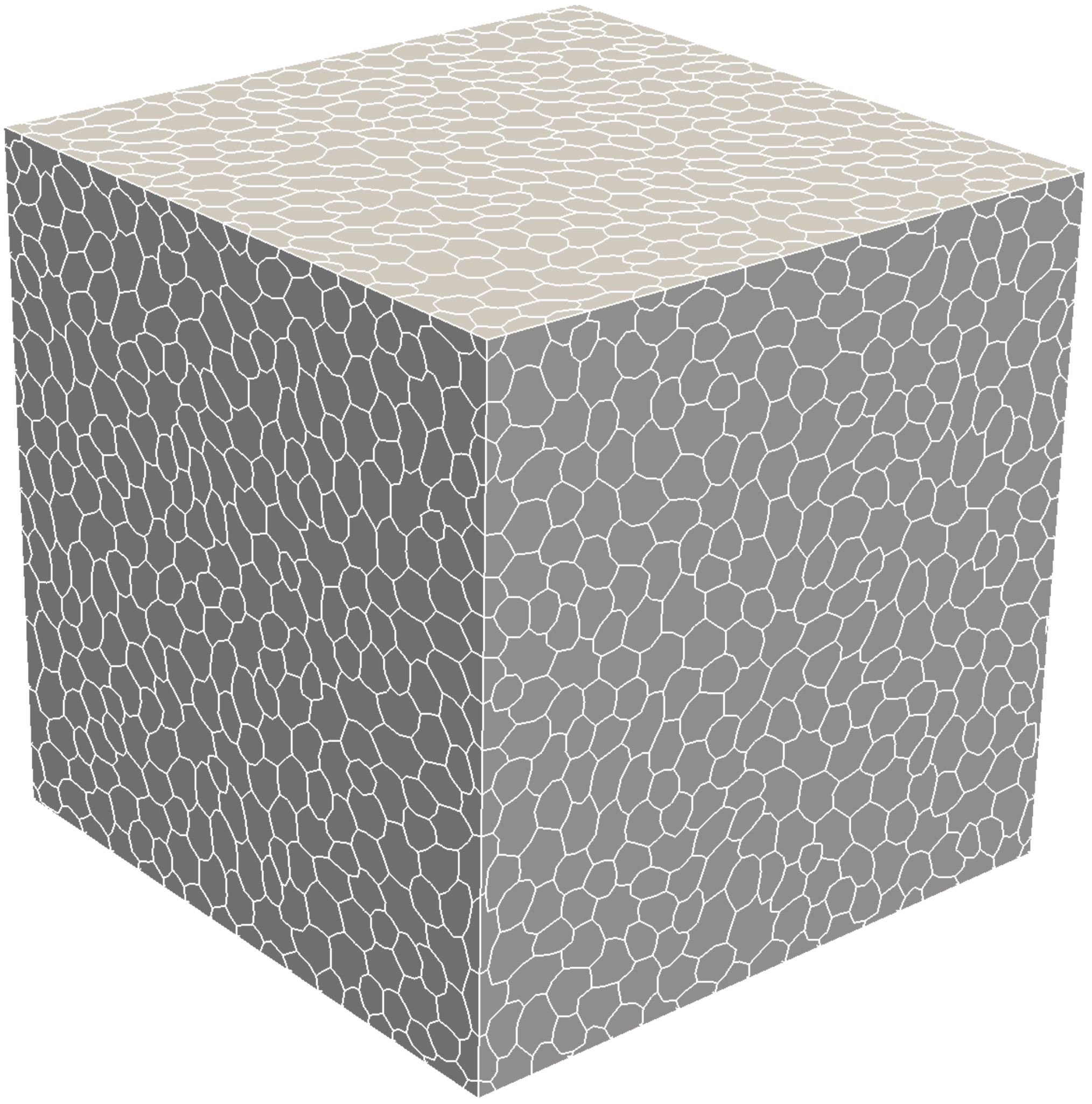}
                \caption{}
        \end{subfigure}
\caption{RVE geometry and polyhedral particle distribution: (a) 25 mm (b) 50 mm (c) 100 mm}
\label{RVEgeom}
\end{figure}

In this section, the nonlinear response of the RVE is investigated under different strain and curvature loading conditions. Three different RVE sizes, $D=$25, 50, and 100 mm, and 7 different spherical aggregate placement inside the RVE are considered for each case. Typical polyhedral particle systems and geometry of each RVE size are shown in Figure \ref{RVEgeom}. The nonlinear homogenized behavior of the RVE is studied under the effect of uniaxial strain tension and compression, hydrostatic compression, bending and torsional curvatures. In the following numerical examples, concrete mix design and model parameters are the same as the ones used in the elastic analysis.

\subsubsection{Nonlinear Analysis of RVE subject to components of the strain tensor}\label{non-ten} 
\begin{figure}[h]
        \centering
        \begin{subfigure}[b]{0.3\textwidth}
                \centering
                \includegraphics[width=\textwidth]{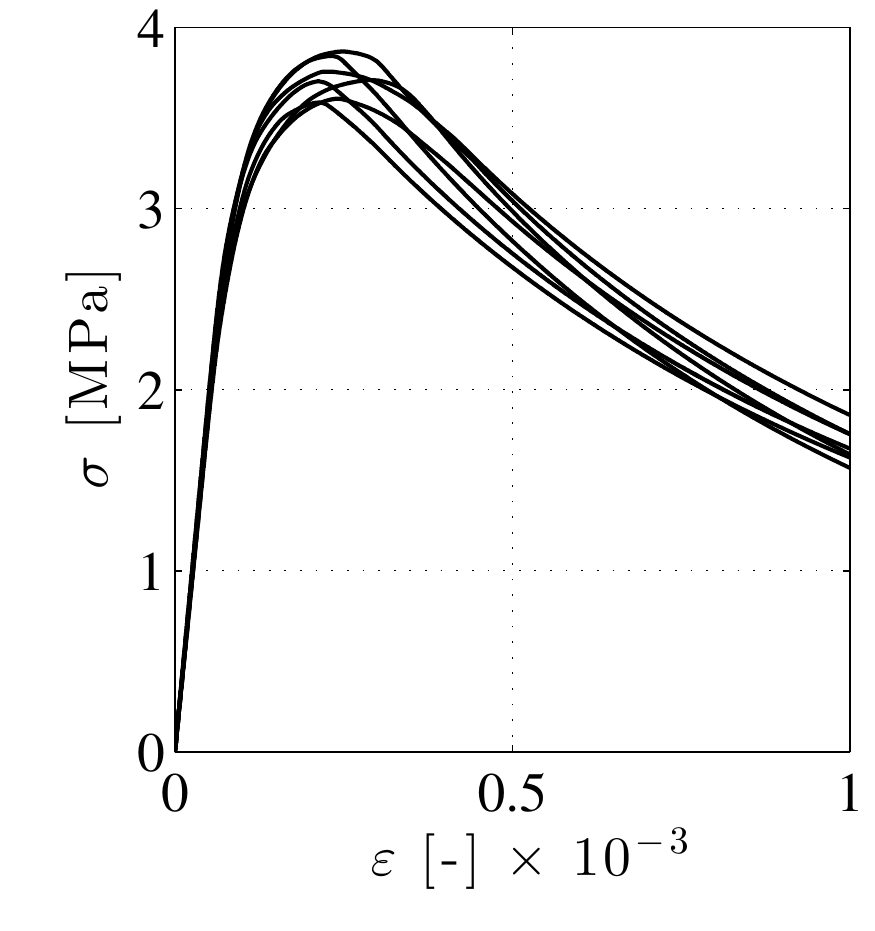}
                \caption{25 mm}
                \label{T25mm}
        \end{subfigure}
        \begin{subfigure}[b]{0.3\textwidth}
                \centering
                \includegraphics[width=\textwidth]{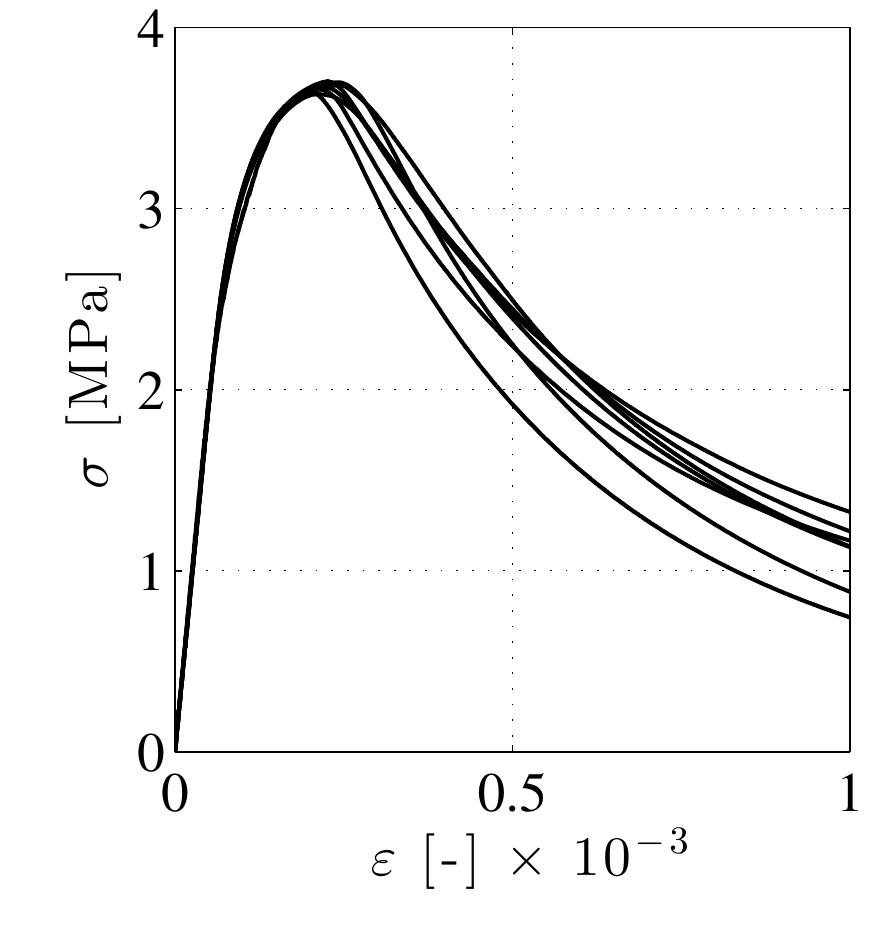}
                \caption{50 mm}
                \label{T50mm}
        \end{subfigure}
        \begin{subfigure}[b]{0.3\textwidth}
                \centering
                \includegraphics[width=\textwidth]{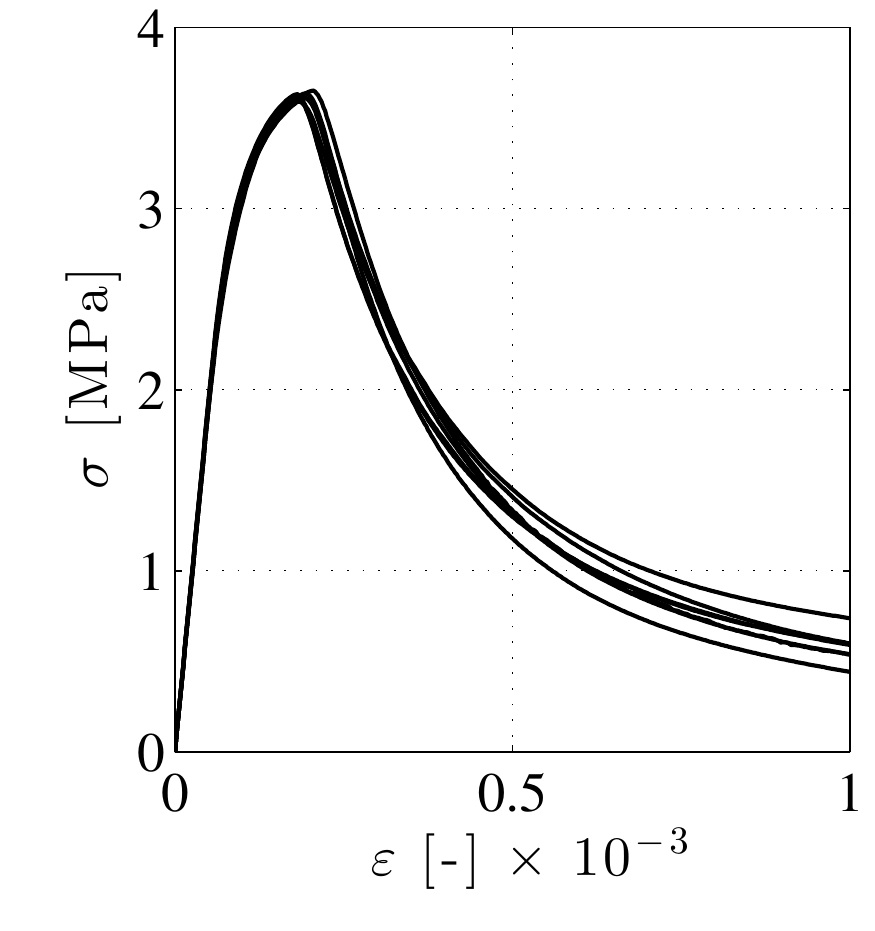}
                \caption{100 mm}
                \label{T100mm}
        \end{subfigure}
        \caption{Macroscopic stress-strain curve for three different RVE size: 25mm, 50mm, 100mm under uni-axial tension}
        \label{stress-strainNonlinear-tens}
\end{figure}

\begin{figure}[h]
        \centering
        \begin{subfigure}[b]{0.3\textwidth}
                \centering
                \includegraphics[width=\textwidth]{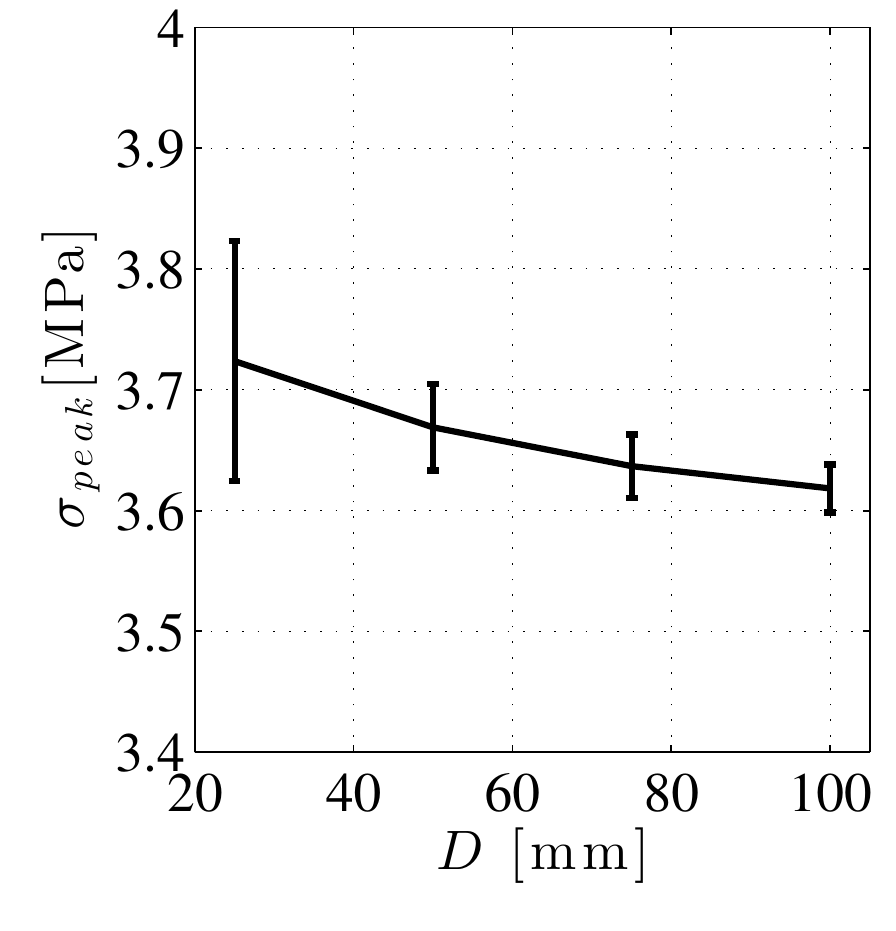}
                \caption{}
                \label{PeakStress}
        \end{subfigure}
        \begin{subfigure}[b]{0.3\textwidth}
                \centering
                \includegraphics[width=\textwidth]{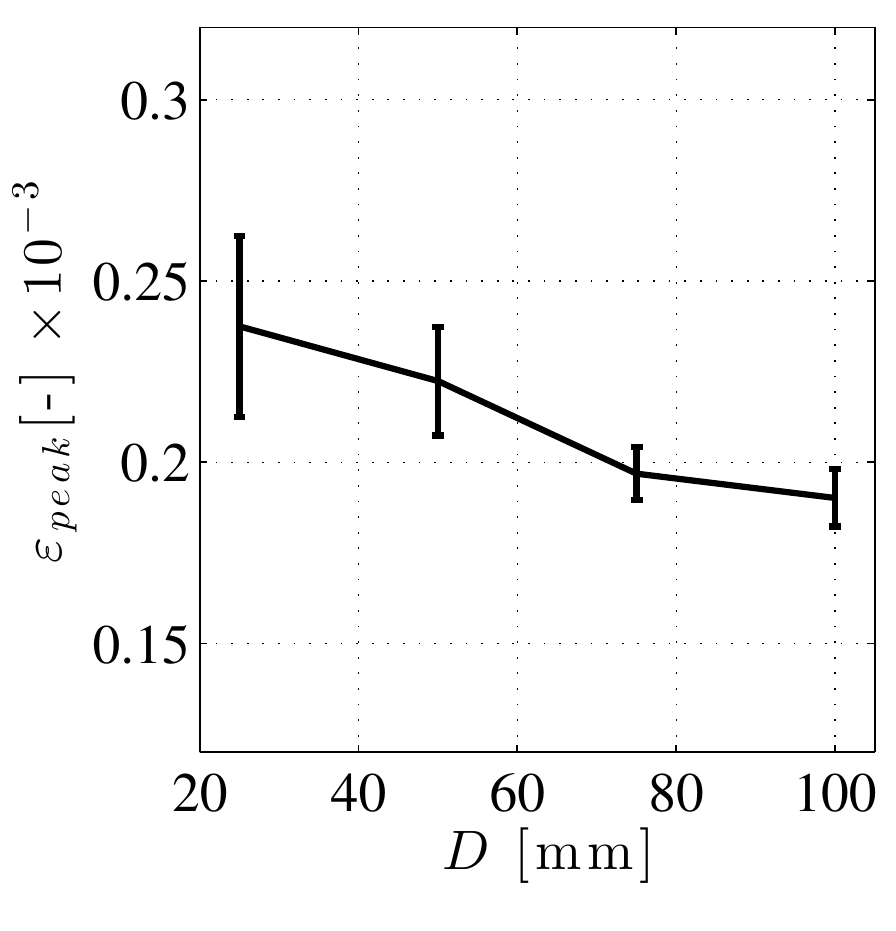}
                \caption{}
                \label{PeakStrain}
        \end{subfigure}
        \caption{Variation of (a) average peak stress and (b) average peak strain, with respect to the RVE size.}
        \label{PeakStat}
\end{figure}

Figure \ref{stress-strainNonlinear-tens} shows the homogenized stress-strain curves for different RVE sizes and polyhedral particle realizations relevant to RVEs subjected to uniaxial tensile strain. The results illustrate that the different polyhedral particle realizations do not affect the linear elastic and nonlinear pre-peak responses, but on the other hand,  it clearly influences on the post-peak softening response. One can notice that the post-peak response of smaller RVE sizes is more scattered, while fine-scale randomness effect on the homogenized response diminishes for the larger RVEs \cite{Gitman-1,Nguyen-1}. Therefore, one can conclude that the mesh realization is a more influential factor on the post-peak softening response of the RVEs of smaller sizes. Average of peak stress and strain values of different mesh realizations are calculated for each RVE size, and its variation with respect to the RVE size is plotted in Figures \ref{PeakStress} and \ref{PeakStrain}. As one can see these quantities as well as mesh realization effect decrease as the size of the RVE increases.

\begin{figure}
	\centering
	\begin{subfigure}[b]{0.3\textwidth}
		\centering
        \includegraphics[width=\textwidth]{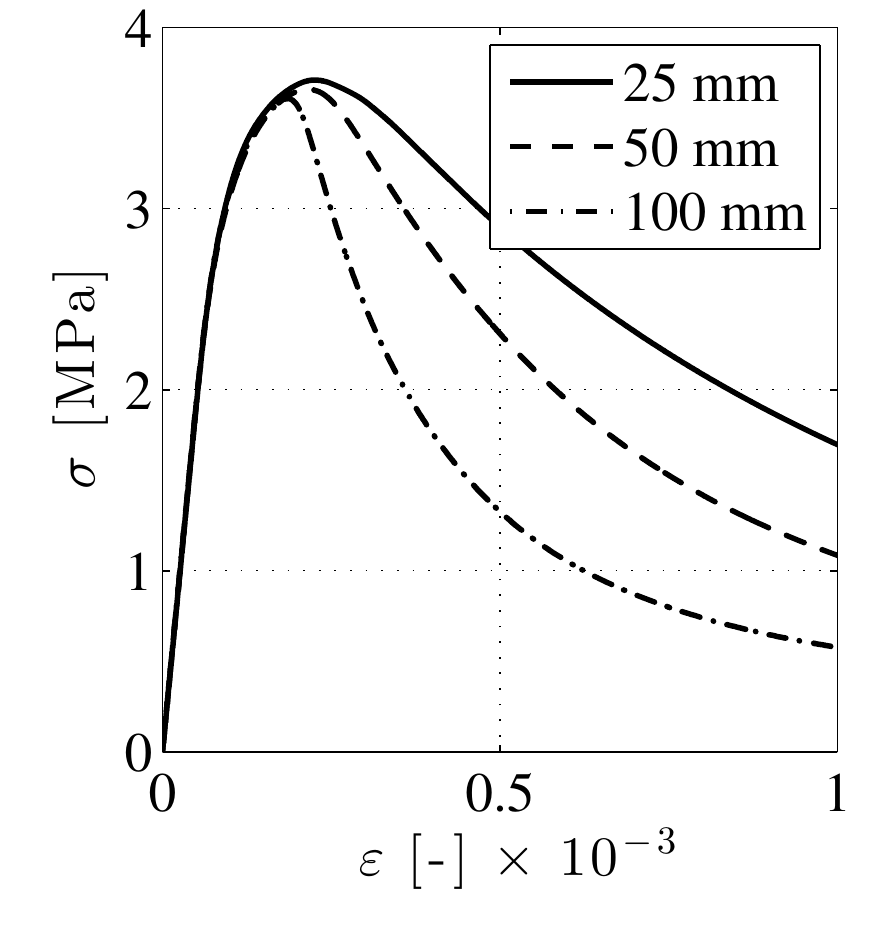}
        \caption{}
        \label{AverageSigmaTens}
    \end{subfigure}
    \begin{subfigure}[b]{0.3\textwidth}
        \centering
        \includegraphics[width=\textwidth]{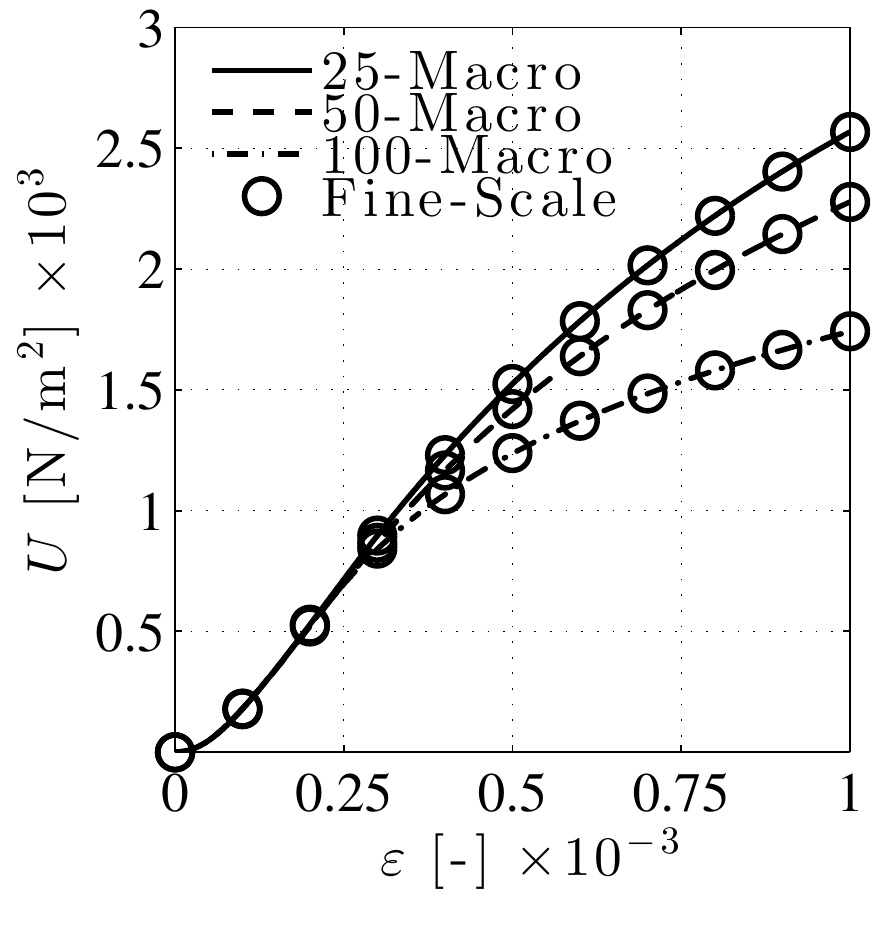}
        \caption{}
        \label{TensileEne}
    \end{subfigure}
    \caption{(a) Average tensile stress-strain curves for three different RVE sizes. (b) Coarse- and fine-scale strain energy density for different RVE sizes.}
    \label{TenSigEneAvg}
\end{figure}

Furthermore, the average stress-strain curves of different polyhedral particle configurations for each RVE size are calculated and plotted in Figure \ref{AverageSigmaTens}. As one can see clearly, increasing size of the RVE affects the post-peak behavior and increases the brittleness of the response. This is consistent with the well-known size effect associated to damage localization in quasi-brittle materials \cite{Bazant-Book}. 

\begin{figure}[h]
\centering 
{\includegraphics[width=0.9\textwidth]{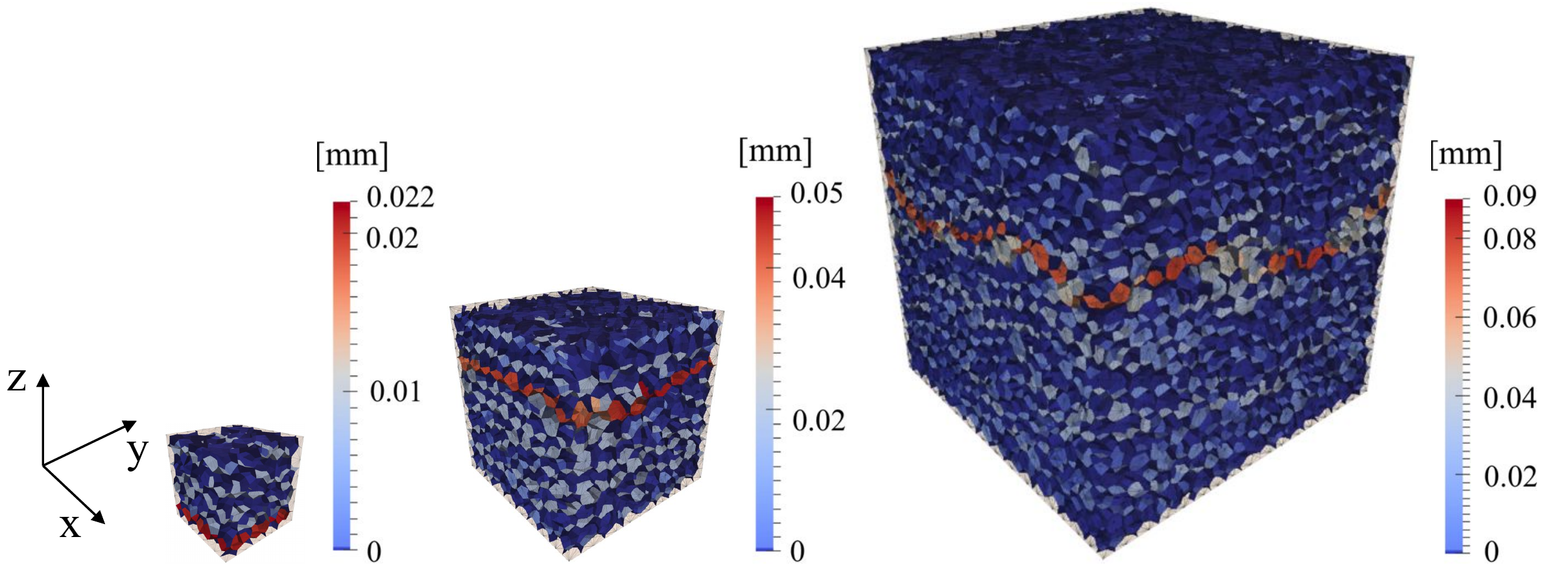}} 
\caption{Crack opening contour of damaged RVEs at tensile strain equal to 0.001 (left) 25 mm (middle) 50 mm (right) 100 mm}
\label{rve-damaged-tens}
\end{figure}

This phenomenon is depicted in Figure \ref{rve-damaged-tens}, which shows damaged RVEs of different sizes at the end of the tensile loading process. The contour plots present meso-scale crack opening distributions corresponding to macroscopic imposed uniaxial strain equal to $10^{-3}$. One can easily notice that the damaged area does not scale with the RVE size leading to the post peak size dependency on the RVE size. 

\begin{figure}[h]
\centering 
{\includegraphics[width = 1\textwidth]{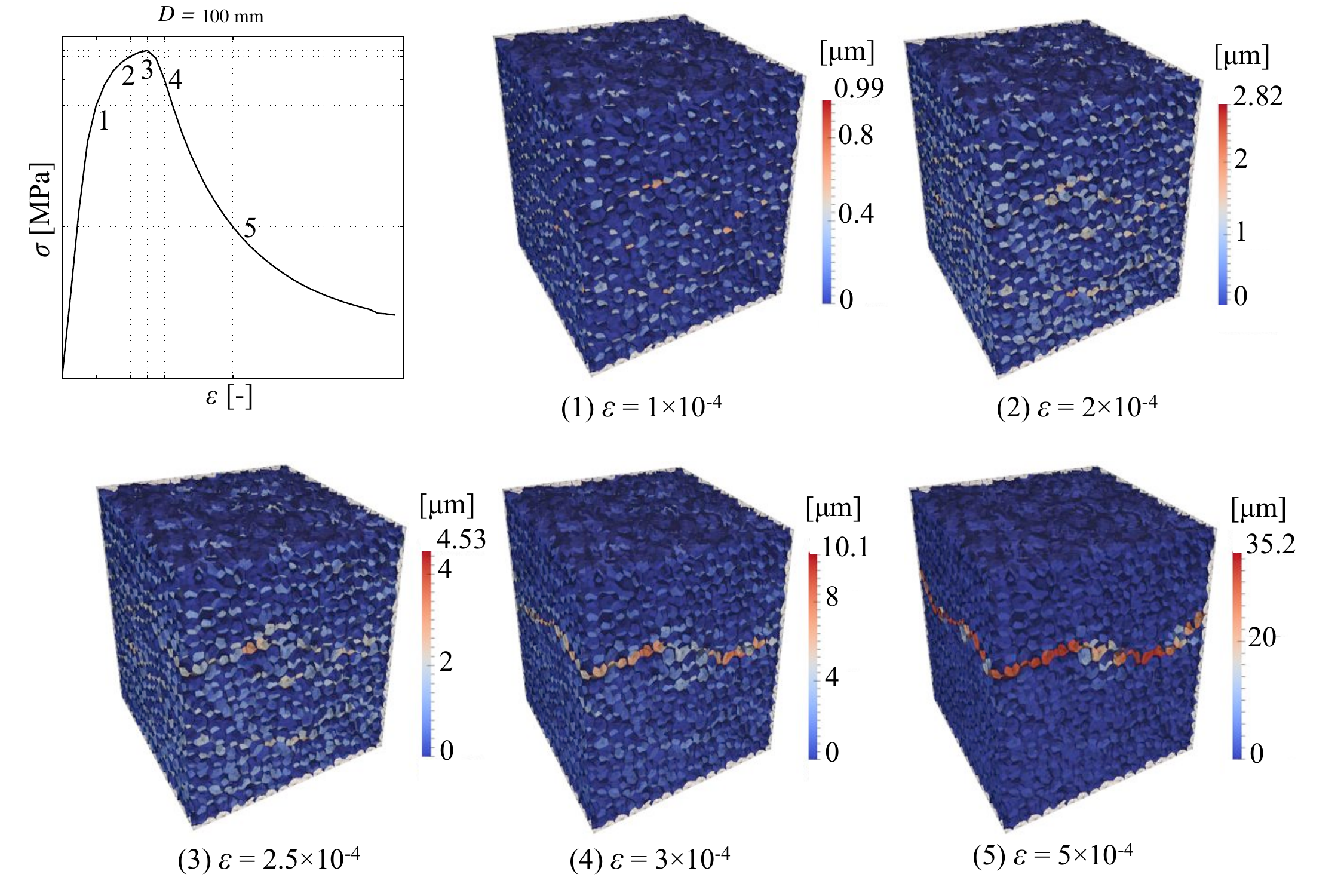}}
\caption{Damage evolution of 100 mm RVE in tension test}
\label{Curve100}
\end{figure}

Evolution of damage for a 100 mm RVE is also shown in Figure \ref{Curve100} at five different macroscopic strain levels. Strain levels (1) and (2) are in pre-peak regime, in which damage is distributed throughout the RVE, which corresponds to the fact that homogenized response is not size dependent in the pre-peak regime. At strain level (3) which corresponds to the peak of the stress-strain curve, damage is still distributed over the RVE; However, as the material undergoes softening, damage localization initiates. Strain levels (4) and (5) are relevant to the softening branch of the response, in which damage localization is clearly visible. The size dependence of the homogenized softening RVE response leads to mesh-dependence of the macroscopic response. This issue has been investigated by some authors \cite{Gitman-1,Nguyen-1,Gitman-2} with reference to continuum-based fine scale models. The complete analysis of this aspect with reference to the current LDPM-based homogenization scheme will be pursed in future work by the writers.

Finally, in Figure \ref{TensileEne}, the Hill-Mandel condition is verified by comparing the RVE strain energy density calculated through fine-scale and macroscopic quantities.

\begin{figure}[h]
        \centering
        \begin{subfigure}[b]{0.3\textwidth}
                \centering
                \includegraphics[width=\textwidth]{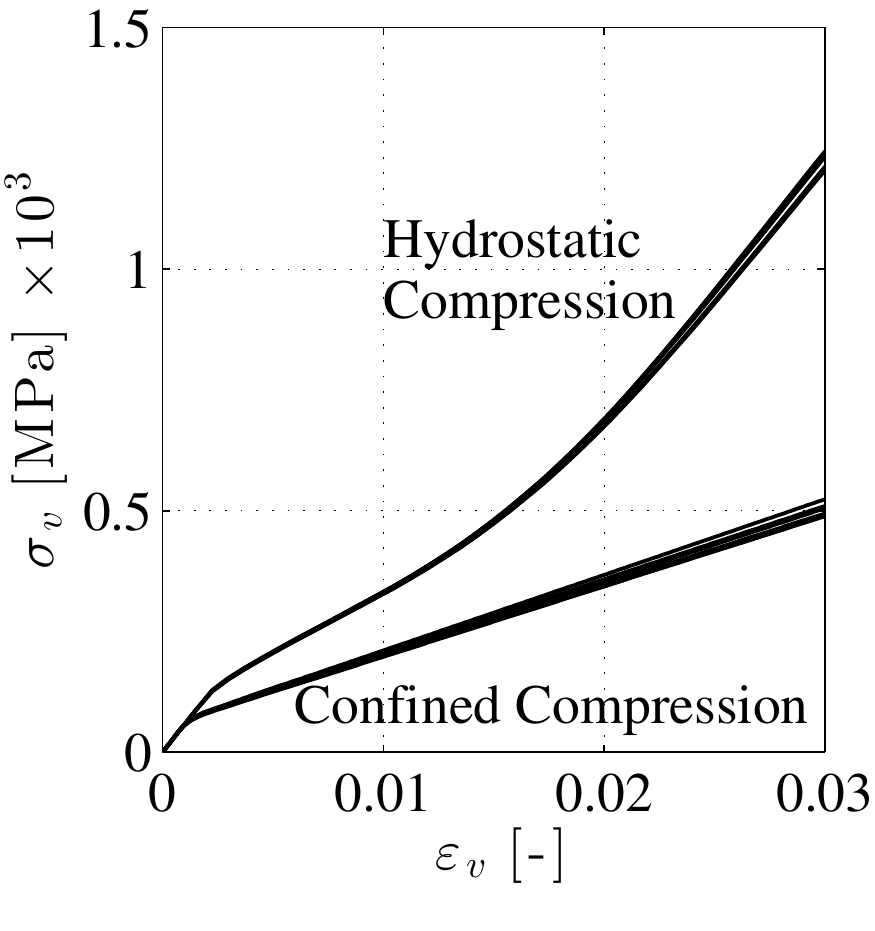}
                \caption{D = 25 mm}
                \label{C25}
        \end{subfigure}
        \begin{subfigure}[b]{0.3\textwidth}
                \centering
                \includegraphics[width=\textwidth]{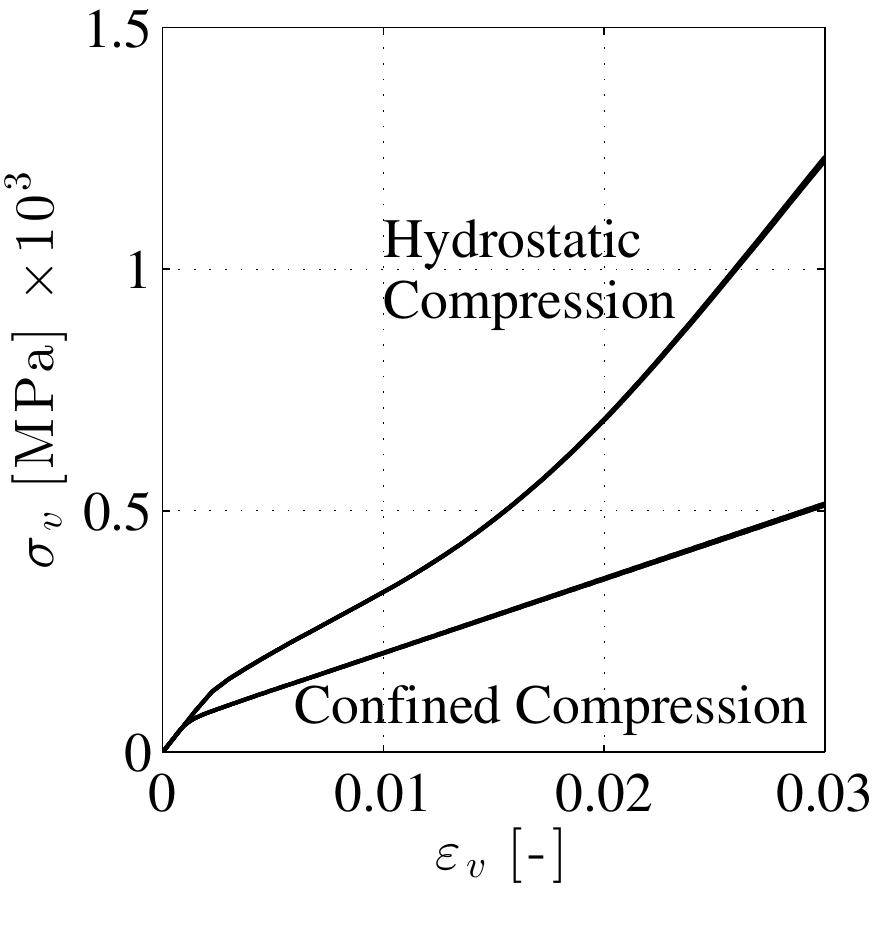}
                \caption{D = 50 mm}
                \label{C50}
        \end{subfigure}
        \begin{subfigure}[b]{0.3\textwidth}
                \centering
                \includegraphics[width=\textwidth]{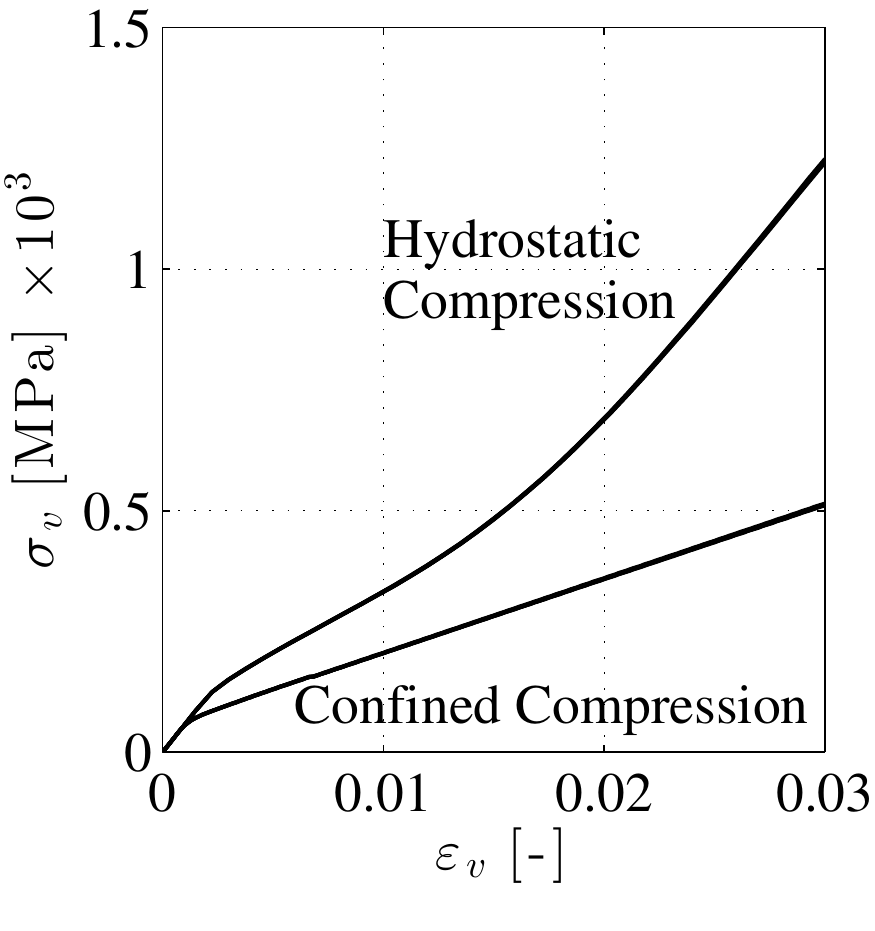}
                \caption{D = 100 mm}
                \label{C100}
        \end{subfigure}
        \caption{Volumetric stress-strain curve for three different RVE sizes under confined compression and hydrostatic compression}
        \label{stress-strainNonlinear-comp}
\end{figure}

Next, the nonlinear homogenized behavior of the RVE is studied under confined (uniaxial strain) and hydrostatic compression. For the confined compression test, a strain tensor with a longitudinal component up to -0.03 is considered, whereas for the hydrostatic compression case, all normal components of the strain tensor are set equal and with value up to -0.03. Figure \ref{stress-strainNonlinear-comp} shows the nonlinear response of RVEs of different sizes and 7 different polyhedral particle configurations. 

\begin{figure}[h]
        \centering
        \begin{subfigure}[b]{0.3\textwidth}
                \centering
                \includegraphics[width=\textwidth]{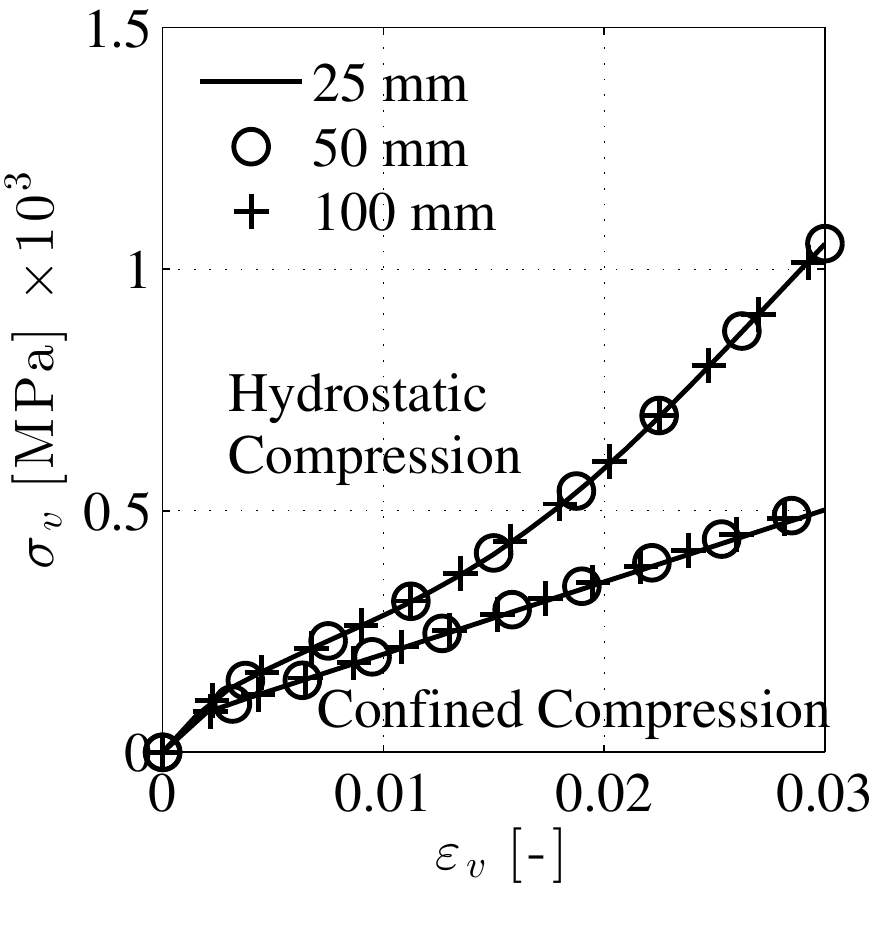}
                \caption{}
                \label{AverageSigmaComp}
        \end{subfigure}
        \begin{subfigure}[b]{0.3\textwidth}
                \centering
                \includegraphics[width=\textwidth]{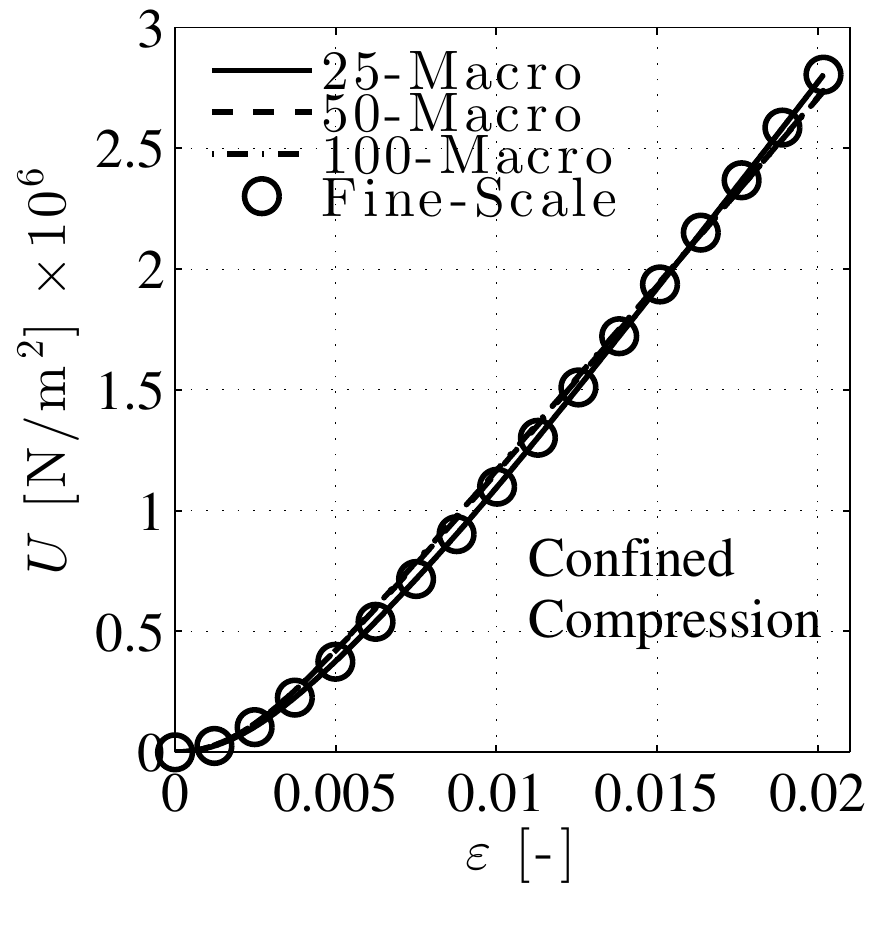}
                \caption{}
                \label{compEne}
        \end{subfigure}
        \caption{(a) Average compressive volumetric stress-strain curves for three different RVE sizes. (b) Coarse- and fine-scale strain energy density for different RVE sizes.}
        \label{AverageSigmaComp-compEne}
\end{figure}

In this case, due to the confinement, the stress-strain response is strain-hardening, and as one can see the different polyhedral particle realizations do not affect significantly the homogenized response in both the elastic and inelastic regime. In addition, the average of different mesh realization stress-strain responses is calculated and plotted for each RVE size in Figure \ref{AverageSigmaComp}. The nonlinear compressive response does not depend on the RVE size, which is consistent with the fact that plastic deformations are distributed through out the specimen, and strain localization does not take place. Finally, the Hill-Mandel condition is verified with reference to the confined compression test, and the fine- and coarse-scale strain energy density of different RVE sizes are plotted in Figure \ref{compEne}. 

\subsubsection{Nonlinear Analysis of RVE subject to components of the curvature tensor}\label{non-cur}
\begin{figure}[h]
\centering 
{\includegraphics[width=1\textwidth]{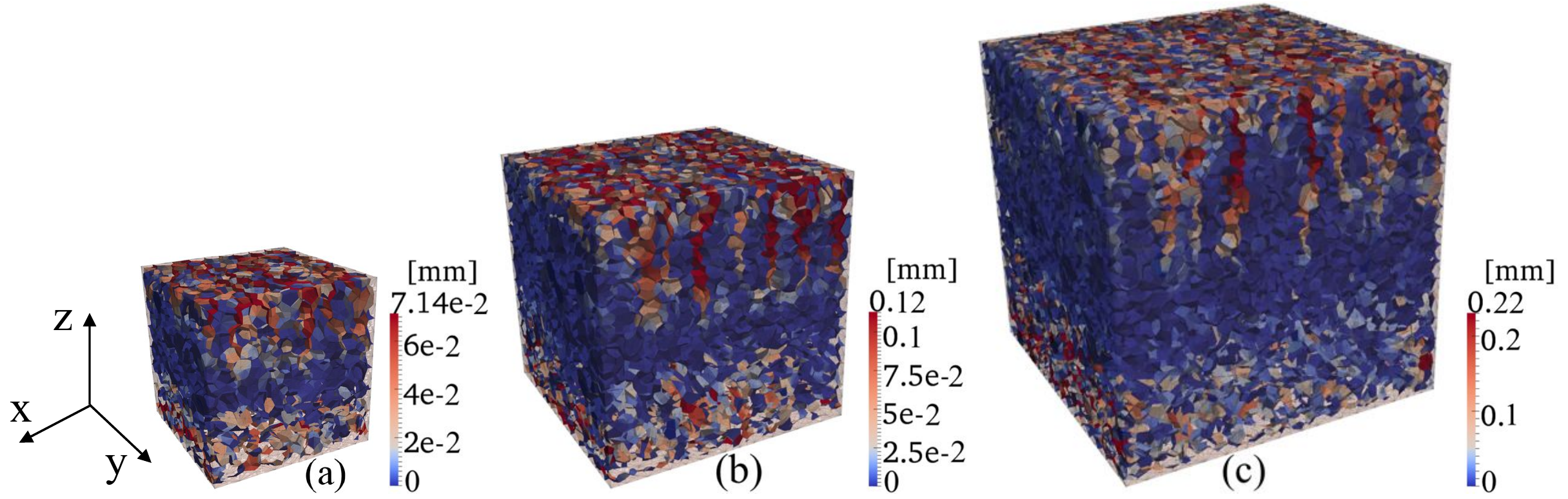}} 
\caption{Crack opening contour of damaged RVEs at bending curvature equal to 0.5 (a) 50 mm (b) 75 mm (c) 100 mm}
\label{rve-damaged-curv}
\end{figure}

\begin{figure}[h]
\centering 
{\includegraphics[width=1\textwidth]{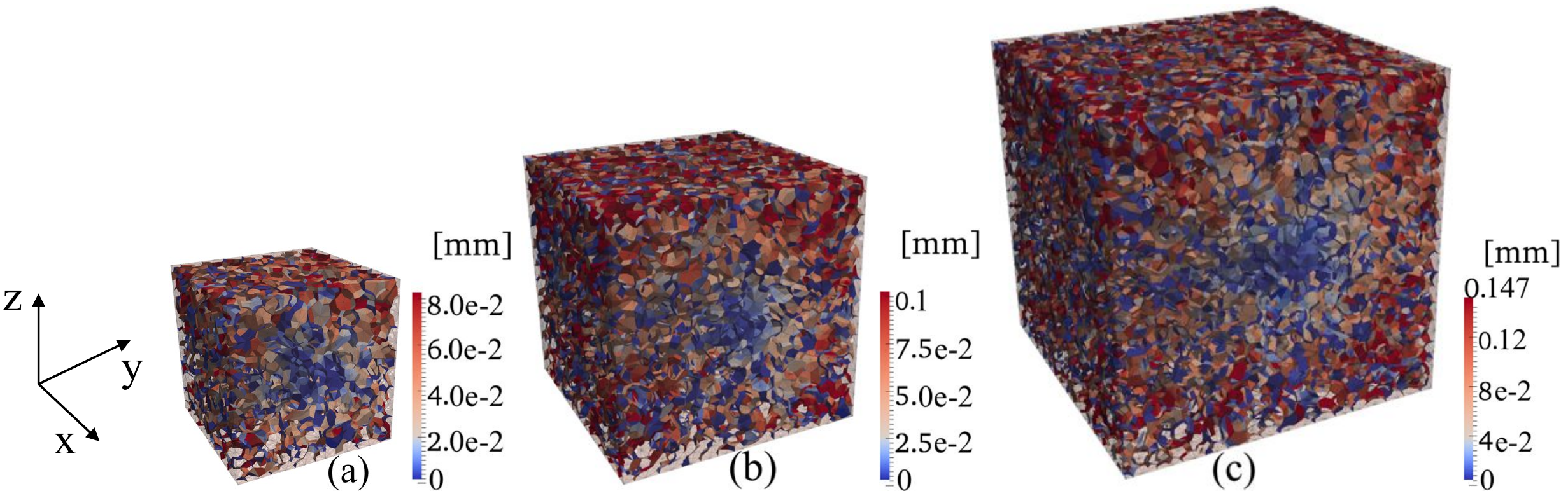}} 
\caption{Crack opening contour of damaged RVEs at torsional curvature equal to 0.5 (a) 50 mm (b) 75 mm (c) 100 mm}
\label{rve-damaged-tor}
\end{figure}

In this section, the nonlinear homogenized behavior of RVEs of 3 different sizes, 50, 75, and 100 mm and 5 five different mesh configurations for each size, is studied under the effect of components of macroscopic curvature tensor. Bending and torsional behavior of the RVEs are investigated by applying macroscopic curvature tensors with the only non-zero components of $\kappa_{12} = 1$ and $\kappa_{11} = 1$, respectively. Figure \ref{rve-damaged-curv} shows crack opening contour of damaged RVEs at the macroscopic curvature  for $\kappa_{12} = 0.5$. The resulting crack pattern conforms with the fracture mode that one may expect from bending theories. Multiple crack lines are generated in the tensile strain domain, which is the top half of the RVEs, while half bottom part in under compression. As more strain is applied in compressive part, splitting cracks take place in the latter region due to transverse tensile stress. Typical crack pattern of RVEs under torsion are plotted in Figure \ref{rve-damaged-tor} for $\kappa_{11} = 0.5$.  Crack opening contours show that the amount of damage close to the RVE center is negligible, while it increases as the facets are placed at further positions. This corresponds to the deformation mechanism and strain distribution in solids subject to torsion.

\begin{figure}[h]
        \centering
        \begin{subfigure}[b]{0.3\textwidth}
                \centering
                \includegraphics[width=\textwidth]{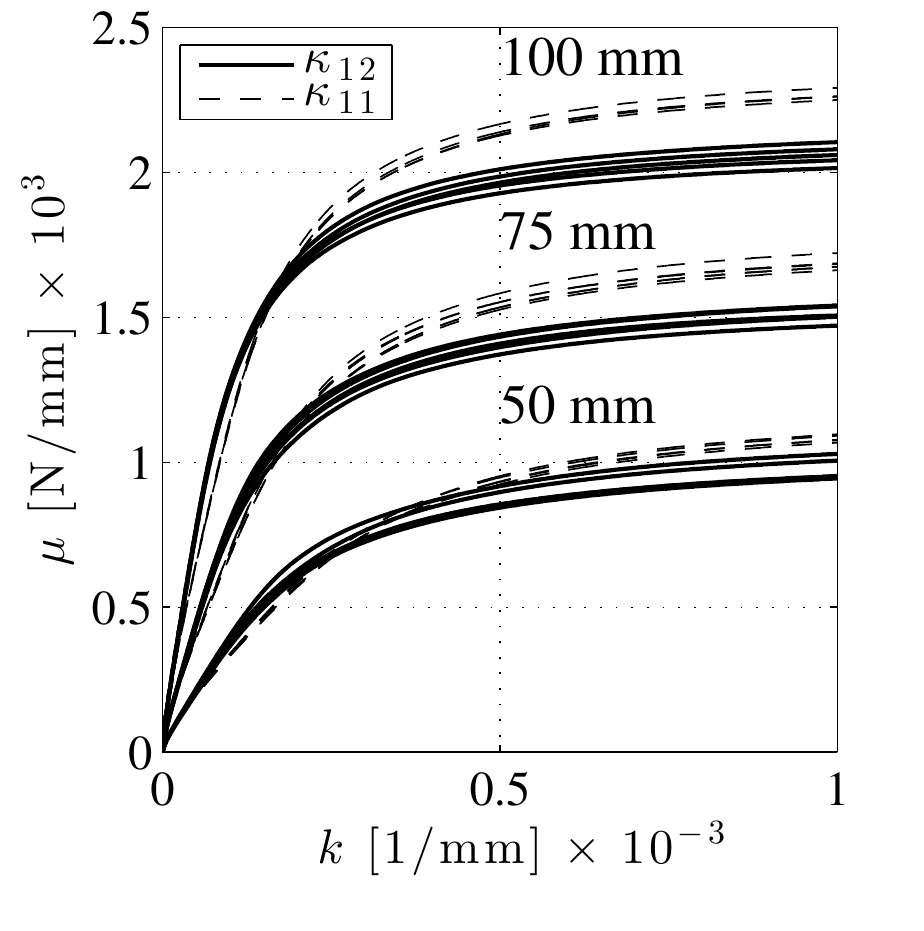}
                \caption{}
                \label{mu_all}
        \end{subfigure}
        \begin{subfigure}[b]{0.3\textwidth}
                \centering
                \includegraphics[width=\textwidth]{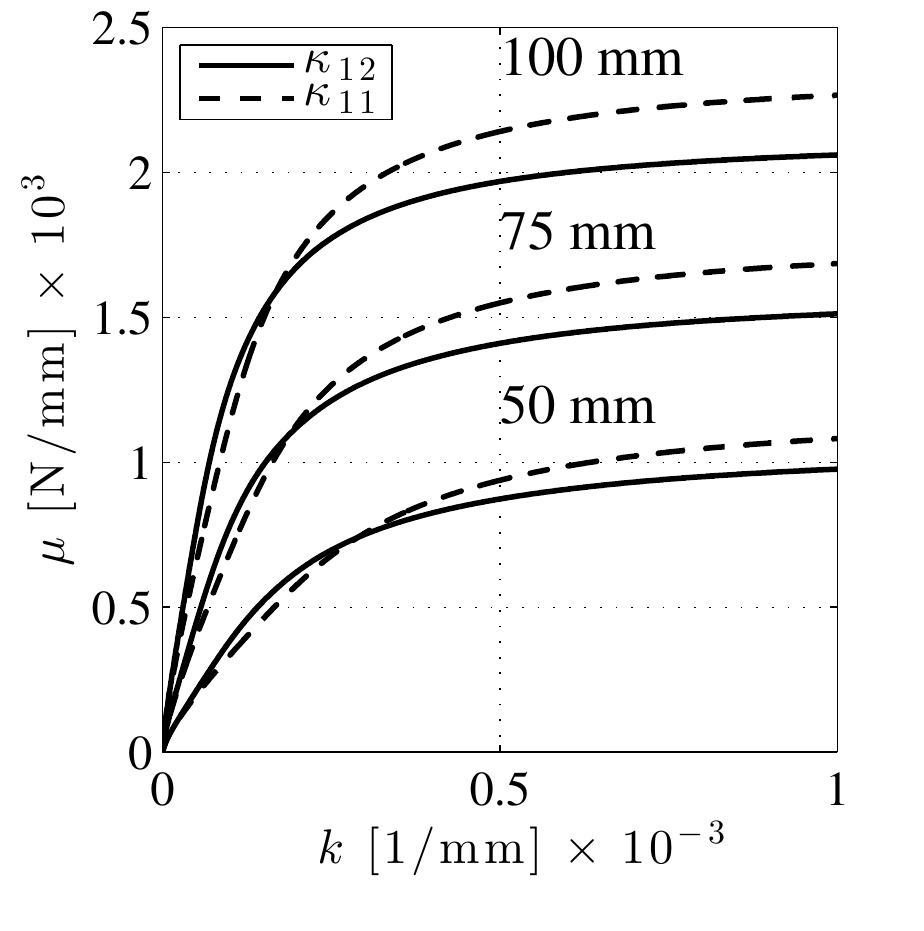}
                \caption{}
                \label{mu_all_avg}
        \end{subfigure} 
        \begin{subfigure}[b]{0.3\textwidth}
                \centering
                \includegraphics[width=\textwidth]{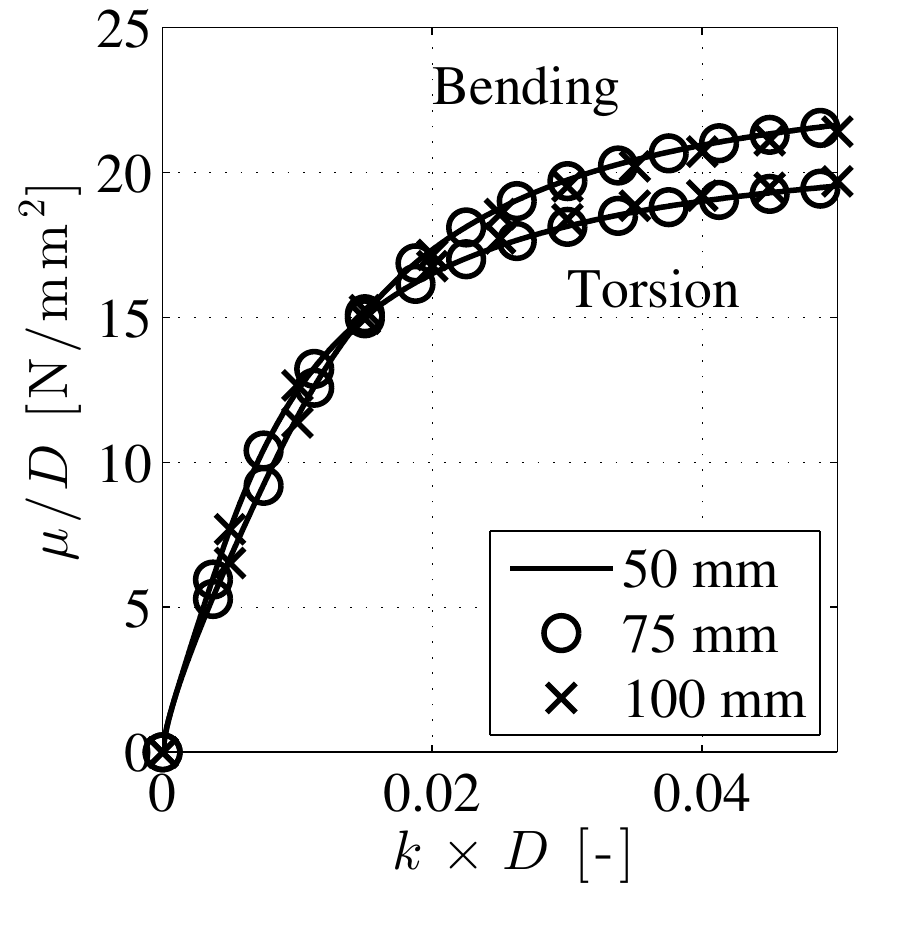}
                \caption{}
                \label{mu_avg_scaled}
        \end{subfigure} 
        \begin{subfigure}[b]{0.3\textwidth}
                \centering
                \includegraphics[width=\textwidth]{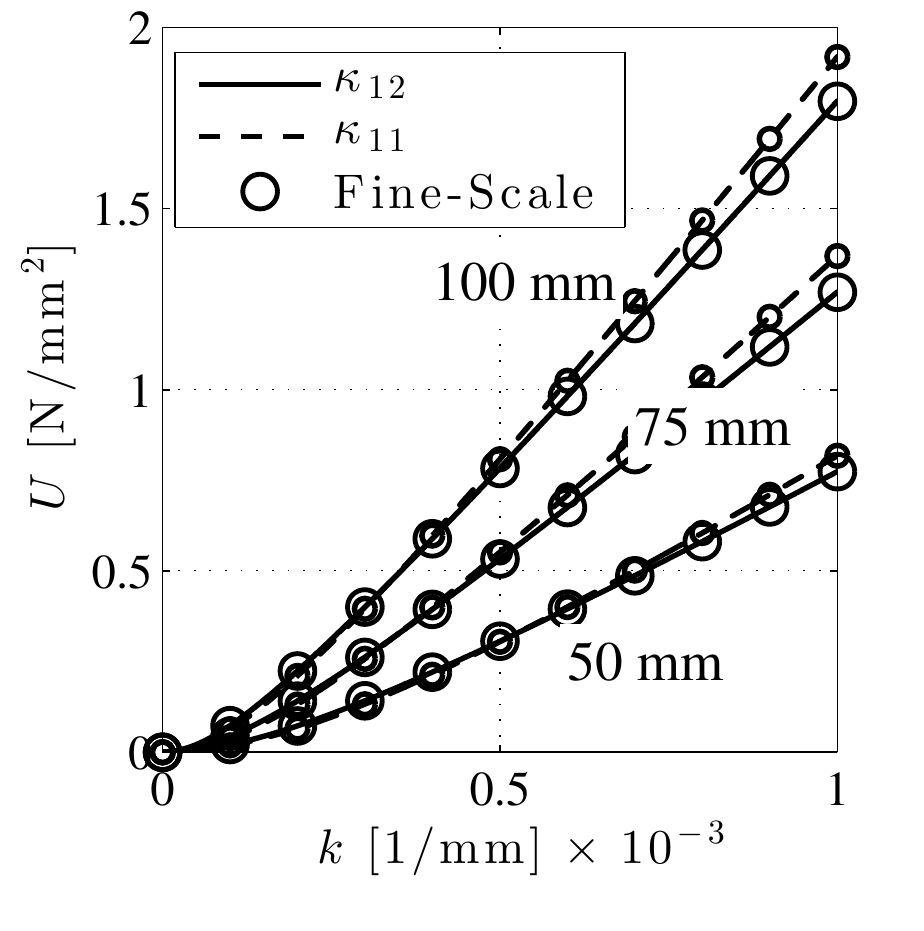}
                \caption{}
                \label{Ene_all}
        \end{subfigure}
        \begin{subfigure}[b]{0.3\textwidth}
                \centering
                \includegraphics[width=\textwidth]{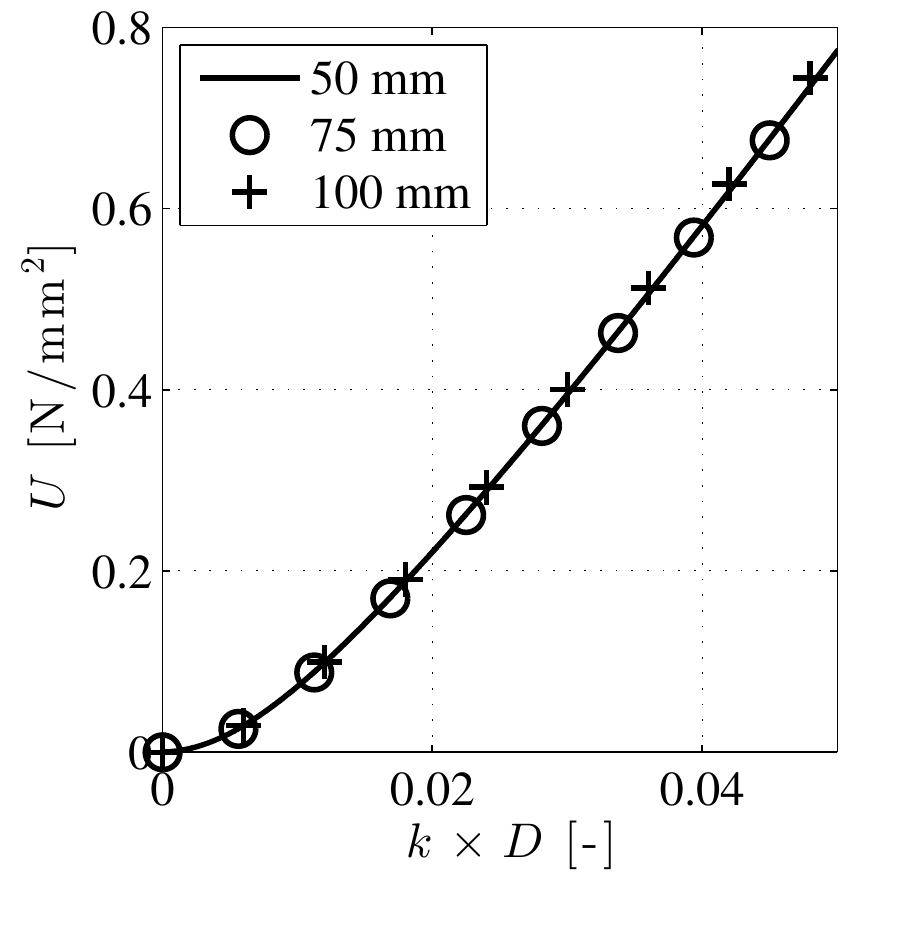}
                \caption{}
                \label{Ene_12_scaled}
        \end{subfigure}
        \begin{subfigure}[b]{0.3\textwidth}
               \centering
               \includegraphics[width=\textwidth]{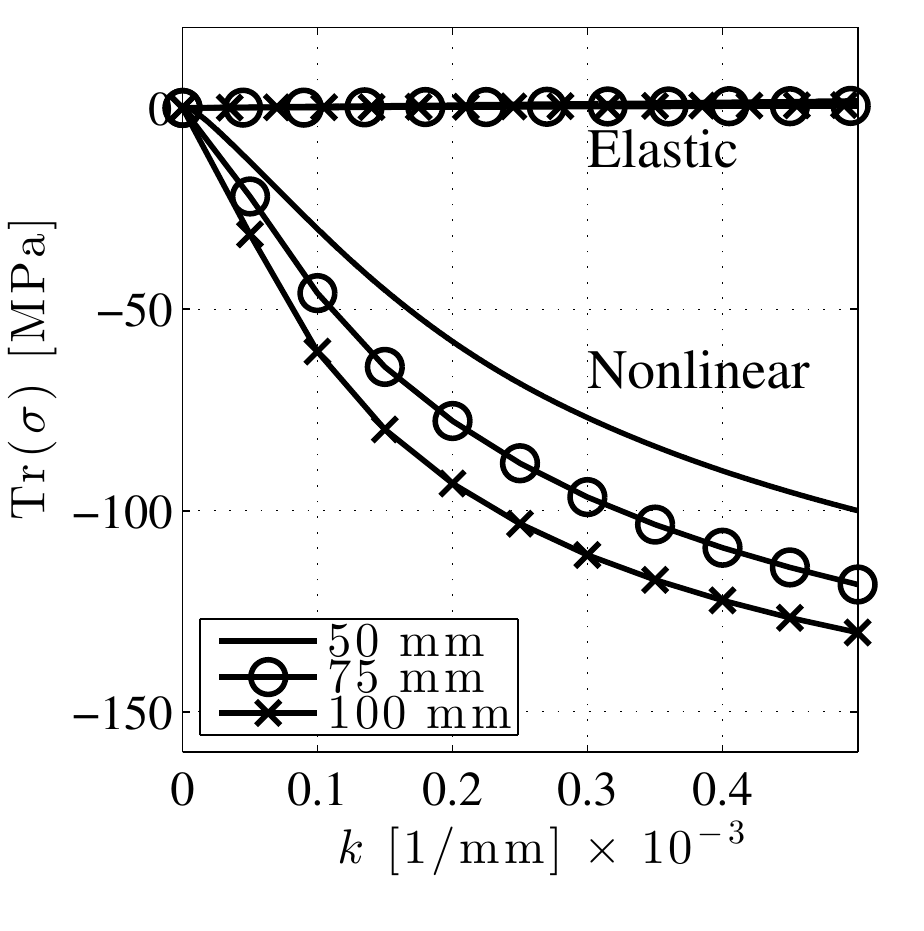}
               \caption{}
               \label{trace}
        \end{subfigure}
        \caption{(a) Homogenized couple stresses $\mu_{11}$ and $\mu_{12}$ versus curvature $\kappa_{11}$ and $\kappa_{12}$ for five different polyhedral particle configurations for each RVE size. (b) Average of the homogenized couple stress of different polyhedral particle configurations for each RVE size for  $\kappa_{11}$ and  $\kappa_{12}$ cases. (c) Scaled couple stress versus curvature curves. (d) Macro and Fine-Scale strain energy density evolution. (e) Scaled strain energy density evolution for the case $\kappa_{12}$. (f) Trace of stress tensor due to elastic and nonlinear analysis of RVE under macroscopic $\kappa_{12}$.}
        \label{NC}
\end{figure}

Homogenized moment stress components $\mu_{12}$ and $\mu_{11}$ versus macroscopic curvature tensor components $\kappa_{12}$ and $\kappa_{11}$ of RVEs of different sizes and polyhedral particle configurations are plotted in Figure \ref{mu_all}. One can see that effect of different polyhedral particle realizations on the homogenized response is negligible, which is due to the occurrence of distributed damage inside the RVE. Homogenized response of RVEs with different polyhedral particle realizations are averaged for each size and plotted in Figure \ref{mu_all_avg}. It can be seen that the homogenized response consists of an initial elastic part and a hardening branch, which is related to the confinement due to the fact that all components of the strain tensor are zero, and the RVE cannot expand laterally. It is illustrated that, at any level of macroscopic curvature, magnitude of the moment stress for larger RVE sizes is bigger compared to the smaller ones. Size dependency of moment stress was discussed in Section \ref{Elastic Analysis}, and it was shown that the elastic Cosserat coefficients are proportional to the RVE size squared. In order to study size dependency in the nonlinear regime, the normalized quantities $\hat{\mu}_{ij} = {\mu}_{ij}/D$ and $\hat{k}_{ij} = k_{ij} \times D$ are plotted in Figure \ref{mu_avg_scaled}. One can see that the normalized curves of three different RVE sizes are unique for both bending and torsion. This implies that the proportionality of the homogenized micropolar properties to the RVE size squared is still valid in the nonlinear regime. In Figure \ref{Ene_all}, the Hill-Mandel condition is verified, and coarse- and fine-scale strain energy density are plotted for each RVE size for both the aforementioned cases. 


Finally, the existence of coupling effect characterized by the dependency of the homogenized stress tensor on the curvature tensor is investigated in both elastic and nonlinear regimes. The macroscopic curvature $\kappa_{12}=1$ is applied on the RVEs, and the trace of the homogenized stress tensor is calculated and plotted in Figure \ref{trace} for different RVE sizes. One can observe that the trace of the stress tensor for the case of elastic RVE behavior is zero throughout the analysis. On the other hand, for the case of nonlinear behavior, it increases monotonically with the curvature. This implies that in elastic regime, stresses and strains are totally uncoupled from couple stresses and curvatures, whereas these quantities are strongly coupled in the nonlinear case. This aspect has been investigated very little in the literature where fully uncoupled behavior has been always postulated.

\subsection{Tension Test on a Concrete Prism with Parallel Elastic Bars}\label{prismExample}
\begin{figure}
        \centering
        \begin{subfigure}[b]{0.2\textwidth}
                \centering
                \includegraphics[scale=0.265]{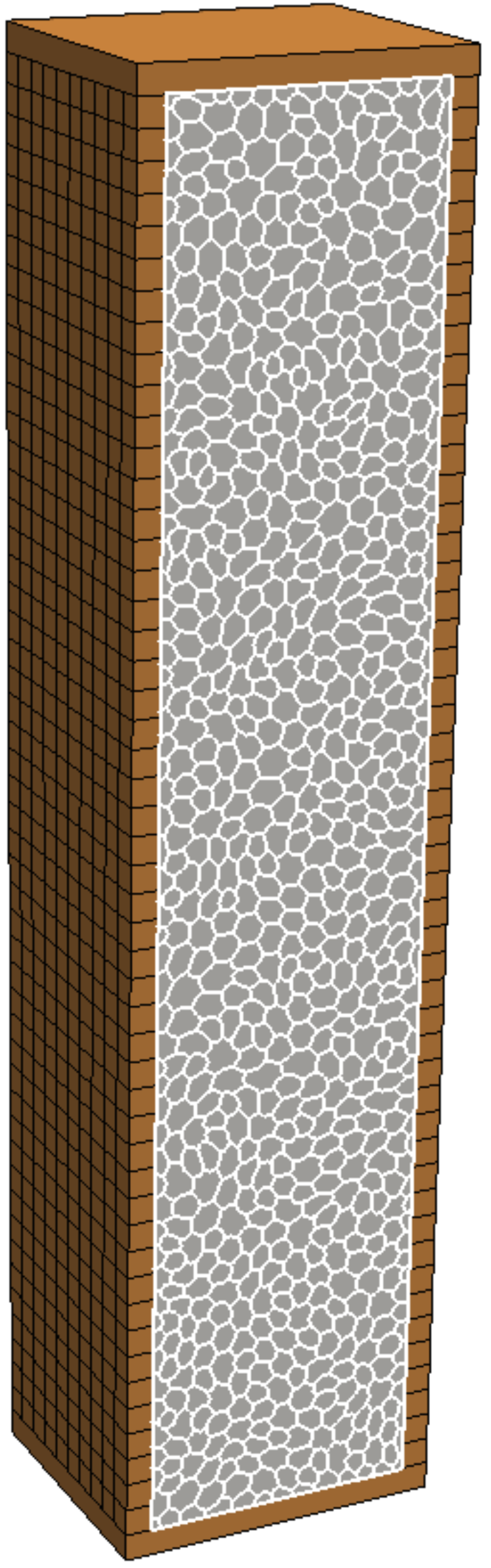}
                \caption{}
                \label{fullldpm-Cbar}
        \end{subfigure}
        \begin{subfigure}[b]{0.3\textwidth}
                \centering
                \includegraphics[scale=0.38]{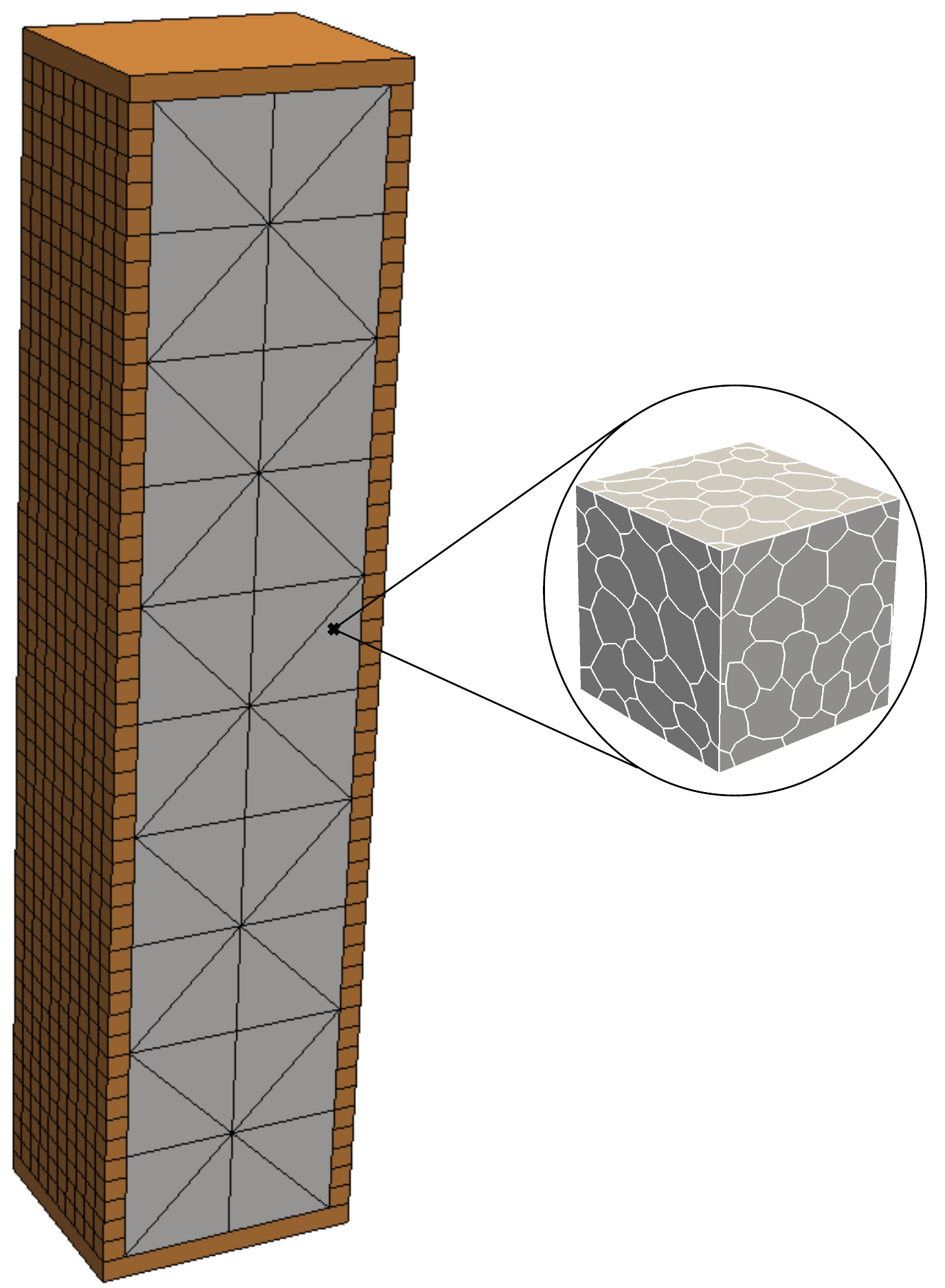}
                \caption{}
                \label{homogenization-Cbar}
        \end{subfigure}
        \begin{subfigure}[b]{0.4\textwidth}
                \centering
                \includegraphics[scale=0.7]{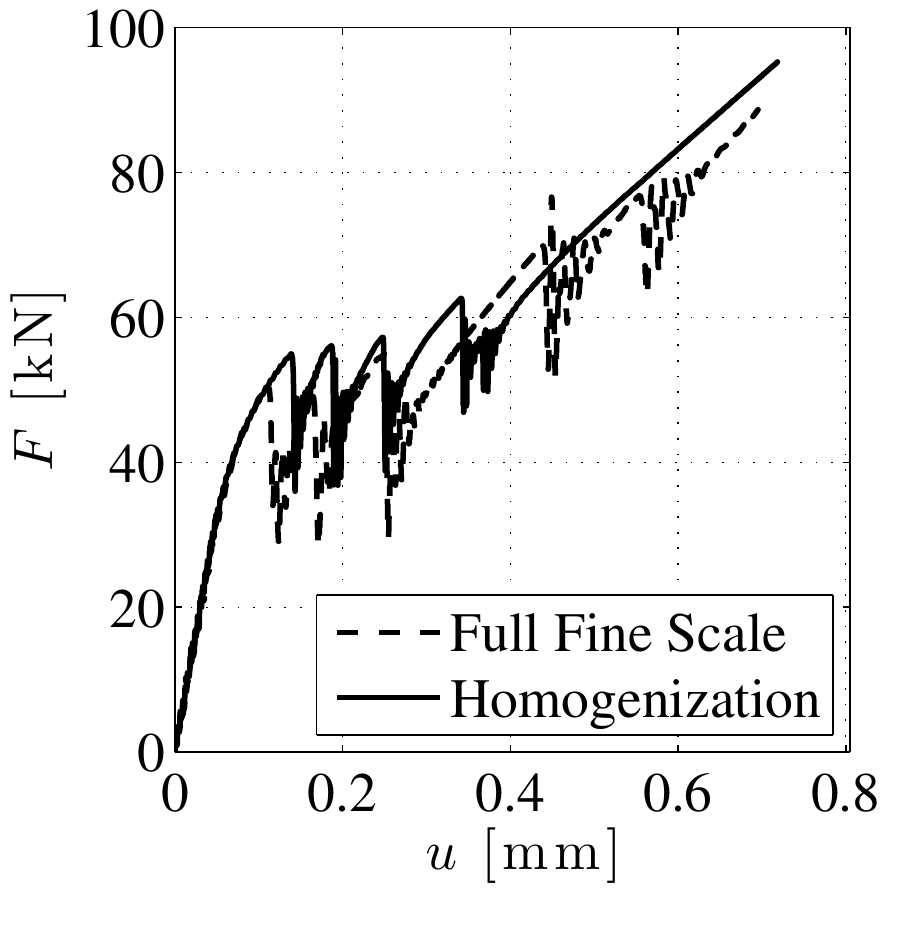}
                \caption{}
                \label{stress-strain-curve}
        \end{subfigure}        
        \caption{(a) Full LDPM concrete prism and attached elastic bars. (b) FE model of the concrete prism and attached elastic bars. (c) Force-Displacement curves obtained by homogenization and full fine-scale simulation.}
\end{figure}

In this section, the behavior of a reinforced concrete prism under tension is studied in a full fine-scale simulation, and  the obtained results are compared to the solution of the same problem through a two-scale homogenization algorithm, in which the concrete prism is modeled as a homogeneous continuum with a meso-scale material RVE assigned to every macroscopic integration point. Figure \ref{fullldpm-Cbar} shows the concrete prism and the two elastic bars attached to it, which are simulated by LDPM and solid finite elements, respectively.  The same specimen in a two-scale homogenization problem is depicted in Figure \ref{homogenization-Cbar}, in which concrete prism is modeled by tetrahedral finite elements. Cross section of the concrete prism is 100 mm $\times$ 100 mm, and its height is 500 mm. Two rigid loading plates are attached at the top and the bottom of the whole specimen cross section to apply the boundary condition. Young modulus and Poisson's ratio of the elastic bars are 28 GPa and 0.18, respectively. The same LDPM parameters used in the previous sections are adopted here. The specimen is pulled in the longitudinal direction up to a displacement equal to 0.7 mm. The RVE size is chosen to be 30 mm which approximately corresponds to the volume of each tetrahedral FE in the coarse mesh. This is done to mitigate the mesh-dependence due to the softening behavior of the RVE. The concrete prism and the elastic bars are connected through a master-slave algorithm. The numerical simulations of the coarse scale are performed by neglecting the couple stresses which are expected to be negligible for this particular application.


\begin{figure}[h!]
        \centering
        \begin{subfigure}[b]{0.2\textwidth}
                \centering
                \includegraphics[scale=0.4]{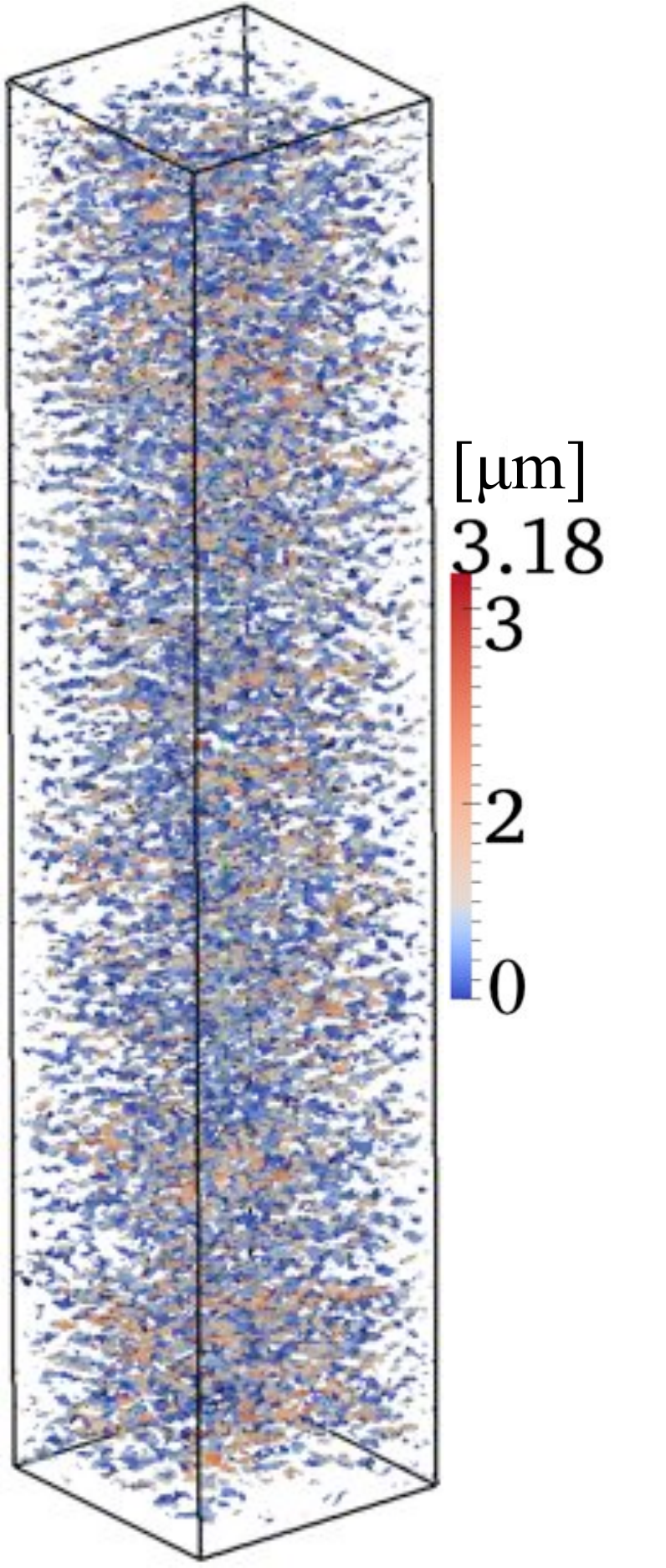}
                \caption{}
                \label{fulla}
        \end{subfigure}
        \begin{subfigure}[b]{0.2\textwidth}
                \centering
                \includegraphics[scale=0.4]{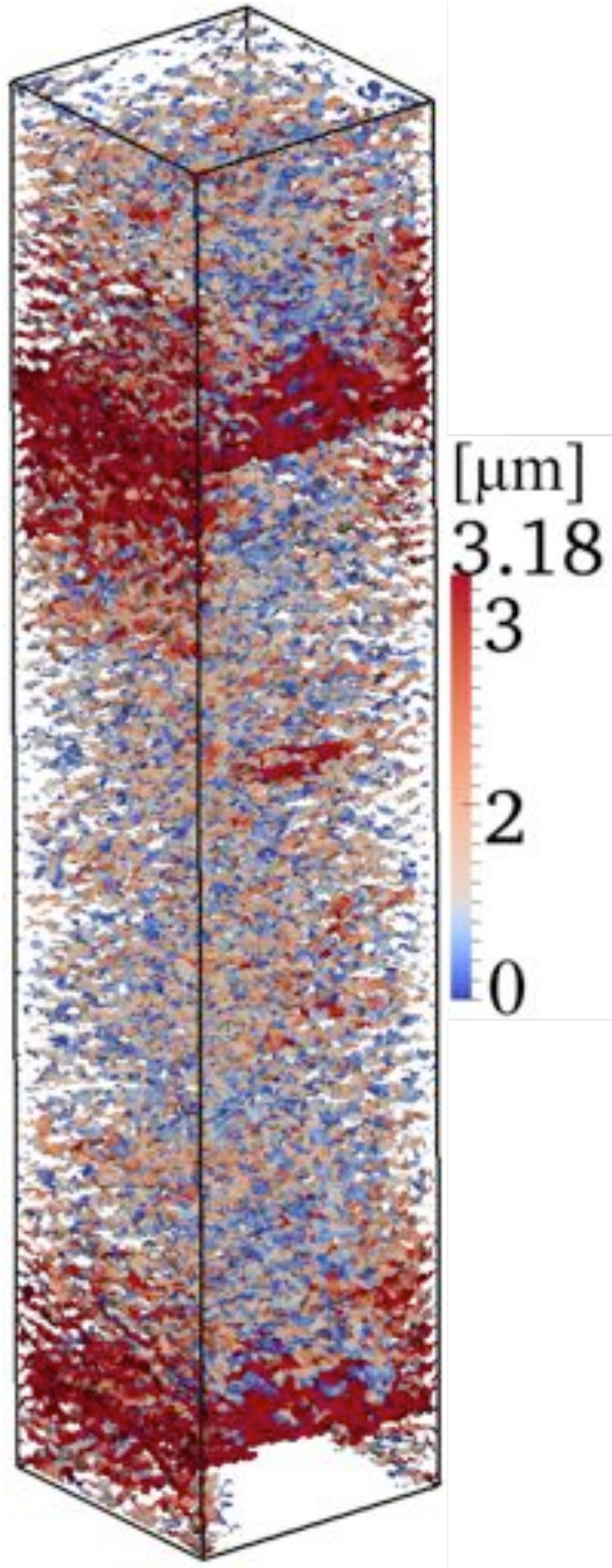}
                \caption{}
                \label{fullb}
        \end{subfigure}
        \begin{subfigure}[b]{0.2\textwidth}
                \centering
                \includegraphics[scale=0.4]{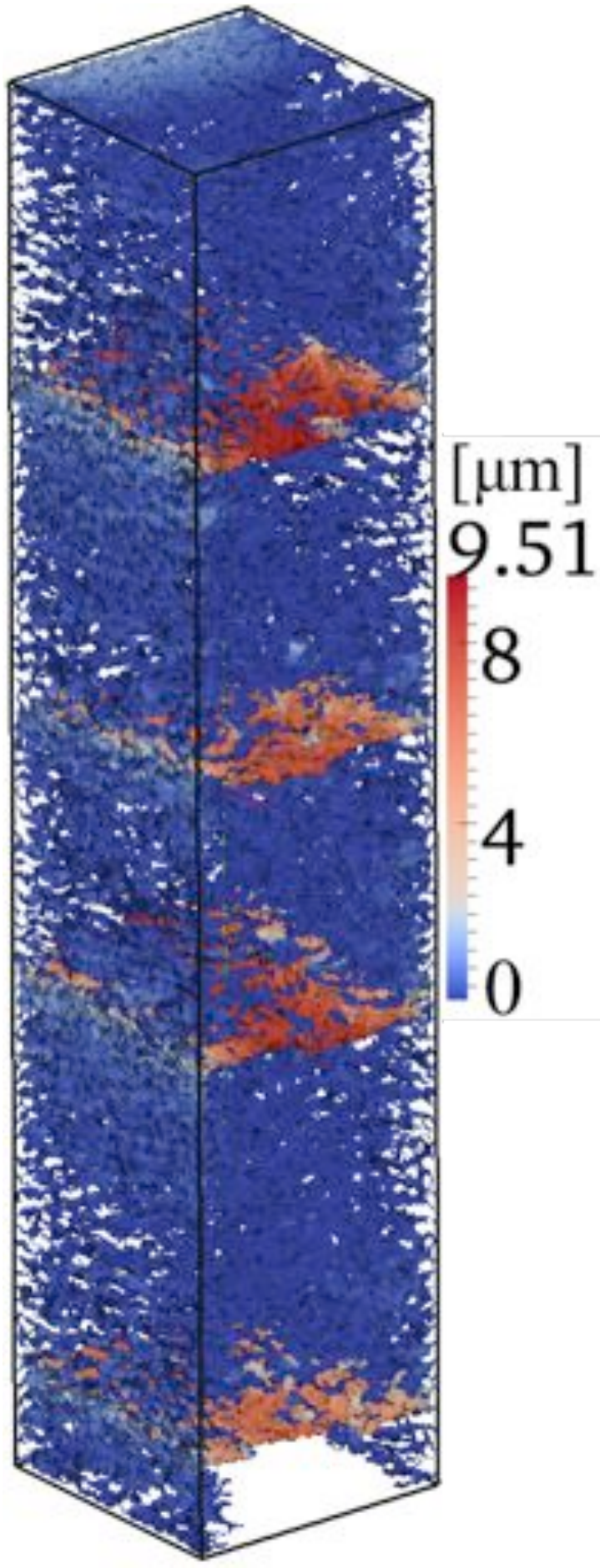}
                \caption{}
                \label{fullc}
        \end{subfigure}        
        \begin{subfigure}[b]{0.2\textwidth}
                \centering
                \includegraphics[scale=0.4]{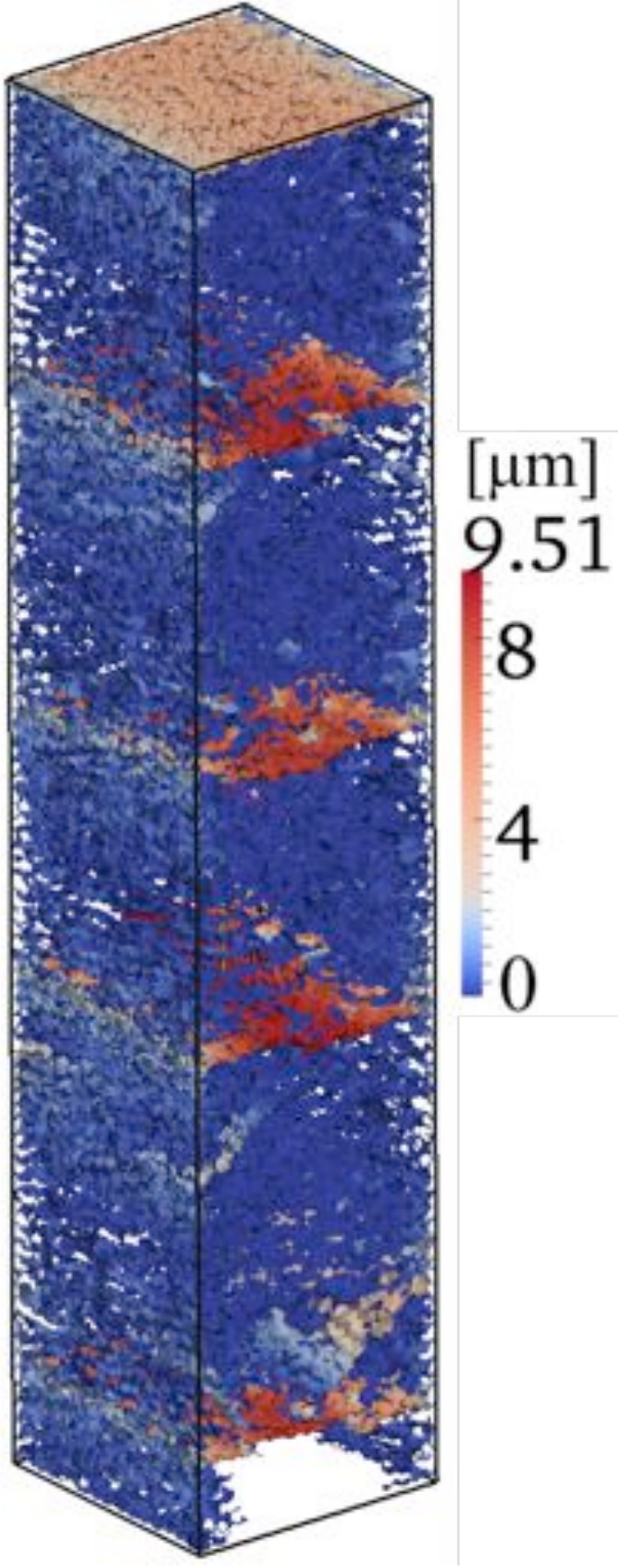}
                \caption{}
                \label{fulld}
        \end{subfigure}   
        \begin{subfigure}[b]{0.2\textwidth}
                \centering
                \includegraphics[scale=0.6]{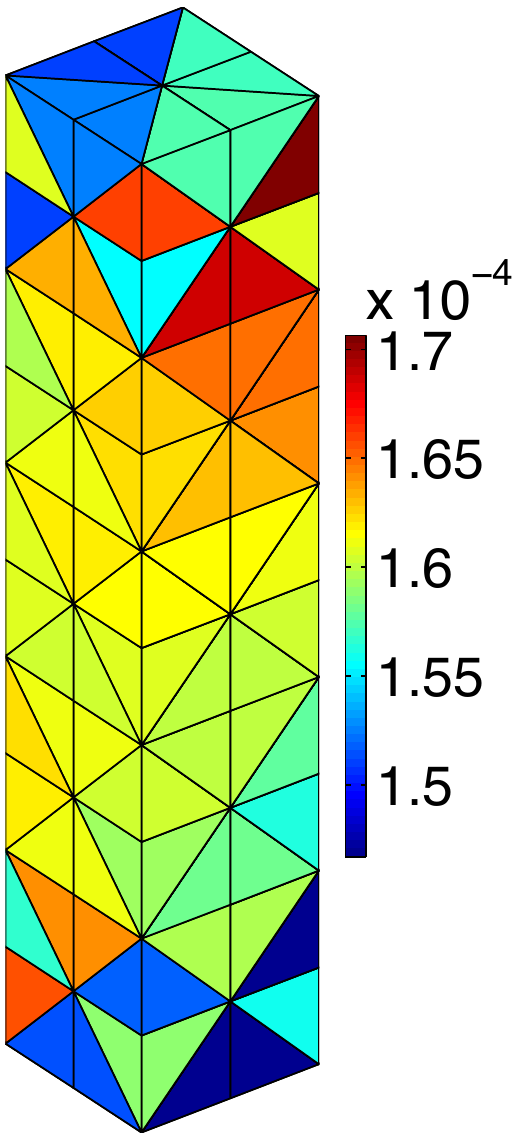}
                \caption{}
                \label{homoa}
        \end{subfigure}
        \begin{subfigure}[b]{0.2\textwidth}
                \centering
                \includegraphics[scale=0.6]{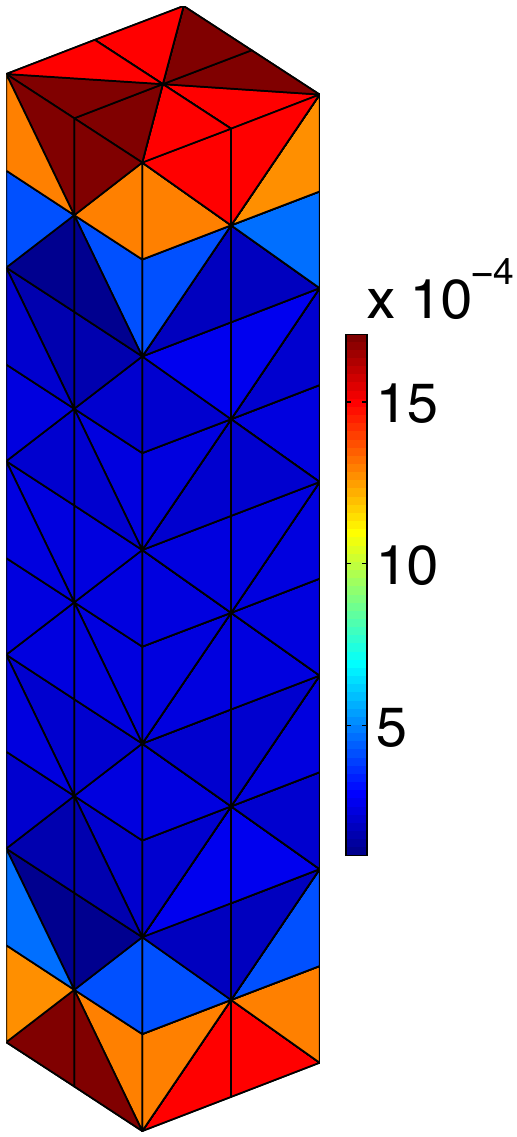}
                \caption{}
                \label{homob}
        \end{subfigure}      
        \begin{subfigure}[b]{0.2\textwidth}
                \centering
                \includegraphics[scale=0.6]{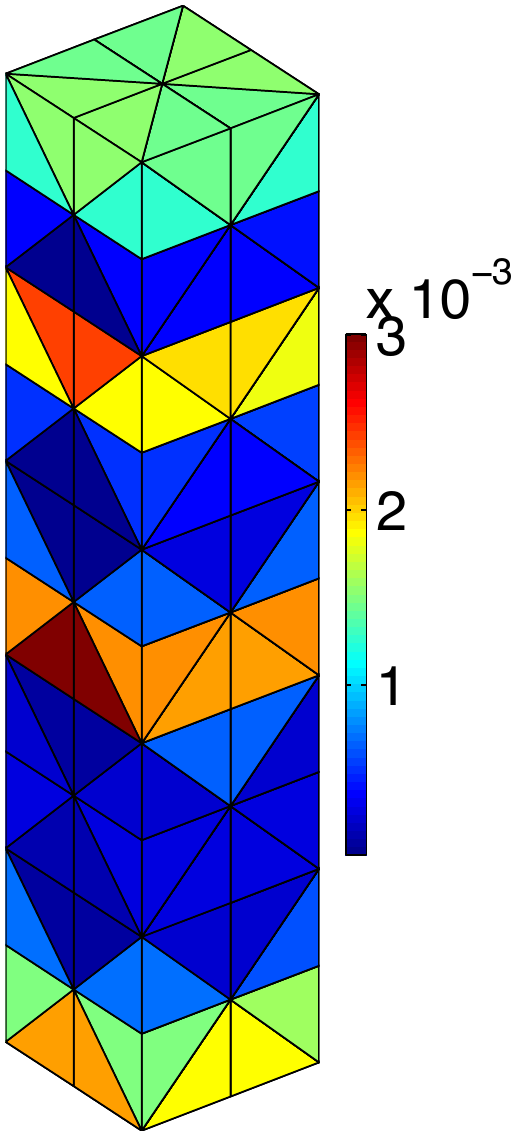}
                \caption{}
                \label{homoc}
        \end{subfigure}        
        \begin{subfigure}[b]{0.2\textwidth}
                \centering
                \includegraphics[scale=0.6]{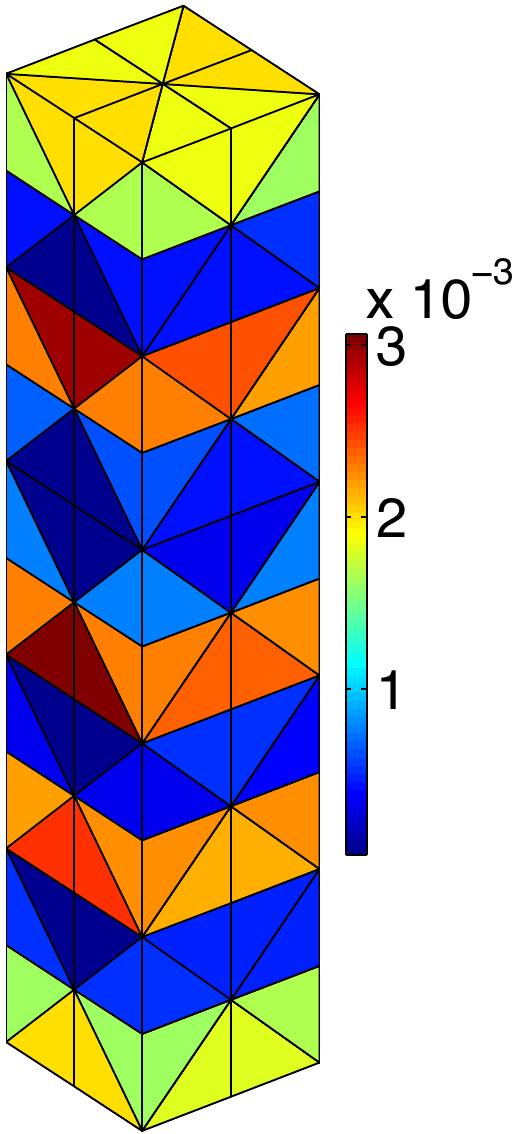}
                \caption{}
                \label{homod}
        \end{subfigure}   
        \caption{(Top row) Crack opening contour at different loading from full fine-scale simulation (Bottom row) Strain distribution contour at different loading steps from homogenization algorithm}
        \label{prism}
\end{figure}

The global force-displacement response of the full fine-scale and homogenization problems are plotted in Figure \ref{stress-strain-curve}. Since concrete prism and elastic bars are tied and deform together during the loading process, distributed damage takes place through the whole specimen during the initial stages of the loading process, see Figure \ref{prism}a. This damage state represents the linear elastic and the first hardening segment of the stress-strain response of the structure. The same damage state is captured through the homogenization procedure. Figure \ref{prism}e shows finite elements normal strain distribution along the loading direction through the specimen. One can see that the strain values are all in the same range, and no localization has occurred. The response of the full fine-scale and homogenization problems show excellent agreement in the elastic and the first hardening segment. As further deformation is applied on the structure, damage localizes in one section of the concrete bar and this causes a sudden drop in the global force-displacement curve. Subsequently, since the elastic bars and the concrete prism are forced to deform in parallel, the overall system can carry more load leading to a rehardening of the global response. Analysis of Figure \ref{stress-strain-curve} shows that five damage localization events occur during the deformation process which corresponds to five sudden drops in the load-displacement curve. Crack pattern of the specimen is plotted after the formation of two, four, and five damage localization in Figure \ref{prism}b, c, and d. It is interesting to show that the homogenization framework is able to generate the same damage distribution pattern. Figures \ref{prism}f, g, and h show that two, four and five strain localization band appear in the specimen, which corresponds to the damage configuration obtained from full fine-scale problem.  The global load-displacement curve of the homogenization problem also shows five sudden drops which conforms to the full fine-scale response, see Figure \ref{stress-strain-curve}. The homogenized response captures well the displacement at which the first three localization events occur, while it underestimates its value for later events. This is likely due to the relatively coarse mesh adopted at the macroscopic scale.


\section{Conclusions} 
This paper presents the asymptotic expansion homogenization of fine-scale periodic discrete systems featuring independent translational and rotational degrees of freedom. Employing consistent asymptotic expansion of displacement and rotation fields, a rigorous analytical derivation was performed for elastic behavior, and it was extended to the nonlinear case upon making reasonable assumptions on the rigid body motions of a RVE. Based on this work, the following general conclusions can be drawn.

\begin{itemize}
\item The equivalent homogenized continuum is of Cosserat-type characterized by  nonsymmetric stress and couple tensors energetically conjugate to nonsymmetric strain and curvature tensors, respectively. The classical linear and rotational momentum balance equations can be derived from the homogenization of the fine-scale equilibrium equations.  
\item The fine-scale kinematic quantities, namely facet strains and curvatures, are demonstrated to be related to the projection of the coarse-scale strains and curvatures into the local facet system of reference. This allows a straightforward implementation of the RVE problem into any computational framework.
\item Similarly to previous research, the derived formula linking the fine-scale response to the coarse-scale stress tensor corresponds to the virial stress formulation commonly used for atomistic systems.
\item The derived formula linking the fine-scale response to the coarse-scale couple tensor is shown to consist of three terms with clear physical meaning. The first term is associated with the fine-scale couple tractions and it can be related to the facet size, which, in turn can be associated with the size of weak spots in the material internal structure. The second term arises from the moment of the fine-scale stress tractions with respect to the particle node. As such, it depends on the size of the fine-scale particles and it can be related to the spacing or characteristic distance of the weak spots in the material internal structure. Finally, the third term is the effect of the moment of fine-scale stress tractions with respect to the center of the RVE and, consequently, it depends on the RVE size.
\end{itemize}

The developed framework was then implemented in a computational software and applied to the upscaling of LDPM. Specific to this fine-scale model, the numerical results demonstrate the following interesting features of the equivalent homogenized continuum.
\begin{itemize}
\item The macro-scale elastic parameters relating the stress tensor to the strain tensor become independent on RVE size and on the random position of the polyhedral particles inside the RVE for RVE sizes larger than about 5 times the maximum spherical aggregate size. On the contrary, the macro-scale parameters relevant to the relationship between curvature and couple tensors are shown to depend on the RVE size squared and they become independent on the random position of the polyhedral particles inside the RVE for RVE sizes larger than about 5 times the maximum spherical aggregate size.
\item The non-symmetric part of the macro-scale stress tensor is negligible since the relevant parameter is at least one order of magnitude smaller than the one governing the symmetric part. As a consequence, the linear and rotational momentum balance equations are decoupled.
\item In the elastic regime the stress-strain and couple-curvature constitutive equations are completely uncoupled.
\item In the non linear regime, for tensile loading and because the fine-scale behavior is strain-softening, the response is RVE-size-dependent. This is an expected result, although very often not acknowledged by most authors in the literature, associated with strain localization induced by softening. On the contrary, such dependence is not observed for compressive dominated loading conditions because the LDPM fine-scale behavior in compression is strain-hardening. 
\item The coarse-scale couple-curvature constitutive equations scale with the square of the RVE size in the nonlinear range also but, contrarily to the elastic case, they show a strong coupling with the stress-strain constitutive equations. Such coupling, never considered in the current literature of Cosserat media, will be studied in future work by the authors.
\end{itemize}

\section{ACKNOWLEDGEMENTS}
This material is based upon work supported by the National Science Foundation under grant no. CMMI-1201087.

\newpage
\appendix
\numberwithin{equation}{section}

\section{Short Review of the Lattice Discrete Particle Model (LDPM) Geometrical Construction and Constitutive Equations}\label{LDPM}
LDPM model generation procedure and governing constitutive equations are explained in the following two sections.

\subsection{LDPM model construction}\label{LDPM-Construction}
\begin{figure}[t]
        \centering
        \begin{subfigure}[b]{0.3\textwidth}
                \centering
                \includegraphics[scale=1]{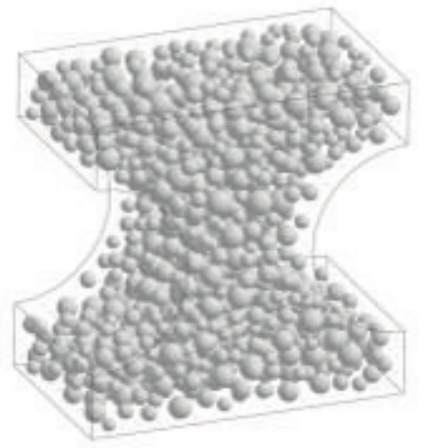}
                \caption{}
                \label{DogbonePRTC}
        \end{subfigure}
        \begin{subfigure}[b]{0.3\textwidth}
                \centering
               \includegraphics[scale=1]{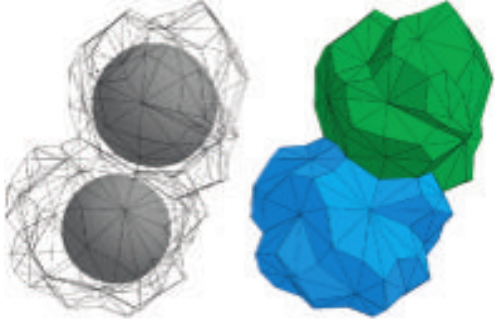}
                \caption{}
                \label{PolyCellsGeom}
        \end{subfigure}   
        \begin{subfigure}[b]{0.3\textwidth}
                \centering
                \includegraphics[scale=0.205]{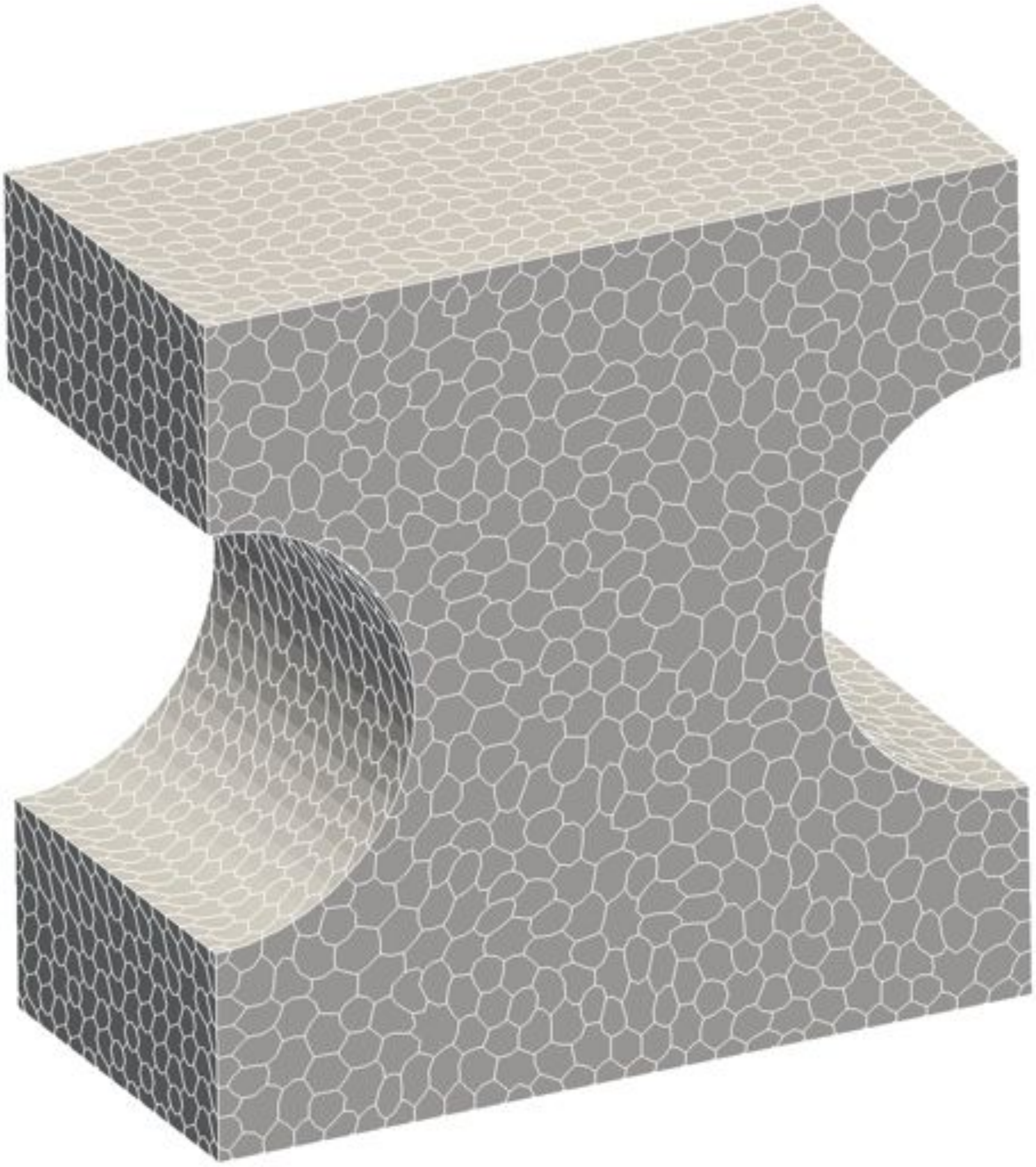}
                \caption{}
                \label{DogboneCells}
        \end{subfigure}           
        \caption{(a) Spherical aggregate system for a typical dogbone specimen. (b) LDPM polyhedral particles for two adjacent spherical aggregate particle. (c) LDPM cell distretization for a typical dogbone specimen.}
        \label{LDPMFigures}
\end{figure}

Concrete meso-scale structure is modeled by LDPM through the following steps:

\begin{itemize}
\item Spherical aggregate generation is the first step which is carried out assuming that each aggregate piece can be approximated as a sphere. Under this assumption, the following spherical aggregate size distribution function proposed by Stroeven \cite{Stroeven-1} is considered

\begin{equation}\label{psd}
f(d) = \frac{qd_0^q}{[1-(d_0/d_a)^q]d^{q+1}}
\end{equation}

in which $d_a$ and $d_0$ are the maximum and minimum spherical aggregate size, respectively, and $q$ is a material parameter. It can be shown \cite{Stroeven-1} that Equation \ref{psd} is associated with a sieve curve (percentage of spherical aggregate by weight retained by a sieve of characteristic size $d$) in the form

\begin{equation}\label{sieve}
f(d) = \bigg(\frac{d}{d_a}\bigg)^{n_f}
\end{equation}

where $n_f = 3-q$. For $n_f=0.5$ Equation \ref{sieve} corresponds to the classical Fuller curve which for its optimal packing properties, is extensively used in concrete technology. Considering concrete cement content $c$, water-to-cement ratio $w/c$, specimen volume, maximum $d_a$ and minimum $d_0$ spherical aggregate size along with the considered distribution function Equation \ref{sieve}, the spherical aggregate system can be generated using a random number generator.   

\item By using a try-and-error random procedure, spherical aggregate pieces are introduced into the concrete volume from the largest to the smallest size. Figure \ref{DogbonePRTC} shows the spherical aggregate system generated for a typical dogbone specimen. 

\item Delaunay tetrahedralization of the spherical aggregate piece centers is employed to define the interactions of the spherical aggregate system (Figure \ref{PolyCellsGeom}).

\item Finally, a three-dimensional domain tessellation anchored to the Delaunay tetrahedralization is carried out to create a system of polyhedral particles interacting through triangular facets, and a lattice system composed of the line segments connecting the spherical aggregate centers. Figure \ref{DogboneCells} shows the final polyhedral particle discretization of a typical dogbone specimen.

\end{itemize}
%

\subsection{LDPM Kinematics}\label{LDPM-Kinematics}
The triangular facets forming the rigid polyhedral particles are assumed to be the potential material failure locations. Each facet is shared between two polyhedral particle and is characterized by a unit normal vector $\bf{n}$ and two tangential vectors $\bf{m}$ and $\bf{l}$. Accordingly, three strain components are defined on each triangular facet using Equations \ref{eps} and \ref{curvature}, which for LDPM gives

\begin{equation} \label{LDPMstr}
\epsilon_{N} = \frac{\mathbf{n}^T \llbracket {\mathbf{u}_{C}} \rrbracket}{r};  \hspace{0.25 in} \epsilon_{M} = \frac{\mathbf{m}^T \llbracket {\mathbf{u}_{C}} \rrbracket}{r}; \hspace{0.25 in} \epsilon_{L} = \frac{\mathbf{l}^T \llbracket {\mathbf{u}_{C}} \rrbracket}{r}
\end{equation}

where $\llbracket {\mathbf{u}_{C}} \rrbracket$ is the displacement jump vector calculated at the facet centroid. One should consider that the LDPM constitutive equations explained in the next section are independent of facet curvatures. 

\subsection{LDPM constitutive equations}\label{LDPM-Constitutive}
This section reviews the specific constitutive equations governing the response of LDPM. First of all, it must be mentioned that LDPM assumes zero couple stresses at the meso-scale in both elastic and inelastic regime. This implies $m_\alpha=0$ for $\alpha=N,M,L$

 In the elastic regime, the normal and shear stresses are proportional to the corresponding strains: $t_{N}= E_N \epsilon_{N};~ t_{M}= E_T \epsilon_{M};~ t_{L}= E_T \epsilon_{L}$, where $E_N=E_0$, $E_T=\alpha E_0$, $E_0=$ effective normal modulus, and $\alpha=$ shear-normal coupling parameter. Beyond the elastic regime, the vectorial constitutive relations are meant to reproduce three distinct sources of nonlinearity as described below. 

\subsubsection{Fracture and cohesion due to tension and tension-shear}\label{LDPM-tens}
For tensile loading ($\epsilon_N>0$), fracturing and cohesive behavior due to tension and tension-shear are formulated through an effective strain, $\epsilon = \sqrt{\epsilon _{N}^{2}+\alpha (\epsilon _{M}^{2} + \epsilon _{L}^{2})}$, and stress, $t = \sqrt{{ t _{N}^2+  (t_{M}+t_{L})^2 / \alpha}}$, which define the normal and shear stresses as \mbox{$t _{N}= \epsilon _{N}(t / \epsilon)$}; \mbox{$t _{M}=\alpha \epsilon_{M}(t / \epsilon)$}; \mbox{$t _{L}=\alpha \epsilon_{L}(t / \epsilon)$}. The effective stress $t$ is incrementally elastic ($\dot{t}=E_0\dot{\epsilon}$) and must satisfy the inequality $0\leq t \leq \sigma _{bt} (\epsilon, \omega) $ where $\sigma_{bt} = \sigma_0(\omega) \exp \left[-H_0(\omega)  \langle \epsilon-\epsilon_0(\omega) \rangle / \sigma_0(\omega)\right]$, $\langle x \rangle=\max \{x,0\}$, and $\tan(\omega) =\epsilon _N / \sqrt{\alpha} \epsilon _{T}$ = $t_N \sqrt{\alpha} / t_{T}$. The post peak softening modulus is defined as $H_{0}(\omega)=H_{t}(2\omega/\pi)^{n_{t}}$, where $n_t$ is the softening exponent, $H_{t}$ is the softening modulus in pure tension ($\omega=\pi/2$) expressed as $H_{t}=2E_0/\left(l_t/l_e-1\right)$; $l_t=2E_0G_t/\sigma_t^2$; $l_e$ is the length of the tetrahedron edge; and $G_t$ is the mesoscale fracture energy. LDPM provides a smooth transition between pure tension and pure shear ($\omega=0$) with parabolic variation for strength given by $\sigma_{0}(\omega )=\sigma _{t}r_{st}^2\Big(-\sin(\omega)+ \sqrt{\sin^2(\omega)+4 \alpha \cos^2(\omega) / r_{st}^2}\Big)/  [2 \alpha \cos^2(\omega)]$, where $r_{st} = \sigma_s/\sigma_t$ is the ratio of shear strength to tensile strength. 
  
\subsubsection{Compaction and pore collapse from compression}\label{LDPM-comp}
For compressive loading ($\epsilon_N<0$), the normal stress evolves incrementally elastically and is subjected to the inequality $-\sigma_{bc}(\epsilon_D, \epsilon_V)\leq t_N \leq 0$ where $\sigma_{bc}$ is a strain-dependent boundary function of the volumetric strain, $\epsilon_V$, and the deviatoric strain, $\epsilon_D$. The function expressing $\sigma_{bc}$ models pore collapse for $-\epsilon_V \leq\epsilon_{c1} = \kappa_{c0} \epsilon_{c0}=\kappa_{c0} \sigma_{c0}/E_0$, and it is formulated as  $\sigma_{bc} = \sigma_{c0} + \langle-\epsilon_V-\epsilon_{c0}\rangle H_c(r_{DV})$ where $H_c(r_{DV})=H_{c0}/(1 + \kappa_{c2} \left\langle r_{DV} - \kappa_{c1} \right\rangle)$, $r_{DV}=\epsilon_D/\epsilon_V$, $\sigma_{c0}$ is the mesoscale compressive yield stress; and $\kappa_{c0}$, $\kappa_{c1}$, $\kappa_{c2}$ and $H_{c0}$ are material parameters. Compaction and rehardening occur beyond pore collapse for $-\epsilon_V \geq \epsilon_{c1}$. In this case one has $\sigma_{bc} = \sigma_{c1}(r_{DV})$ $\exp \left[( -\epsilon_{V}-\epsilon_{c1} ) H_c(r_{DV})/\sigma_{c1}(r_{DV}) \right]$ and $\sigma_{c1}(r_{DV}) = \sigma_{c0} + (\epsilon_{c1}-\epsilon_{c0}) H_c(r_{DV})$. 

\subsubsection{Friction due to compression-shear}\label{LDPM-shear}
The evolution of shear stresses simulate frictional behavior due to compression-shear. The incremental shear stresses are computed as  $\dot{t}_M=E_T(\dot{\epsilon}_M-\dot{\epsilon}_M^p)$ and \mbox{$\dot{t}_L=E_T(\dot{\epsilon}_L-\dot{\epsilon}_L^p)$}, where  \mbox{$\dot{\epsilon}_M^p=\dot{\lambda} \partial \varphi / \partial t_M$}, \mbox{$\dot{\epsilon}_L^p=\dot{\lambda} \partial \varphi / \partial t_L$}, and $\lambda$ is the plastic multiplier with loading-unloading conditions  $\varphi \dot{\lambda} \leq 0$ and $\dot{\lambda} \geq 0$. The plastic potential is defined as \mbox{$\varphi=\sqrt{t_M^2+t_L^2} - \sigma_{bs}(t_N)$}, where the nonlinear frictional law for the shear strength is assumed to be $\sigma_{bs} = \sigma_s + (\mu_0 - \mu_\infty)\sigma_{N0}[1 - \exp(t_N / \sigma_{N0})] - \mu_\infty t_N$; $\sigma_{N0}$ is the transitional normal stress; $\mu_0$ and $\mu_\infty$ are the initial and final internal friction coefficients.  

Detailed description of model behavior in the nonlinear range can be found in Ref. \cite{cusatis-ldpm-1}.

\subsection{Concrete Mix-Design and Model Parameters Used in the Numerical Simulations}
Minimum and maximum spherical aggregate size are $d_0=$ 4 mm and $d_a=$ 8 mm, respectively; cement content c = 612 $\text{kg/m}^\text{3}$; water to cement ratio w/c = 0.4; aggregate to cement ratio a/c = 2.4; Fuller curve coefficient $n_f$ = 0.42.

The following LDPM parameters are used: $E_N = 60$ GPa, $\sigma_t = 3.45$ MPa, $\sigma_{c0} = 150$ MPa, $\alpha = 0.25$, $n_t=0.4$, $l_t=500$ mm, $r_{st}=2.6$, $H_{c0}/E_0 = 0.4$, $\mu_0 = 0.4$, $\mu_\infty = 0$, $k_{c1} = 1$, $k_{c2}=5$, $\sigma_{N0} = 600$ MPa, $\alpha=E_T/E_N=0.25$.

\section{Asymptotic Expansion of Strains and Curvatures} \label{exp-strains-details}

In order to obtain multiple scale definition of facet strain vector, one should first plug macroscopic Taylor series expansion of displacement and rotation of particle $J$ around particle $I$, Eqs. \ref{taylor-1-J} and \ref{taylor-2-J}, into facet strain definition, Equation \ref{eps}. In addition, equation $\mbf{x} = \eta\mbf{y}$ is used to change the length type variables to fine-scale quantities; $\eta y^{IJ}_j = x^{IJ}_j$, $\eta \bar{c}^{I}_k = c^{I}_k$ and $\eta \bar{c}^{J}_k = c^{J}_k$. Equation \ref{eps} writes

\begin{eqnarray}\label{eps-expansion-1}
\begin{aligned}
\epsilon_{\alpha}=\eta^{-1} \bar{r}^{-1} \bigg[& u_i^J+ \eta u^J_{i,j} y^{IJ}_j + \eta^2 \frac{1}{2}u^J_{i,jk}  y^{IJ}_j y^{IJ}_k -  u_i^I  \\ 
& +\eta \varepsilon_{ijk} \bigg( \theta_j^{J}+\eta \theta_{j,m}^J y^{IJ}_m +\frac{1}{2} \eta^2 \theta^J_{j,mn} y^{IJ}_m y^{IJ}_n \bigg) \bar{c}_{k}^{J} -\eta \varepsilon_{ijk} \theta_j^I \bar{c}_{k}^{I} \bigg] e^{IJ}_{\alpha i}
\end{aligned}
\end{eqnarray}

Spatial derivatives of displacement and rotation in equation above are partial derivative with respect to $x$. So, first and second order partial derivative of displacement and rotation asymptotic expansions, Eqs. \ref{disp-expansion} and \ref{rot-expansion}, with respect to $x$ are as follows

\begin{equation}
u_{i,j} \approx u^0_{i,j} +\eta u^1_{i,j} \hspace{0.25in} u_{i,jk} \approx u^0_{i,jk} +\eta u^1_{i,jk}
\label{disp-asymp-der}
\end{equation}

\begin{equation}
\theta_{i,j} \approx \eta^{-1} \omega^0_{i,j} + \varphi^0_{i,j} +  \omega^1_{i,j} + \eta \varphi^1_{i,j} \hspace{0.25in}  \theta_{i,jk} \approx \eta^{-1} \omega^0_{i,jk} + \varphi^0_{i,jk} +  \omega^1_{i,jk} + \eta \varphi^1_{i,jk}
\label{rot-asymp-der}
\end{equation}
 
Using asymptotic expansion of displacement and rotation of a particle, Eqs. \ref{disp-expansion} and \ref{rot-expansion}, along with their macro-scale derivatives, Eqs. \ref{disp-asymp-der}, \ref{rot-asymp-der}, and replacing them into Equation \ref{eps-expansion-1}, one obtains 

\begin{eqnarray}\label{eps-expansion-2}
\begin{aligned}
\epsilon_{\alpha} = \eta^{-1} \bar{r}^{-1} \bigg[& u_i^{0J} + \eta u_i^{1J} + \eta  u^{0J}_{i,j} y^{IJ}_j + \eta^2 u^{1J}_{i,j} y^{IJ}_j + \eta^2 \frac{1}{2}u^{0J}_{i,jk} y^{IJ}_j y^{IJ}_k + \eta^3 \frac{1}{2}u^{1J}_{i,jk} y^{IJ}_j y^{IJ}_k - u^{0I}_i - \eta u^{1I}_i \\ 
& + \eta \varepsilon_{ijk} \bigg( \eta^{-1} \omega_j^{0J} + \varphi_j^{0J} + \omega_j^{1J} + \eta \varphi_j^{1J} + \omega_{j,m}^{0J} y^{IJ}_m + \eta \varphi_{j,m}^{0J} y^{IJ}_m + \eta \omega_{j,m}^{1J} y^{IJ}_m + \eta^2 \varphi_{j,m}^{1J} y^{IJ}_m \\ 
&  ~~~~~~~~~~~ + \eta \frac{1}{2} \omega^{0J}_{j,mn} y^{IJ}_m y^{IJ}_n + \eta^2 \frac{1}{2} \varphi^{0J}_{j,mn} y^{IJ}_m y^{IJ}_n + \eta^2 \frac{1}{2} \omega^{1J}_{j,mn} y^{IJ}_m y^{IJ}_n + \eta^3 \frac{1}{2} \varphi^{1J}_{j,mn} y^{IJ}_m y^{IJ}_n\bigg) \bar c_{k}^{J} \\
&  - \eta \varepsilon_{ijk} \bigg( \eta^{-1} \omega_j^{0I} + \varphi_j^{0I} + \omega_j^{1I} + \eta \varphi_j^{1I} \bigg) \bar c_{k}^{I} \bigg) \bigg] e^{IJ}_{\alpha i}
\end{aligned}
\end{eqnarray}

Regrouping terms of the same order in above equation, one would get multiple scale definition of facet strain
\begin{eqnarray}\label{eps-expansion-3}
\begin{aligned}
\epsilon_{\alpha} = \bar{r}^{-1} \bigg[ & \eta^{-1} \bigg( u_i^{0J} - u^{0I}_i + \varepsilon_{ijk} \omega_j^{0J} \bar c_{k}^{J} - \varepsilon_{ijk} \omega_j^{0I} \bar c_{k}^{I} \bigg) \\
& + \eta^0 \bigg( u_i^{1J} + u^{0J}_{i,j} y^{IJ}_j - u^{1I}_i + \varepsilon_{ijk} \bigg( \varphi_j^{0J} + \omega_j^{1J} + \omega_{j,m}^{0J} y^{IJ}_m  \bigg) \bar c_{k}^{J} -  \varepsilon_{ijk} \bigg( \varphi_j^{0I} + \omega_j^{1I} \bigg) \bar c_{k}^{I} \bigg) \\
& + \eta \bigg( u^{1J}_{i,j} y^{IJ}_j + \frac{1}{2}u^{0J}_{i,jk} y^{IJ}_j y^{IJ}_k + \varepsilon_{ijk} \bigg( \varphi_j^{1J} + \varphi_{j,m}^{0J} y^{IJ}_m + \omega_{j,m}^{1J} y^{IJ}_m + \frac{1}{2} \omega^{0J}_{j,mn} y^{IJ}_m y^{IJ}_n \bigg) \bar c_{k}^{J} \\
& ~~~~~~ - \varepsilon_{ijk} \varphi_j^{1I} \bar c_{k}^{I} \bigg)
\bigg] e^{IJ}_{\alpha i}
\end{aligned}
\end{eqnarray}

In equation above, terms of order two and higher are neglected. Multiple scale definition of facet strain is derived, which consists of three classes of terms of $\mathcal{O}(-1)$, $\mathcal{O}(0)$, and $\mathcal{O}(1)$. Multiple scale definition of facet curvature vector will be obtained subsequently. Taylor series definition of rotation of particle $J$ with respect to particle $I$ in macro coordinate system, Equation \ref{taylor-2-J}, should be inserted into definition of facet curvature, Equation \ref{curvature}

\begin{equation} \label{cur-exp-1}
\chi_{\alpha}=\eta^{-1} \bar{r}^{-1} \left[ \theta_i^J+\eta \theta_{i,j}^J y^{IJ}_j + \frac{1}{2} \eta^2 \theta_{i,jk}^J y^{IJ}_j y^{IJ}_k - \theta^I_i \right] {e}^{IJ}_{\alpha i}
\end{equation}

Asymptotic expansion of rotation, Equation \ref{rot-expansion}, along with its macroscopic first and second order derivatives, Equation \ref{rot-asymp-der}, are inserted into Equation \ref{cur-exp-1}

\begin{eqnarray} \label{cur-exp-2}
\begin{aligned}
\chi_{\alpha}= \eta^{-1} \bar{r}^{-1} \bigg[ & \eta^{-1} \omega_i^{0J} + \varphi_i^{0J} + \omega_i^{1J} + \eta \varphi_i^{1J} + \omega_{i,m}^{0J} y^{IJ}_m + \eta \varphi_{i,m}^{0J} y^{IJ}_m + \eta \omega_{i,m}^{1J} y^{IJ}_m + \eta^2 \varphi_{i,m}^{1J} y^{IJ}_m \\
& \eta \frac{1}{2} \omega^{0J}_{i,mn} y^{IJ}_m y^{IJ}_n + \eta^2 \frac{1}{2} \varphi^{0J}_{i,mn} y^{IJ}_m y^{IJ}_n + \eta^2 \frac{1}{2} \omega^{1J}_{i,mn} y^{IJ}_m y^{IJ}_n + \eta^3 \frac{1}{2} \omega^{1J}_{i,mn} y^{IJ}_m y^{IJ}_n \\
& - \eta^{-1} \omega_i^{0J} - \varphi_i^{0J} - \omega_i^{1J} - \eta \varphi_i^{1J}
\bigg] {e}^{IJ}_{\alpha i}
\end{aligned}
\end{eqnarray}

Collecting the terms of the same order and neglecting the ones of order more than zero, one can restate above equation as 

\begin{eqnarray} \label{cur-exp-3}
\begin{aligned}
\chi_{\alpha}=\bar{r}^{-1} \bigg[ & \eta^{-2} \bigg( \omega_i^{0J} - \omega_i^{0I} \bigg) +\\
& \eta^{-1} \bigg( \varphi_i^{0J}+ \omega_i^{1J} + \omega_{i,j}^{0J} y^{IJ}_j - \varphi_i^{0I} - \omega_i^{1I} \bigg) \\
& + \eta^0 \bigg(\varphi_i^{1J} + \omega_{i,j}^{1J} y^{IJ}_j + \varphi_{i,j}^{0J} y^{IJ}_j + \frac{1}{2} \omega_{i,jk}^{0J} y^{IJ}_j y^{IJ}_k - \varphi_i^{1I} \bigg)
\bigg] {e}^{IJ}_{\alpha i}
\end{aligned}
\end{eqnarray}

Equation \ref{cur-exp-3} is the multiple scale definition of facet curvature vector, which consists of terms of $\mathcal{O}(-2)$, $\mathcal{O}(-1)$, and $\mathcal{O}(0)$. 

\section{Asymptotic Expansion of Facet Strain and Curvature using definition of rigid body motion of RVE} \label{Revised-strain-curvature}
Multiple scale definition of facet strain, Equation \ref{eps-expansion-3}, and facet curvature, Equation \ref{cur-exp-3}, can be rewritten regarding the definition of $\mbf{u}^0$, Equation \ref{U0}. One can calculate first and second partial derivative of  $\mathbf{u}^0$ with respect to $\mathbf{x}$ as follows

\begin{equation}\label{UJ0-der}
u^{0J}_{i,j} = v^{0J}_{i,j} +\varepsilon_{imn} \omega ^{0J}_{m,j} y^J_n \hspace{0.5 in} u^{0J}_{i,jk} = v^{0J}_{i,jk} +\varepsilon_{imn} \omega ^{0J}_{m,jk} y^J_n
\end{equation}

Using Eqs. \ref{U0}, \ref{UJ0-der} along with the fact that $\mbf{v}^0$, $\mb{\omega}^0$ and  $\mb{\varphi}^0$ are constant over the RVE:  $\mathbf{v}^{0I} = \mathbf{v}^{0J} = \mathbf{v}^{0}$, $\mb {\omega}^{0I} = \mb{\omega}^{0J} = \mb{\omega}^{0}$ and $\mb {\varphi}^{0} = \mb{\omega}^{0}$, one can revise Equation \ref{eps-expansion-3}  

\begin{eqnarray}\label{eps-expansion-4}
\begin{aligned}
\epsilon_{\alpha}=\bar{r}^{-1} \bigg[ & \eta^{-1} \bigg( v_i^{0} - v^{0}_i + \varepsilon_{ijk} \omega_j^{0} ( \bar y_{k}^{J} - \bar y_{k}^{I} ) + \varepsilon_{ijk} \omega_j^{0} ( \bar c_{k}^{J} - \bar c_{k}^{I} ) \bigg) \\
&+ \eta^0 \bigg( u_i^{1J} - u^{1I}_i + \varepsilon_{ijk}  \omega_j^{1J} \bar c_{k}^{J} - \varepsilon_{ijk} \omega_j^{1I} \bar c_{k}^{I}  \\ 
& ~~~~~~~ + v^{0}_{i,j} y^{IJ}_j + \varepsilon_{ijk} \omega_j^{0} ( \bar c_{k}^{J} - \bar c_{k}^{I} ) + \varepsilon_{imn} \omega ^{0}_{m,j} y^{IJ}_j y^J_n + \varepsilon_{ijk}  \omega_{j,m}^{0} y^{IJ}_m \bar c_{k}^{J}  \bigg) \\
& + \eta \bigg( u^{1J}_{i,j} y^{IJ}_j + \frac{1}{2}v^{0}_{i,jk} y^{IJ}_j y^{IJ}_k + \frac{1}{2} \varepsilon_{imn} \omega^{0}_{m,jk} y^{IJ}_j y^{IJ}_k y^{J}_n + \frac{1}{2} \varepsilon_{ijk} \omega^{0J}_{j,mn} y^{IJ}_m y^{IJ}_n  \bar c_{k}^{J} \\
& ~~~~~~ + \varepsilon_{ijk} \bigg( \varphi_j^{1J} + \omega_{j,m}^{0} y^{IJ}_m + \omega_{j,m}^{1J} y^{IJ}_m  \bigg) \bar c_{k}^{J} 
- \varepsilon_{ijk} \varphi_j^{1I} \bar c_{k}^{I} \bigg) \bigg] e^{IJ}_{\alpha i}
\end{aligned}
\end{eqnarray}

Using $\mathbf{y}^{IJ} = \mathbf{y}^{J} - \mathbf{y}^{I}$ and $\mathbf{y}^{IJ} = \mathbf{\bar c}^{I} - \mathbf{\bar c}^{J}$ in above equation along with $\mbf{y}^J+\mbf{\bar c}^J=\mbf{y}^c$  , one would get

\begin{eqnarray}\label{eps-expansion-5}
\begin{aligned}
\epsilon_{\alpha}=\bar{r}^{-1} \bigg[ & \eta^0 \bigg( u_i^{1J} - u^{1I}_i + \varepsilon_{ijk}  \omega_j^{1J} \bar c_{k}^{J} - \varepsilon_{ijk} \omega_j^{1I} \bar c_{k}^{I}  + v^{0}_{i,j} y^{IJ}_j - \varepsilon_{ijk} \omega_j^{0} y^{IJ}_k + \varepsilon_{ijk}  \omega_{j,m}^{0} y^{IJ}_m y_{k}^{c} \bigg) \\
& + \eta \bigg( u^{1J}_{i,j} y^{IJ}_j + \varepsilon_{ijk} \varphi_j^{1J} \bar c_{k}^{J} + \varepsilon_{ijk} \omega_{j,m}^{1J} y^{IJ}_m \bar c_{k}^{J} 
- \varepsilon_{ijk} \varphi_j^{1I} \bar c_{k}^{I} \\
& ~~~~~~ + \frac{1}{2}v^{0}_{i,jk} y^{IJ}_j y^{IJ}_k + \frac{1}{2} \varepsilon_{ijk} \omega^{0}_{j,mn} y^{IJ}_m y^{IJ}_n  y_{k}^{c} + \varepsilon_{ijk} \omega_{j,m}^{0} y^{IJ}_m \bar c_{k}^{J}  \bigg) \bigg] e^{IJ}_{\alpha i}
\end{aligned}
\end{eqnarray}

Multiple scale definition of facet curvature can also be revised by using $\mb {\omega}^{0I} = \mb{\omega}^{0J} = \mb{\omega}^{0}$ and $\mb {\varphi}^{0} = \mb{\omega}^{0} $ along with $\mathbf{y}^{IJ} = \mathbf{y}^{J} - \mathbf{y}^{I}$ , one can rewrite Equation \ref{cur-exp-3}

\begin{eqnarray} \label{cur-exp-4}
\begin{aligned}
\chi_{\alpha}=\bar{r}^{-1} \bigg[ & \eta^{-1} \bigg( \omega_i^{1J} + \omega_{i,j}^{0J} y^{IJ}_j - \omega_i^{1I} \bigg) \\
& + \eta^0 \bigg(\varphi_i^{1J} + \omega_{i,j}^{1J} y^{IJ}_j - \varphi_i^{1I} + \omega_{i,j}^{0} y^{IJ}_j + \frac{1}{2} \omega_{i,jk}^{0J} y^{IJ}_j y^{IJ}_k \bigg)
\bigg] {e}^{IJ}_{\alpha i}
\end{aligned}
\end{eqnarray}

\section{Macroscopic Translational and Rotational Equilibruim Equations} \label{MacroEquil-Derivation}
In order to derive macroscopic RVE translational equation of motion, one should consider the terms of $\mathcal{O}(1)$ in Equation \ref{motion-1-sep}

\begin{equation}\label{macro-trans-derv-1}
\bar{M}_u^I\ddot {u}_i^{0I}  = \sum_{\mathcal{F}_I}{\bar{A}\, t^1_{\alpha} {e}_{\alpha i}^{IJ}} + \bar{V}^I b_i^0
\end{equation}

Scaling back Equation \ref{macro-trans-derv-1} by multiplying both sides of the equation by $\eta^3$ and using the definition of $t^1_{\alpha}$ presented in Equation \ref{ZeroOne-Terms-Def}, one can get

\begin{equation} \label{macro-trans-derv-2}
{M}_u^I \ddot {u}_i^{0I} = \eta \sum_{\mathcal{F}_I}{{A} \frac {\partial {t}^{IJ}_{i}}{\partial \epsilon^0_{\alpha}} \epsilon^1_{\alpha}} + {V}^I {b}^{0}_i
\end{equation}

where $t_i^{IJ}=t^0_\beta e_{\beta i}^{IJ}$. Equation \ref{macro-trans-derv-2} represents the $\mathcal{O}(1)$ translational equilibrium equation for each particle inside the RVE. One can derive the RVE macroscopic translational equilibrium equation by summing up Equation \ref{macro-trans-derv-2} over all RVE particles and dividing by the RVE volume $V_0$

\begin{equation} \label{macro-trans-derv-3}
\frac{1}{V_0}\sum_I {M}_u^I (\ddot {v}_i^{0I} + \varepsilon_{imn} \ddot {\omega} ^{0I}_{m} y^I_n ) = \frac{1}{V_0}\sum_I \sum_{\mathcal{F}_I}{\eta A \frac {\partial {t}^{IJ}_{i}}{\partial \epsilon^0_{\alpha}} \epsilon^1_{\alpha}} + \frac{1}{V_0}\sum_I {V}^I {b}^{0}_i
\end{equation}

In above equation ${u}_i^{0I}$ is replaced by its definition, Equation \ref{U0}. Considering the fact that ${v}_i^{0I}$ and ${\omega}_m^{0I}$ are equal for all RVE particles and the body force $b_i^0$ is considered to be constant over the RVE, Equation \ref{macro-trans-derv-3} can be written as

\begin{equation} \label{macro-trans-derv-4}
\ddot {v}_i^{0} \bigg(\frac{1}{V_0}\sum_I {M}_u^I \bigg) + \varepsilon_{imn} \ddot {\omega} ^{0}_{m} \bigg(\frac{1}{V_0}\sum_I {M}_u^I y^I_n \bigg) = \frac{1}{V_0}\sum_I \sum_{\mathcal{F}_I}{\eta A \frac {\partial {t}^{IJ}_{i}}{\partial \epsilon^0_{\alpha}} \epsilon^1_{\alpha}} + {b}^{0}_i \bigg(\frac{1}{V_0}\sum_I {V}^I \bigg)
\end{equation}

Second term on the left hand side of the Equation \ref{macro-trans-derv-4} is equal to zero considering the assumption that the local system of reference is the mass center of the particle system within the RVE; $\sum_I {M}_u^I y^{I}_i=0$. Final form of the Equation \ref{macro-trans-derv-4} is presented in Equation \ref{macro-1-1-averaged} in Section \ref{macro-derivation}.

Macroscopic RVE rotational equation of motion can be derived by considering the terms of $\mathcal{O}(1)$ in Equation \ref{motion-2-sep}. To have a consistent formulation for all particles and RVEs, one should consider the moment of all forces  with respect to a fixed point in space, say the origin of a global coordinate system as shown in Figure \ref{TwoScaleAnalysis}b,  which implies that the moment of Equation \ref{macro-trans-derv-1} should be taken into account. Therefore, one can write $\mathcal{O}(1)$ moment equilibrium equation of particle $I$ as 

\begin{equation} \label{macro-rot-derv-1}
\bar{M}_u^I \varepsilon_{ijk} Y^{I}_j \ddot {u}_k^{0I} + \bar {M_\theta^I}\ddot{\omega}_i^{0I} = \sum_{\mathcal{F}_I} \bar{A}\, (p^{1}_{\alpha}{e}_{\alpha i}^{IJ} + q^{1}_{\alpha}{e}_{\alpha i}^{IJ}) + \bar{V}^I \varepsilon_{ijk} Y^{I}_j {b}_k^{0} 
\end{equation}

where $Y^I_j$ is the position vector of particle $I$ in the fine-scale global coordinate system $\mathbf{Y}=\mathbf{X}/\eta$;  $p^{1}_{\alpha} {\mathbf e}_\alpha^{IJ} = {\mathbf{Y}}^C \times t^{1}_{\alpha} {\mathbf e}_\alpha^{IJ}$ is the moment of the facet traction with respect to the origin of the fine-scale global coordinate system, in which ${\mathbf{Y}}^C = {\mathbf{X}}^C/\eta$ is the position vector of the contact point $C$ between particles $I$ and $J$ in the global coordinate system. Scaling back Equation \ref{macro-rot-derv-1} by multiplying both sides of the equation by $\eta^4$ and using the definition of $p^1_{\alpha}$ and $q^1_{\alpha}$ presented in Equation \ref{ZeroOne-Terms-Def}, one can get

\begin{equation} \label{macro-rot-derv-2}
{M}_u^I \varepsilon_{ijk} X^{I}_j  \ddot {u}_k^{0I} + \eta^{-1}  {M_\theta^I} \ddot{\omega}_i^{0I} = \eta \sum_{\mathcal{F}_I} A \left( {\frac {\partial {w}_{i}^{IJ}}{\partial \epsilon^0_{\alpha}} \epsilon^1_{\alpha}} + \frac {\partial {m}_{i}^{IJ}}{\partial \psi^{0}_{\alpha}} \psi^1_{\alpha} \right) + {V}^I \varepsilon_{ijk} X^{I}_j {b}_k^{0}
\end{equation}

Equation \ref{macro-rot-derv-2} represents the $\mathcal{O}(1)$ rotational equilibrium equation for each particle inside the RVE. RVE macroscopic rotational equilibrium equation can be obtained by summing up Equation \ref{macro-rot-derv-2} over all RVE particles and dividing by the RVE volume $V_0$

\begin{equation} \label{macro-rot-derv-3}
\begin{aligned}
\frac{1}{V_0}\sum_I {M}_u^I \varepsilon_{ijk} X^{I}_j (\ddot {v}_k^{0I} + & \varepsilon_{kmn} \eta^{-1} \ddot {\omega} ^{0I}_{m} x^I_n )  + \frac{1}{V_0}\sum_I \eta^{-1}  {M_\theta^I} \ddot{\omega}_i^{0I} = \\
& \frac{\eta }{V_0}\sum_I \sum_{\mathcal{F}_I} A \left( {\frac {\partial {w}_{i}^{IJ}}{\partial \epsilon^0_{\alpha}} \epsilon^1_{\alpha}} + \frac {\partial {m}_{i}^{IJ}}{\partial \psi^{0}_{\alpha}} \psi^1_{\alpha} \right) + \frac{1}{V_0}\sum_I {V}^I \varepsilon_{ijk} X^{I}_j {b}_k^{0}
\end{aligned}
\end{equation}

In above equation ${u}_i^{0I}$ is replaced by its definition, Equation \ref{U0}. Considering equality of ${v}_i^{0I}$ and ${\omega}_m^{0I}$ for all RVE particles along with $X^I_j =  X_j + x^I_j$, Equation \ref{macro-rot-derv-3} can be written as
 
\begin{equation} \label{macro-rot-derv-4}
\begin{aligned}
& \frac{1}{V_0}\sum_I {M}_u^I \varepsilon_{ijk} X_j \ddot {v}_k^{0} + \frac{1}{V_0}\sum_I \left( {M_\theta^I} \delta_{im}+ {M}_u^I \varepsilon_{ijk} \varepsilon_{kmn} x^I_j  x^I_n\right) \eta^{-1}  \ddot{\omega}_m^{0} \\
& + \frac{1}{V_0}\sum_I {M}_u^I \varepsilon_{ijk} x^I_j \ddot {v}_k^{0} + \frac{1}{V_0}\sum_I {M}_u^I \varepsilon_{ijk} \varepsilon_{kmn} X_j x^I_n  \eta^{-1} \ddot {\omega} ^{0}_{m} =  \\ 
& \frac{\eta }{V_0}\sum_I \sum_{\mathcal{F}_I} A \left( {\frac {\partial {w}_{i}^{IJ}}{\partial \epsilon^0_{\alpha}} \epsilon^1_{\alpha}} + \frac {\partial {m}_{i}^{IJ}}{\partial \psi^{0}_{\alpha}} \psi^1_{\alpha} \right) + \frac{1}{V_0}\sum_I {V}^I \varepsilon_{ijk} X_j {b}_k^{0} + \frac{1}{V_0}\sum_I {V}^I \varepsilon_{ijk} x^I_j {b}_k^{0}
\end{aligned}
\end{equation}

Considering $\sum_I {M}_u^I x^{I}_i=0$ along with the equality of ${v}_i^{0}$, ${\omega}_m^{0}$, $X_j$ for all RVE particles, one can conclude that the third and the forth terms on the left hand side and the last term on the right hand side of the Equation \ref{macro-rot-derv-4} is equal to zero. Final form of the Equation \ref{macro-rot-derv-4} is presented in Equation \ref{macro-2-averaged-init} in Section \ref{macro-derivation}.

\end{document}